\documentclass[aps,prc,reprint,nofootinbib]{revtex4-1}

\usepackage{amsmath,amssymb,slashed}
\usepackage{dsfont}
\usepackage{float}
\usepackage{graphicx}
\usepackage{subcaption}
\usepackage{epstopdf}  
\usepackage{slashed}
\usepackage{hyperref}

\begin{document}

\title{The electromagnetic form factors of the transition from the spin-3/2 Sigma to the Lambda hyperon}
\author{Olov Junker}
\author{Stefan Leupold}
\author{Elisabetta Perotti}
\author{Timea Vitos}

\affiliation{Institutionen f\"or fysik och astronomi, Uppsala Universitet, Box 516, S-75120 Uppsala, Sweden}
\date{\today}

%
%
%

\begin{abstract}
The three electromagnetic form factors for the transition from a $3/2^+$ $\Sigma^*$ hyperon to the ground-state $\Lambda$ hyperon 
are studied.
At low energies, combinations of the transition form factors can be deduced from Dalitz decays of the $\Sigma^*$ hyperon to 
$\Lambda$ plus an electron-positron pair. It is pointed out how more information can be obtained 
with the help of the self-analyzing weak decay of the $\Lambda$. In particular it is shown that these transition form factors are complex quantities already in this kinematical region. 
Such measurements are feasible at hyperon factories as for instance the Facility for Antiproton and Ion Research (FAIR).
At higher energies, the transition form factors can be measured in electron-positron collisions. 
The pertinent relations between the transition form factors and the decay distributions and differential cross sections 
are presented.
Using dispersion theory, the low-energy electromagnetic form factors for the $\Sigma^*$-to-$\Lambda$ transition are related 
to the pion vector 
form factor. The additionally required input, i.e.\ the two-pion--$\Sigma^*$--$\Lambda$ amplitudes 
are determined from relativistic next-to-leading-order (NLO) baryon chiral perturbation theory 
including the baryons from the octet and the decuplet. A poorly known NLO parameter is fixed to the experimental value of the $\Sigma^*\to\Lambda\gamma$ decay width. 
Pion rescattering is taken into account by dispersion theory solving a Muskhelishvili-Omn\`es equation. 
Subtracted and unsubtracted dispersion relations are discussed. However, in view of the fact that the transition form factors are complex quantities, the current data situation does not allow for a full determination of the subtraction constants.  To reduce the number of free parameters, unsubtracted dispersion relations are used to make predictions for the transition form factors in the low-energy space- and timelike regions. 
\end{abstract}
%

\maketitle

\section{Introduction and summary}
\label{sec:intro}

Electromagnetic form factors have become an important tool to study the structure of strongly interacting objects, see e.g.\   
\cite{Landsberg:1986fd,Czerwinski:2012ry,Miller:2007uy,Punjabi:2015bba,Korner:1976hv,Carlson:1985mm,%
Pascalutsa:2006up,Kaxiras:1985zv,Kubis:2000aa,Ablikim:2019vaj,Granados:2017cib,Alarcon:2017asr} and references therein. 
Depending on the invariant mass of the virtual photon, one achieves different 
resolutions and different degrees of freedom become relevant. At very large energies, one ``sees'' the minimal quark content of 
the probed object \cite{Brodsky:1974vy,Lepage:1979zb,Carlson:1985mm}. Asymptotic freedom causes a suppression of the influence
of any non-minimal quark or gluon content of the probed state. At low energies, the dynamics of pions has an important influence
on the shape of form factors. Dynamical chiral symmetry breaking causes the appearance of 
Goldstone bosons \cite{Scherer:2012xha}, the pions. Because they are much lighter than any other hadron, pions can be excited 
with energies so low that all other degrees of freedom are still frozen. Both aspects, dominance of minimal quark content at high
energies and universal pion dynamics at low energies, are model-independent consequences of Quantum Chromodynamics (QCD). 
A complete description of a form factor must include both of these aspects.

The purpose of \cite{Granados:2017cib} and of the present and future work is to provide the low-energy input for such a complete 
description of form factors in the hyperon sector. Here we extend previous work of the 
Uppsala group \cite{Granados:2017cib,Leupold:2017ngs} where dispersion theory is used to relate in a model-independent way 
isovector form factors of baryons to pion-baryon scattering amplitudes. In a second step, these scattering amplitudes are 
approximated by relativistic chiral perturbation theory ($\chi$PT) including the baryon octet and decuplet as active degrees 
of freedom. Conceptually this is close in spirit to \cite{Alarcon:2017asr,Alarcon:2017ivh,Alarcon:2017lhg,Alarcon:2018irp}. 
On a more technical level, the rescattering of pions 
is treated differently in \cite{Alarcon:2017asr} and in \cite{Granados:2017cib}. In \cite{Alarcon:2017asr} an N/D method is used;
in \cite{Granados:2017cib} a Muskhelishvili-Omn\`es (MO) equation is solved. As has been demonstrated 
in \cite{Leupold:2017ngs} for the nucleon case, solving an MO equation with input from $\chi$PT 
up to next-to-leading order leads to 
better results when compared to a fully dispersive calculation \cite{Hoferichter:2015hva,Hoferichter:2016duk}. 
For the case of hyperons, the use of dispersively reconstructed pion-baryon amplitudes is not an option because there
are no direct data on pion-hyperon scattering. Therefore, we rely also in the present work on input from $\chi$PT \cite{Jenkins:1991es,Dashen:1993as,Pascalutsa:2006up,Ledwig:2014rfa,Holmberg:2018dtv} and solve an MO equation. 
A combined use of dispersion theory and $\chi$PT has been pioneered in \cite{Donoghue:1990xh}; 
see also \cite{Donoghue:1996kw} for a brief review.

A general motivation for studying hyperon form factors has been provided in \cite{Granados:2017cib} in great detail. 
With the present work, we extend the approach of \cite{Granados:2017cib} to electromagnetic form factors of hyperons 
with spin 3/2. 
Our framework is suited for the determination of isovector form factors. Therefore we focus in the present work on the only 
electromagnetic form factors of hyperons that are purely isovector (and involve a spin 3/2 state). 
These are the form factors for the transition of the lowest lying spin 3/2 decuplet state $\Sigma^{*0}$ to the 
spin 1/2 ground state $\Lambda$. 

In the timelike region, these transition form factors (TFFs) can be measured at low energies via the Dalitz decay 
$\Sigma^{*0} \to \Lambda \, e^+ e^-$. 
It can be expected that these Dalitz decays will be addressed in the near future by the collaborations 
HADES \cite{Ramstein:2019kaz} and PANDA \cite{Lutz:2009ff}
at the Facility for Antiproton and Ion Research (FAIR). Therefore we regard our present work as very timely. 

Concerning the $\Sigma^{*}$-$\Lambda$ transition, two distinct qualitative aspects are noteworthy; one is more case specific, 
one is universal. We start with the latter. Whenever pions are excited, they rescatter. In the isovector channel, the p-wave 
pion phase shift shows a relatively broad, essentially elastic resonance, 
the $\rho$ meson \cite{Colangelo:2001df,GarciaMartin:2011cn}. Thus, the universal pion dynamics gives rise to the coupling of the
virtual photon to the $\rho$-meson. Phenomenologically, this is covered by the concept of vector meson 
dominance \cite{sakuraiVMD}. In the dispersive framework this is covered by the pion phase shift and the pion vector form factor.
We will explore the quantitative importance of these effects in the present paper.

There is a second aspect, however, which is also covered by our dispersive framework, but is typically missing in a vector meson
dominance approach. This is the aspect that we called ``more case specific''. 
Being resonances, the $\Sigma^*$ hyperons are unstable.
In particular, the $\Sigma^{*0}$ can decay to $\Sigma^\pm \, \pi^\mp$. This pair can rescatter into a $\Lambda$ and a real or 
virtual photon. Therefore, the TFFs are complex quantities in {\em all} kinematically allowed regimes:
in the spacelike scattering region of $\Sigma^{*0} \, e^- \leftrightarrow \Lambda \, e^-$; 
at the photon point $\Sigma^{*0} \to \Lambda \, \gamma$; in the low-energy timelike Dalitz decay region of 
$\Sigma^{*0} \to \Lambda \, e^+ e^-$; and in the high-energy production region of $e^+ e^- \to \Sigma^{*0} \, \bar \Lambda$. 
This is in contrast to TFFs for hadrons that are stable with respect to the strong interaction. 
For stable hadrons, the TFFs are essentially real in all regimes except for the 
production region. 

Complex form factors allow for non-trivial 
interference patterns between them. Those can be measured, e.g., with the help of the 
self-analyzing weak decays of the ``stable'' hyperons. In practice, this means that in the succession of the two decays
$\Sigma^{*0} \to \Lambda \, e^+ e^-$ and $\Lambda \to p \, \pi^-$, the angular distribution of the second decay contains 
interesting information about the interference of the TFFs. This information is accessible without involving 
the production process or the spin orientation of the $\Sigma^{*0}$ and without determining the spin orientation of the 
proton \cite{Faldt:2016qee,Perotti:2018wxm}. On the other hand, in a strict vector meson dominance scenario, the $\Sigma^{*0}$ 
couples just via a pointlike interaction to $\rho \, \Lambda$. There, a form factor can only become complex where the $\rho$ 
becomes unstable. This happens essentially only above the two-pion threshold. But in the Dalitz decay region of 
$\Sigma^{*0} \to \Lambda \, e^+ e^-$, the maximally possible dielectron invariant mass (``$\rho$-meson invariant mass'') is 
$m_{\Sigma^*}-m_\Lambda < 2 m_\pi$ \cite{pdg}. 
Thus in reality, the TFFs are complex but in a simple vector meson dominance
approach they are real in the Dalitz decay region. 
We will also explore the quantitative importance of these effects. 
One peculiarity we observe is that even if the imaginary part of a TFF at the photon point is very small, it gets larger for the  transition radius.

Ideally we would like to use subtracted dispersion relations, but the available experimental input  is too scarce to allow for it. For the case at hand there are three TFFs and therefore three complex valued subtraction constants. For the time being we choose to use unsubtracted dispersion relations; being aware of the large uncertainties they carry, we still expect to obtain results of the correct order of magnitude. 

In the first part of the paper we define the $\Sigma^{*0}$-$ \Lambda$ TFFs and relate them to several observables, accessible in different kinematical regions. Directly after we enter the core of the theoretical work: we derive the appropriate dispersion relations for pion-hyperon scattering amplitudes and TFFs. Finally the results are presented. There are several appendices with various purposes. Appendix \ref{sec:disc} clarifies how the individual contributions of meson and baryon dynamics influence the final results. The others complement the main text with technical details.

\section{Transition form factors and observables}
\label{sec:tffobs}

Following essentially \cite{Korner:1976hv} we define three TFFs via
\begin{equation}
  \label{eq:defTFF}
  \langle 0 \vert j_\mu \vert \Sigma^*  \bar\Lambda \rangle 
  = e \, \bar v_\Lambda(p_\Lambda,\lambda) \, \Gamma_{\mu\nu}(p_{\Sigma^*},p_{\Lambda}) \, u_{\Sigma^*}^\nu(p_{\Sigma^*},\sigma) 
\end{equation}
with
\begin{eqnarray}
  \Gamma^{\mu\nu}(p_{\Sigma^*},p_{\Lambda}) 
  & := & -(\gamma^\mu q^\nu - \not \!q \, g^{\mu\nu}) \, m_{\Sigma^*} \, \gamma_5 \, F_1(q^2)   \nonumber \\ && {}
  + (p_{\Sigma^*}^\mu q^\nu - p_{\Sigma^*} \cdot q \, g^{\mu\nu}) \, \gamma_5 \, F_2(q^2)   \nonumber \\ && {}
  + (q^\mu q^\nu - q^2 \, g^{\mu\nu}) \, \gamma_5 \, F_3(q^2) 
  \label{eq:defTFF2}
\end{eqnarray}
and $q:=p_{\Sigma^*}+p_\Lambda$. Conventions for the spin-3/2 spinor $u^\mu$ are provided in Appendix \ref{sec:proj}. 
The neutral spin-3/2 Sigma hyperon is denoted by $\Sigma^*$. 
The helicities (not spins!) of $\Sigma^*$ and $\bar\Lambda$ are called $\sigma$ and $\lambda$, respectively.

The TFFs defined via (\ref{eq:defTFF}) are appropriate for a dispersive representation where we study formally the 
reaction $\Sigma^* \bar\Lambda \to \pi^+ \pi^- \to \gamma^*$. Physically, however, we study the reactions 
$e^+ e^- \to \gamma^* \to \bar\Sigma^* \Lambda$ and $\Sigma^* \to \Lambda \, \gamma^* \to \Lambda \; e^+ e^-$. In addition,
if one wants to compare the results of the electromagnetic form factors for the transition $\Sigma^* \to \Lambda$ with the ones 
for $\Delta \to N$ it is convenient to adapt to the conventions used in the $\Delta$ sector where mostly electroproduction 
is studied \cite{Carlson:1985mm,Pascalutsa:2006up} and not Dalitz decays. Thus one should also look at the reaction 
$e^- \Lambda \to e^- \Sigma^*$ or more formally $\Lambda \, \gamma^* \to \Sigma^*$. Therefore we present the transition form 
factors also for other kinematical regimes. 

In principle, the reactions $\Sigma^* \bar\Lambda \to \gamma^*$, 
$\gamma^* \to \bar\Sigma^* \Lambda$ and $\Sigma^* \to \Lambda \, \gamma^*$ are related by crossing symmetry. 
For $\Lambda \, \gamma^* \to \Sigma^*$ one might involve charge conjugation and then again crossing symmetry. 

For the amplitude relevant for the Dalitz decay, $\Sigma^* \to \Lambda \, \gamma^*$, one finds
\begin{equation}
  \label{eq:defTFFDalitz}
  \langle \Lambda \vert j_\mu \vert \Sigma^* \rangle 
  = e \, \bar u_\Lambda(p_\Lambda,\lambda) \, \Gamma_{\mu\nu}(p_{\Sigma^*},-p_{\Lambda}) \, u_{\Sigma^*}^\nu(p_{\Sigma^*},\sigma) \,.
\end{equation}
In practice this leads to the very same expression as on the right-hand side of \eqref{eq:defTFF2} but with 
$q:=p_{\Sigma^*}-p_\Lambda$. 

For the production amplitude $\gamma^* \to \bar\Sigma^* \Lambda$ one has to specify the meaning of the two-fermion bra state:
\begin{equation}
  \label{eq:subtle2f}
  \langle \bar\Sigma^*  \Lambda \vert := \vert \bar\Sigma^*  \Lambda \rangle^\dagger \,.
\end{equation}
The structure corresponding to \eqref{eq:defTFF2} would be $\Gamma^{\mu\nu}(-p_{\Sigma^*},-p_{\Lambda})$, but it is not convenient to 
define $q$ as $-p_{\Sigma^*}-p_{\Lambda}$. Therefore we rather provide a fully explicit version of the TFFs adapted to the 
production process:
\begin{equation}
  \label{eq:defTFFtinv}
  \langle \bar\Sigma^*  \Lambda \vert j_\mu \vert 0 \rangle 
  = -e \, \bar u_\Lambda(p_\Lambda,\lambda) \, \tilde\Gamma_{\mu\nu}(p_{\Sigma^*},p_{\Lambda}) \, v_{\Sigma^*}^\nu(p_{\Sigma^*},\sigma)
  \,,
\end{equation}
\begin{eqnarray}
  \tilde\Gamma^{\mu\nu}(p_{\Sigma^*},p_{\Lambda})
  &:=&  (\gamma^\mu q^\nu - \not \!q \, g^{\mu\nu}) \, m_{\Sigma^*} \, \gamma_5 \, F_1(q^2)   \nonumber \\ && {}
  + (p_{\Sigma^*}^\mu q^\nu - p_{\Sigma^*} \cdot q \, g^{\mu\nu}) \, \gamma_5 \, F_2(q^2)   \nonumber \\ && {}
  + (q^\mu q^\nu - q^2 \, g^{\mu\nu}) \, \gamma_5 \, F_3(q^2) 
  \label{eq:defTFF2prod}
\end{eqnarray}
with $q:=p_{\Sigma^*}+p_\Lambda$.

Finally we obtain for the excitation process:
\begin{equation}
  \label{eq:defTFFelprod}
  \langle \Sigma^*  \vert j_\mu \vert \Lambda \rangle 
  = -e \, \bar u_{\Sigma^*}^\nu(p_{\Sigma^*},\sigma) \, \tilde\Gamma_{\mu\nu}(p_{\Sigma^*},-p_{\Lambda}) \, u_\Lambda(p_\Lambda,\lambda) \,.
\end{equation}
Here the pertinent expression for $\tilde\Gamma^{\mu\nu}$ agrees with the right-hand side of \eqref{eq:defTFF2prod} provided one 
defines $q:=p_{\Sigma^*}-p_\Lambda$.

Next we introduce linear combinations of $F_1$, $F_2$ and $F_3$, which correspond to TFFs with fixed helicity combinations.
We denote them by $G_m$ ($m=\sigma-\lambda=0,\pm 1$) and find:
\begin{eqnarray}
  G_{-1}(q^2) &:=&  (-m_\Lambda (m_\Lambda+m_{\Sigma^*}) + q^2) \, F_1(q^2)  \nonumber \\ && {}
  +\frac12 \, (m_{\Sigma^*}^2-m_\Lambda^2+q^2) \, F_2(q^2) + q^2 \, F_3(q^2) \nonumber \\ &&  \mbox{for}
  \quad \sigma=-\frac12 \,, \lambda=+\frac12  \,,
  \label{eq:F-1def}
\end{eqnarray}
\begin{eqnarray}
  G_{0}(q^2) &:=& m^2_{\Sigma^*} \, F_1(q^2) + m^2_{\Sigma^*} \, F_2(q^2) \nonumber \\ && {}
  + \frac12 \, (m_{\Sigma^*}^2-m_\Lambda^2+q^2) \, F_3(q^2)   \nonumber \\ &&  \mbox{for}
  \quad \sigma=+\frac12 \,, \lambda=+\frac12 \,,
  \label{eq:F0def}
\end{eqnarray}
and 
\begin{eqnarray}
  G_{+1}(q^2) &:=& m_{\Sigma^*} (m_\Lambda+m_{\Sigma^*}) \, F_1(q^2) \nonumber \\ && {}
  + \frac12 \, (m_{\Sigma^*}^2-m_\Lambda^2+q^2) \, F_2(q^2) + q^2 \, F_3(q^2)   \nonumber \\ &&   \mbox{for}
  \quad \sigma=+\frac32 \,, \lambda=+\frac12 \,.
  \label{eq:F+1def}
\end{eqnarray}

In the following we adopt the reference frame from \cite{Granados:2017cib} where the virtual photon is at rest, i.e.\ the $\Sigma^*$-$\bar\Lambda$ center-of-mass system,
and where the $\Sigma^*$ is moving in the $z$-direction.
In this frame the three-momentum of the $\Sigma^*$ is given by $\vec p_{\Sigma^*} = p_z \, \vec e_z$ with
\begin{eqnarray}
  \label{eq:defpz}
  p_z = \frac{\sqrt{\lambda(q^2,m_{\Sigma^*}^2,m_{\Lambda}^2)}}{2\sqrt{q^2}}
\end{eqnarray}
where we have introduced the K\"all\'en function
\begin{eqnarray}
  \label{eq:kallenfunc}
  \lambda(a,b,c):=a^2+b^2+c^2-2(ab+bc+ac) \,.
\end{eqnarray}

We find
\begin{eqnarray}
  \label{eq:expect0}
  && \bar v_\Lambda(-p_z,1/2) \, \Gamma_{3 \, \nu} \, u_{\Sigma^*}^\nu(p_z,+1/2) \nonumber \\
  &&  = \bar v_\Lambda \gamma_5 u^3_{\Sigma^*} \, \frac{2 \, q^2}{m_{\Sigma^*}^2-m_\Lambda^2+q^2} \, G_{0}(q^2)   \,,
\end{eqnarray}
\begin{equation}
  \label{eq:expect-1}
  \bar v_\Lambda(-p_z,1/2) \, \Gamma_{1 \, \nu} \, u_{\Sigma^*}^\nu(p_z,-1/2) = \bar v_\Lambda \gamma_5 u^1_{\Sigma^*} 
  \,  G_{-1}(q^2)   \,, \phantom{m}
\end{equation}
\begin{equation}
  \label{eq:expect+1}
  \bar v_\Lambda(-p_z,1/2) \, \Gamma_{1 \, \nu} \, u_{\Sigma^*}^\nu(p_z,+3/2) = \bar v_\Lambda \gamma_5 u^1_{\Sigma^*} 
  \,  G_{+1}(q^2)   \,.  \phantom{m}
\end{equation}
The spinors on the right-hand side are evaluated with the same arguments as on the respective left-hand side. Note that in these
relations the explicit ``photon'' indices 3 and 1 are covariant, not contravariant as it is the case for the corresponding 
relations in \cite{Granados:2017cib}. This will lead to a sign change in (\ref{eq:def-redampl}) below as compared to the
conventions of \cite{Granados:2017cib}.

To make further contact with the existing literature, we relate our TFFs to the ones introduced in \cite{Carlson:1985mm}.
Therein, the transition from nucleon to $\Delta$ is considered. 
We replace $\Delta \to \Sigma^*$ and $N \to \Lambda$ to obtain our 
case at hand. The conventions for this process are provided in \eqref{eq:defTFFelprod}. 
It is convenient to define $Q^2:=-q^2$. Since one studies now reactions with $Q^2 >0$, it is meaningful to introduce 
also $Q:=\sqrt{Q^2}$. 
The TFFs of Carlson \cite{Carlson:1985mm} (in the following labeled with ``Ca'') are related to our 
TFFs by\footnote{There is a mismatch between the conventions used in \cite{Carlson:1985mm} and here. This is 
essentially based on the fact that we introduce our TFFs via the coupling of a virtual timelike photon to a spin-3/2 
baryon and a spin-1/2 antibaryon where the latter has helicity $+1/2$; see \eqref{eq:expect0}-\eqref{eq:expect+1}. 
In \cite{Carlson:1985mm} the TFFs are introduced 
via the coupling of a virtual spacelike photon to an incoming spin-1/2 baryon and an outgoing spin-3/2 baryon. The former has 
helicity $+1/2$. If one translates our case to the one in \cite{Carlson:1985mm} our antibaryon turns to a baryon with 
helicity $-1/2$ and not $+1/2$. This sign change relates our TFF $G_m$ to Carlson's TFF $G^{\rm Ca}_{-m}$ 
for all $m=0,\pm 1$.}
\begin{eqnarray}
  G_-^{\rm Ca} &=& \frac{Q_-}{2 \, m_\Lambda} \, G_{+1}  \,, \nonumber \\
  G_+^{\rm Ca} &=& \frac{Q_-}{2 \sqrt{3} \, m_\Lambda} \, G_{-1}  \,, \nonumber \\
  G_0^{\rm Ca} &=& \frac{Q \, Q_-}{\sqrt{6} \, m_\Lambda \, m_{\Sigma^*}} \, G_0  
  \label{eq:rel-Carlson}
\end{eqnarray}
with $Q_-:=\sqrt{Q^2+(m_\Lambda -m_{\Sigma^*})^2}$.

In \cite{Pascalutsa:2006up} various conventions for the TFFs are related to each other, including the ones 
from \cite{Carlson:1985mm}. With the help of \eqref{eq:rel-Carlson} and \cite{Pascalutsa:2006up} our TFFs can be easily 
related to any other TFF combinations and conventions. 

At large spacelike momenta, i.e.\ for large $Q^2$, one finds the following asymptotic behavior from perturbative 
QCD \cite{Carlson:1985mm}:
\begin{eqnarray}
  G_{-1}(-Q^2) \sim \frac{1}{Q^4} \,, \quad G_{0,+1}(-Q^2) \sim \frac{1}{Q^6} \,,   \nonumber \\
  F_1(-Q^2) \sim \frac{1}{Q^6} \,, \quad F_{2,3}(-Q^2)  \sim \frac{1}{Q^8}  \,.
  \label{eq:Qscaling}  
\end{eqnarray}
Since we will provide only a low-energy representation for the various TFFs, one cannot expect to reproduce this 
asymptotic behavior without involving physics beyond the low-energy region. In general, this requires too much modeling. 
Nonetheless, it might be reasonable to aim for a representation where the TFFs fall off with 
$1/Q^4$ at least. We will come back to this point below. 

Pion-loop contributions to the TFFs can be most easily addressed for fixed helicity combinations. This 
favors the use of the TFFs (\ref{eq:F-1def}), (\ref{eq:F0def}), (\ref{eq:F+1def}). However, these 
combinations are subject to kinematical constraints, i.e.\
there is a kinematical point where these TFFs are not independent from each other. 
This happens at $q^2=(m_\Lambda+m_{\Sigma^*})^2$ where $G_{+1}=G_{-1}=G_0 \, (m_\Lambda+m_{\Sigma^*})/m_{\Sigma^*}$. 
Dispersion relations should be formulated for constraint-free quantities 
\cite{Bardeen:1969aw,Tarrach:1975tu}, otherwise one might
have to involve additional subtractions. The construction procedure of \cite{Bardeen:1969aw,Tarrach:1975tu} leads to the 
TFFs of (\ref{eq:defTFF2}). Therefore it can be useful to invert the 
relations (\ref{eq:F-1def}), (\ref{eq:F0def}), (\ref{eq:F+1def}), which yields 
\begin{eqnarray}
  F_1(q^2) & = &\frac{G_{+1}(q^2)-G_{-1}(q^2)}{(m_{\Sigma^*}+m_\Lambda)^2-q^2}  \,, \nonumber \\
  F_2(q^2) & = & \frac{2}{\lambda(m_{\Sigma^*}^2,m_\Lambda^2,q^2)}  \nonumber \\ && \times  
  \left[ -2q^2 \, G_0(q^2) \right. \nonumber \\ 
  && \left. \phantom{.m} + (m_{\Sigma^*} \, m_\Lambda - m_\Lambda^2 + q^2) \, G_{+1}(q^2) \right.  \nonumber \\ 
  && \left. \phantom{.m}   + (m_{\Sigma^*}^2 - m_{\Sigma^*} \, m_\Lambda) \, G_{-1}(q^2) \right] \,, \nonumber \\
  F_3(q^2) & = & \frac{2}{\lambda(m_{\Sigma^*}^2,m_\Lambda^2,q^2)}  \nonumber \\ && \times  
  \left[ (m_{\Sigma^*}^2-m_\Lambda^2+q^2) \, G_0(q^2) \right.   \nonumber \\ 
  && \left. \phantom{.m} - m_{\Sigma^*}^2 \left(G_{+1}(q^2) + G_{-1}(q^2) \right)\right]
  \label{eq:F1Gs}
\end{eqnarray}
with the K\"all\'en function given in \eqref{eq:kallenfunc}.

Let us turn now to observable production and decay processes. In terms of the TFFs the decay width of 
$\Sigma^* \to \Lambda \gamma$ is given by
\begin{eqnarray}
  \Gamma & =& \frac{e^2 (m_{\Sigma^*}^2-m_{\Lambda}^2)}{96 \pi m_{\Sigma^*}^3} (m_{\Sigma^*}-m_{\Lambda})^2
              \nonumber \\ 
  && \times \left( 3 |G_{+1}(0)|^2+|G_{-1}(0)|^2 \right)   \,.
  \label{eq:photCase}
\end{eqnarray}
For the differential cross section of the reaction $e^+ e^- \to \bar\Sigma^* \Lambda$ (see also \cite{Korner:1976hv}) we obtain
in the center-of-mass frame and neglecting the electron mass:
\begin{eqnarray}
  && \left( \frac{\text{d}\sigma}{\text{d}\Omega}\right)_{\rm CM}(q^2,\theta)
  = \frac{e^4}{96 \pi ^2 q^6} \, p_z \, \frac{\sqrt{q^2}}{2} (q^2-(m_{\Sigma^*}-m_{\Lambda})^2)  \nonumber \\
  && \phantom{m} \times \Big[(1+\cos^2 \theta) \left( 3|G_{+1}(q^2)|^2+|G_{-1}(q^2)|^2 \right) \nonumber \\
  && \phantom{mmn} {} +\frac{4q^2}{m_{\Sigma^*}^2}\sin^2 \theta \, |G_0(q^2)|^2  \Big] 
  \label{eq:prodTFF}
\end{eqnarray}
with the center-of-mass momentum $p_z$ given in \eqref{eq:defpz}.

For the Dalitz decay distribution of $\Sigma^* \to \Lambda \; e^+ e^-$ we provide one version keeping the electron mass and
one where only the kinematical velocity factor is kept. We introduce the electron velocity by
\begin{eqnarray}
  \label{eq:defbetaelectron}
  \beta_e := \sqrt{1-\frac{4m_e^2}{q^2}} 
\end{eqnarray}
with the electron mass $m_e$. The doubly-differential decay rate is given by
\begin{eqnarray}
  &&\frac{\text{d}\Gamma}{\text{d}q^2 \, \text{d}\cos\theta} =  \nonumber \\
  &&\frac{e^4}{(2\pi)^3 \, 96 m_{\Sigma^*}^3 q^2} \, p_z \, \frac{\sqrt{q^2}}{2} \, \beta_e
  \, \left((m_{\Sigma^*}-m_{\Lambda})^2-q^2\right)  \nonumber \\
  && \times \left[
     \left( 1+\cos^2 \theta + \frac{4m_e^2}{q^2}\, \sin^2 \theta \right) \right. \nonumber \\
  && \phantom{mn} \times \left( 3|G_{+1}(q^2)|^2+|G_{-1}(q^2)|^2 \right) 
  \nonumber \\
  && \phantom{m} \left. 
     {}+4\left(\sin^2 \theta + \frac{4m_e^2}{q^2}\cos^2 \theta \right)\frac{q^2}{m_{\Sigma^*}^2} |G_0(q^2)| ^2\right]
       \nonumber \\[1em]
  && \approx \frac{e^4}{(2\pi)^3 \, 96 m_{\Sigma^*}^3 q^2} \, p_z \, \frac{\sqrt{q^2}}{2} \, \beta_e
  \, \left((m_{\Sigma^*}-m_{\Lambda})^2-q^2\right)  \nonumber \\
  && \phantom{m} \times \Big[
     \left( 1+\cos^2 \theta \right) \left( 3|G_{+1}(q^2)|^2+|G_{-1}(q^2)|^2 \right) 
  \nonumber \\
  && \phantom{mmn} 
     {}+\frac{4q^2}{m_{\Sigma^*}^2}\sin^2 \theta \, |G_0(q^2)| ^2 \Big]  \,.
    \label{eq:Dalitzdistr}
\end{eqnarray}
Here $\theta$ denotes the angle between electron and $\Lambda$ in the rest frame of the electron-positron pair. If one calculates
the integrated decay rate, the integration in $\theta$ ranges from $\pi$ to $0$ such that the $\cos\theta$ integration ranges
from $-1$ to $+1$.

One should note that in the given decay the invariant mass $q^2$ of the photon is limited in the kinematical region 
\begin{eqnarray}
  \label{eq:limitsqsq}
  4m_e^2 \leq q^2 \leq (m_{\Sigma^*}-m_{\Lambda})^2
\end{eqnarray}
and so the factor $(m_{\Sigma^*}-m_{\Lambda})^2-q^2$ will always be non-negative. If one blindly neglected the electron mass,
one would obtain a divergent integrated decay rate. The phase-space factor $\beta_e$ and the proper integration range 
\eqref{eq:limitsqsq} ensure a physical, finite result.

For later use we also introduce a QED version of \eqref{eq:Dalitzdistr}, which is supposed to describe the situation where the structure of hyperons is not resolved. In practice we replace the TFF combinations by their $q^2=0$ expressions and make in this way also contact with the real photon case \eqref{eq:photCase}:
\begin{eqnarray}
  &&\frac{\text{d}\Gamma_{\mathrm{QED}}}{\text{d}q^2 \, \text{d}\cos\theta} :=  \nonumber \\
  &&\frac{e^4}{(2\pi)^3 \, 96 m_{\Sigma^*}^3 q^2} \, p_z \, \frac{\sqrt{q^2}}{2} \, \beta_e
  \, \left((m_{\Sigma^*}-m_{\Lambda})^2-q^2\right)  \nonumber \\
  && \times 
     \left( 1+\cos^2 \theta + \frac{4m_e^2}{q^2}\, \sin^2 \theta \right)  \nonumber \\
  && \phantom{mn} \times \left( 3|G_{+1}(0)|^2+|G_{-1}(0)|^2 \right)\,.
    \label{eq:QEDcase}
\end{eqnarray}

Conceptually, small momenta go along with small $q^2$ and with treating the mass difference $m_{\Sigma^*}-m_\Lambda$ as small. 
By inspecting \eqref{eq:Dalitzdistr}, we see that at small momenta the decay rate is dominated by the combination 
$3 \vert G_{+1}\vert^2 + \vert G_{-1}\vert^2$. In turn, \eqref{eq:F-1def} and \eqref{eq:F+1def} show that for low momenta 
the dominant contribution to $G_{+1}$ and $G_{-1}$ originates from $F_1$. 
At high momenta, $G_{-1}$ is dominant, as can be read off from 
\eqref{eq:Qscaling}; see also \cite{Carlson:1985mm}. We deduce from \eqref{eq:F-1def} and \eqref{eq:Qscaling} that it is again 
$F_1$ that dominates $G_{-1}$. 
Thus in both limiting cases, low and high momenta, the TFF $F_1$ plays the dominant role. 

More detailed access to the TFFs can be obtained by determining the angular distribution 
of the subsequent weak decay of the $\Lambda$; see also \cite{Perotti:2018wxm}. 
To this end consider the decay $\Lambda \to p \pi^-$ 
governed by the amplitude \cite{pdg}
\begin{eqnarray}
  \label{eq:Lambdadecay}
  {\cal M}_{\rm weak} = G_F \, m_\pi^2 \, \bar u_p(p_1) \left(A-B \gamma_5 \right) u_\Lambda(p)   \,.
\end{eqnarray}
It is useful to introduce the asymmetry parameter
\begin{eqnarray}
  \alpha_\Lambda & := & \frac{2 {\rm Re}(T^*_s T_p)}{\vert T_s \vert^2 + \vert T_p \vert^2}
  \label{eq:defasym}
\end{eqnarray}
with the s-wave amplitude $T_s := A$, the p-wave amplitude $T_p := p_{\rm f} \, B/(E_p+m_p)$ 
and mass $m_p$, energy $E_p$ and 
momentum $p_{\rm f}$ of the proton in the rest frame of the decaying $\Lambda$ hyperon, i.e.
\begin{eqnarray}
  \label{eq:protonen}
  E_p = \frac{m_\Lambda^2+m_p^2-m_\pi^2}{2 m_\Lambda}
\end{eqnarray}
and
\begin{eqnarray}
  p_{\rm f} = \frac{\lambda^{1/2}(m_\Lambda^2,m_p^2,m_\pi^2)}{2 m_\Lambda}  \phantom{mi}
\end{eqnarray}
with the K\"all\'en function (\ref{eq:kallenfunc}). 

For the differential decay width of the whole four-body decay $\Sigma^* \to \Lambda \, e^+ e^- \to p \pi^- \, e^+ e^-$
one finds (neglecting again the electron mass where meaningful)
\begin{eqnarray}
  \label{eq:M2av-withweak}  
  &&\frac{\text{d} \Gamma}{\text{d}q^2 \, \text{d} \cos\theta \, \text{d}\Omega_p} \approx  \\
  &&\frac{e^4}{(2\pi)^4 \, 192 m_{\Sigma^*}^3 q^2} \, p_z \, \frac{\sqrt{q^2}}{2} \, \beta_e
  \, \left((m_{\Sigma^*}-m_{\Lambda})^2-q^2\right)  \, {\rm Br}_{\Lambda \to p \pi^-} \nonumber \\
  && \times   \Big[
     (1+\cos^2 \theta) \left(3|G_{+1}(q^2)|^2+|G_{-1}(q^2)|^2 \right)\nonumber \\ 
  && \phantom{n}{}
     + \frac{4q^2}{m_{\Sigma^*}^2} \text{sin}^2\theta \, |G_0(q^2)|^2  \nonumber \\ 
  && \phantom{n}+ \frac{4\sqrt{q^2}}{m_{\Sigma^*}} \, \alpha_\Lambda \, {\rm Im}\left(G_0(q^2) \, G_{-1}^*(q^2) \right)
     \cos\theta \sin\theta \sin\theta_p \sin\phi_p \Big]       \,.  \nonumber
\end{eqnarray}
Here Br$_{\Lambda \to p \pi^-}$ denotes the branching ratio while 
$\theta_p$ and $\phi_p$ are the angles of the proton three-momentum measured in the rest frame of $\Lambda$.
The coordinate system in this frame is defined by $\vec q$ pointing in the negative $z$-direction
(i.e.\ in the rest frame of the virtual photon the $\Sigma^*$ and $\Lambda$ direction defines the positive $z$-axis)
and the electron moves
in the $x$-$z$ plane with positive momentum projection on the $x$-axis. In this frame, $\theta_p$ is the angle of the proton 
momentum relative to the $z$-axis and $\phi_p$ is the angle between the $x$-axis and the projection of the proton momentum
on the $x$-$y$ plane, i.e.
\begin{eqnarray}
  && \vec p_p = p_{\rm f}  \, (\sin\theta_p \cos\phi_p,\sin\theta_p \sin\phi_p,\cos\theta_p) \,, \nonumber \\
  && \vec q = \vert \vec q \vert \, (0,0,-1)    \,, \nonumber \\
  && \vec p_{e^-} \cdot \vec e_y = 0 \,, \quad \vec p_{e^-} \cdot \vec e_x > 0 \,, \quad
     \vec e_y = \frac{\vec p_{e^-} \times \vec q}{\vert \vec p_{e^-} \vert \, \vert \vec q \, \vert}  \,.
  \label{eq:defmoms-Lambdarest}  
\end{eqnarray}
Note the subtlety that $\theta$ is measured in the rest frame of the virtual photon while $\Omega_p$ denotes angles in the
rest frame of the $\Lambda$-hyperon. In terms of Lorentz invariant quantities the angles are related to
\begin{eqnarray}
  p_\Lambda \cdot k_e &=& -\frac12 \, \lambda^{1/2}(m_{\Sigma^*}^2,m_\Lambda^2,q^2) \, \cos\theta \,, \nonumber \\
  \epsilon_{\mu\nu\alpha\beta} \, k_e^\mu \, p_\Lambda^\nu \, p_p^\alpha \, q^\beta
                      &=& - \frac12 \, \sqrt{q^2} \, \lambda^{1/2}(m_{\Sigma^*}^2,m_\Lambda^2,q^2) \nonumber \\
                      &&\times p_{\rm f} \,
                         \sin\theta \sin\theta_p \sin\phi_p
  \label{eq:LIangles}
\end{eqnarray}
with $k_e := p_{e^-}-p_{e^+}$, $q = p_{e^-}+p_{e^+} = p_{\Sigma^*}-p_\Lambda$ and the convention \cite{pesschr}
for the Levi-Civita symbol: 
\begin{equation}
\epsilon_{0123}=-1.
\label{eq:epsilon}
\end{equation}
A peculiar feature of \eqref{eq:M2av-withweak} is the presence of 
the combination Im$(G_0G_{-1}^*)$, which is non-zero even below the two-pion threshold. This is a consequence of the $\Sigma^*$ being unstable with respect to the strong interaction. This property plays a crucial role throughout the development of this paper, and constitutes the main difference from the analogous $\Sigma$-$\Lambda$ case \cite{Granados:2017cib}.

\section{Dispersive machinery}
\label{sec:disp}

Essentially this goes along the same lines as described in \cite{Granados:2017cib,Leupold:2017ngs}. In particular we use the 
same Omn\`es function, 
\begin{eqnarray}
  \Omega(s) = \exp\left\{ s \, \int\limits_{4m_\pi^2}^\infty \frac{\text{d}s'}{\pi} \, \frac{\delta(s')}{s' \, (s'-s-i \epsilon)} \right\}
  \label{eq:omnesele}  
\end{eqnarray}
where $\delta$ denotes the pion p-wave phase shift \cite{Colangelo:2001df,GarciaMartin:2011cn}. 
The pion vector form factor $F^V_\pi$ is taken 
from \cite{Leupold:2017ngs} (see also \cite{Hanhart:2012wi,Hanhart:2013vba,Hoferichter:2016duk}):
\begin{eqnarray}
  \label{eq:FV-Omnes-alphaV}
  F^V_\pi(s) = (1+\alpha_V \, s) \, \Omega(s) \,.
\end{eqnarray}
For the pion phase shift from \cite{GarciaMartin:2011cn}, a value of 
\begin{eqnarray}
  \label{eq:alphaV}
  \alpha_V = 0.12 \, {\rm GeV}^{-2}
\end{eqnarray}
yields an excellent description of the data on the pion vector form factor from tau decays \cite{Fujikawa:2008ma} for 
energies below 1 GeV; see \cite{Leupold:2017ngs}.

\subsection{Dispersion relations}
\label{subsec:dispgen}

Based on the asymptotic behavior \eqref{eq:Qscaling}, the three TFFs introduced in \eqref{eq:defTFF2}
satisfy unsubtracted dispersion relations
\begin{eqnarray}
  \label{eq:disp-verygen}
  F_j(q^2) = \int \frac{\text{d}s}{2\pi i} \, \frac{{\rm disc}F_j(s)}{s-q^2}
\end{eqnarray}
for $j=1,2,3$. Here ``disc'' denotes the discontinuity of the function $F_j$.

How does this translate to the TFFs $G_m$ introduced in \eqref{eq:F-1def}, \eqref{eq:F0def},
\eqref{eq:F+1def}? We can discuss this rather generally:
If one defines two new TFFs, $A$ and $B$, via 
\begin{eqnarray}
  A(q^2) &:=& F_1(q^2) + F_2(q^2)  \,, \nonumber \\
  B(q^2) &:=& F_1(q^2) + \frac{q^2}{s_0} \, F_2(q^2)  \,,
  \label{eq:defAB-super}
\end{eqnarray}
one sees that they are subject to the kinematical constraint
\begin{eqnarray}
  \label{eq:kinemat-constrAB}
  A(s_0) = B(s_0)  \,.
\end{eqnarray}
The dispersion relation for $A$ can be formulated without problems. For $B$ one obtains
\begin{eqnarray}
  B(q^2) &=& 
  \int \frac{\text{d}s}{2\pi i} \, \frac{1}{s-q^2} \, \left({\rm disc}F_1(s) + \frac{q^2}{s_0} \, {\rm disc}F_2(s) \right)
  \nonumber \\
  & = &   \int \frac{\text{d}s}{2\pi i} \, \frac{1}{s-q^2} \, \left({\rm disc}B(s) + \frac{q^2-s}{s_0} \, {\rm disc}F_2(s) \right)
  \nonumber \\
  & = &   \int \frac{\text{d}s}{2\pi i} \, \frac{{\rm disc}B(s)}{s-q^2} 
  - \frac{1}{s_0} \, \int \frac{\text{d}s}{2\pi i} \, {\rm disc}F_2(s)  \,.
  \label{eq:calcBdisp}
\end{eqnarray}
This shows that in general one has to deal with an additional constant in a dispersive calculation of $B$. It is this constant 
that ensures that \eqref{eq:kinemat-constrAB} holds. In addition, we have implicitly assumed that the dispersive integral
over disc$B$ actually converges.

For the TFFs $F_j$ that show the high-energy behavior \eqref{eq:Qscaling}, the situation is actually simpler. This
high-energy behavior provides conditions for the integrals over disc$F_j$. In particular, the condition 
\begin{eqnarray}
  \label{eq:asym-simple}
  \lim\limits_{Q^2 \to \infty} Q^2 \, F_j(-Q^2) = 0
\end{eqnarray}
leads to 
\begin{eqnarray}
  \label{eq:disp-verygen-cond}
  \int \frac{\text{d}s}{2\pi i} \, {\rm disc}F_j(s) = 0  \,.
\end{eqnarray}
Thus, the additional constant in \eqref{eq:calcBdisp} vanishes. One obtains standard unsubtracted dispersion relations
for $A$ and for $B$. In view of the relations \eqref{eq:F-1def}, \eqref{eq:F0def}, \eqref{eq:F+1def} one can therefore
conclude that also all the $G_m$'s satisfy unsubtracted dispersion relations:
\begin{eqnarray}
  \label{eq:disp-verygenGm}
  G_m(q^2) = \int \frac{\text{d}s}{2\pi i} \, \frac{{\rm disc}G_m(s)}{s-q^2}
\end{eqnarray}
for $m=0, \pm 1$.

\subsection{General considerations about the analytic structure}
\label{subsec:genana}

At low energies, it can be expected that the $q^2$ behavior of the TFFs is determined by the lowest-energy states
that can be excited. For the isovector TFFs that we study here, the lowest energetic states are pion pairs.
Therefore we can write in complete analogy to \cite{Granados:2017cib}:
\begin{eqnarray}
  G_m(q^2) &=& \frac{1}{12\pi} \, \int\limits_{4 m_\pi^2}^\infty \frac{\text{d}s}{\pi} \, 
     \frac{T_m(s) \, p_{\rm c.m.}^3(s) \, F^{V*}_\pi(s)}{s^{1/2} \, (s-q^2-i \epsilon)}  \nonumber \\
     && {} + G^{\rm anom}_m(q^2) + \ldots
  \label{eq:genGmdispelips}
\end{eqnarray}
where the ellipsis denotes other intermediate states as for instance four-pion or baryon-antibaryon states.
The ``anomalous'' piece will be determined later. It is related to anomalous thresholds. 

The pion-hyperon scattering amplitudes $T_m$ are obtained in a two-step procedure: 
In line with (\ref{eq:expect0}), (\ref{eq:expect-1}), (\ref{eq:expect+1}) 
we define first the reduced amplitudes 
\begin{eqnarray}
  K_{\pm 1}(s) & := & -\frac{3}{4} \, \int\limits_0^\pi \text{d}\theta \, \sin^2 \theta \, \nonumber \\ && {} \times 
  \frac{{\cal M}(s,\theta,1/2 \pm 1,1/2)}{\bar v_\Lambda(-p_z,1/2) \, \gamma_5 \, u^1_{\Sigma^*}(p_z,1/2 \pm 1) \, p_{\rm c.m.} } \,,
  \nonumber \\
  K_0(s) & := & -\frac{3}{2} \, \frac{m_{\Sigma^*}^2-m_\Lambda^2+s}{2 \, s} \, \int\limits_0^\pi \text{d}\theta \, 
  \sin \theta \, \cos\theta \nonumber \\ && {} \times 
  \frac{{\cal M}(s,\theta,1/2,1/2)}{\bar v_\Lambda(-p_z,+1/2) \, \gamma_5 \, u^3_{\Sigma^*}(p_z,+1/2)\, p_{\rm c.m.} }  \,. \phantom{mm}
  \label{eq:def-redampl}
\end{eqnarray}
Here $p_{\rm c.m.}$ denotes the modulus of the momenta of the pions in the center-of-mass frame. We have introduced
${\cal M}(s,\theta,\sigma,\lambda)$ as the approximation to the Feynman amplitude for the reaction
$\Sigma^* \, \bar\Lambda \to \pi^+ \pi^-$. In practice, ${\cal M}(s,\theta,\sigma,\lambda)$ does not include the rescattering
effect of the pions. This will be taken care of in the second step. 
In addition, we want to distinguish conceptually between processes with
left-hand cut structures and purely polynomial terms. In practice, the reduced amplitudes $K$ originate from the left-hand cut
structures only, while we denote the polynomial terms by $P$. All the formulae presented explicitly for $K$ apply also to $P$.

Pion rescattering is taken into account by solving a Muskhelishvili-Omn\`es equation \cite{zbMATH03081975,Omnes:1958hv}.
The result is
\begin{eqnarray}
  T_m(s) & = & K_m(s) + \Omega(s) \, P_{m} + T_m^{\rm anom}(s) 
  \nonumber \\ && {}+ \Omega(s) \, s \, 
  \int\limits_{4m_\pi^2}^\infty \, \frac{\text{d}s'}{\pi} \, 
  \frac{K_m(s') \, \sin\delta(s')}{\vert\Omega(s')\vert \, (s'-s-i \epsilon) \, s'} \,. \phantom{m}  
  \label{eq:tmandel}
\end{eqnarray}
As already spelled out, $K_m$ takes care of the left-hand cut structures. $P_m$ is a constant (per channel) that can be obtained
ideally from a fit to data or estimated from $\chi$PT. We have used here a once-subtracted dispersion relation.
In principle, one could use more subtractions, which brings in a polynomial instead of a constant. But this would worsen the 
high-energy behavior. In the following, we will occasionally suppress the index $m$ when presenting generic formulae.

If there is an anomalous threshold, there might be an extra piece $T^{\rm anom}(s)$ that is added to the amplitude.
Such a situation can occur if the mass $m_{\rm exch}$ of the exchanged state in the t/u-channel is ``too light''. 
For our reaction the condition to have an anomalous threshold is \cite{Karplus:1958zz}
\begin{eqnarray}
  \label{eq:anomthr}
  m_{\rm exch}^2 < \frac12 \, \left(m_{\Sigma^*}^2 + m_\Lambda^2 - 2 m_\pi^2 \right)  \,.
\end{eqnarray}
For the formal reaction $\Sigma^* \bar\Lambda \to 2\pi$ one has to deal with the exchange of states carrying strangeness.
In practice we will take into account the exchange of $\Sigma$ and $\Sigma^*$ hyperons. 
The condition \eqref{eq:anomthr} does not hold for the $\Sigma^*$ exchange,\footnote{It does not hold for any exchange of a
  many-particle state that contains a hyperon. The lightest such state would be a $\Lambda$ and one pion. Using that
  the $\Sigma^*$ is lighter than a $\Lambda$ and two pions, it is easy to check that the condition \eqref{eq:anomthr} is
  not satisfied for $m_{\rm exch} \ge m_\Lambda + m_\pi$.}
but is satisfied for the $\Sigma$ exchange. In the latter case the 
logarithm obtained from the partial-wave projection (\ref{eq:def-redampl}) requires a proper analytic continuation. 
If one takes the partial-wave projection literally (straight-line integral) as given in (\ref{eq:def-redampl}), then the 
obtained logarithm has a cut in the complex $s$ plane that intersects with the right-hand cut (unitarity cut), i.e.\ part
of this cut lies on the physical Riemann sheet. 
To disentangle the cuts, 
one can define the cut of the logarithm such that it connects the branch point to the start of the unitarity cut by a straight
line. The additional contribution $T^{\rm anom}(s)$ takes care of the extra cut. A general discussion is provided 
in Appendix \ref{sec:anom-disp}. 

To be more concrete, we note that the p-wave partial-wave projection of type (\ref{eq:def-redampl}) for a t- or u-channel
exchange process produces a term 
\begin{eqnarray}
  K(s) &=& g(s) - \frac{2 f(s)}{Y(s) \, \kappa^2(s)} \nonumber \\
  &&{}+ f(s) \, \frac{1}{\kappa^3(s)} \, \log\frac{Y(s)+\kappa(s)}{Y(s)-\kappa(s)}
  \label{eq:Klog}
\end{eqnarray}
with the functions $Y$, $\kappa$ and $\sigma$ defined in Appendix \ref{sec:anom-disp} for
$m_1 \to m_{\Sigma^*}$, $m_2 \to m_\Lambda$.
In addition, we have introduced $f(s)$, $g(s)$ as functions without cuts. These functions might have poles at kinematical
thresholds, but they conspire such that no poles show up for $K$ as given in \eqref{eq:Klog}. 
If one expands the log function in powers of $\kappa/Y$ one sees that there are no poles for $\kappa \to 0$.
Concrete formulae are given in Section \ref{sec:results}.

If one considers the standard logarithm with a cut along the real negative axis, then (\ref{eq:Klog}) is ill-defined 
for $Y(s)=0$. This point lies on the unitarity cut if (\ref{eq:anomthr}) is satisfied. To disentangle the two cuts 
one starts with a proper analytic continuation of the logarithm along the unitarity cut. To this end we introduce the 
following four points: 
\begin{itemize}
\item At $s_4:=(m_{\Sigma^*}+m_\Lambda)^2$ we have $\kappa=0$. 
  Above this point, i.e.\ for $s$ real and larger than $s_4$, there is the true scattering 
  region. There, $\kappa$ is real and $Y$ is 
  positive and larger than $\kappa$. The logarithm in (\ref{eq:Klog}) can be defined as the real-valued standard logarithm of 
  positive numbers. 
\item At $s_3:=m_{\Sigma^*}^2 + m_\Lambda^2 + 2 m_\pi^2-2 m_{\rm exch}^2$ we have $Y=0$. 
  For $s$ real and between $s_3$ and $s_4$ the function $\kappa$ is purely imaginary and
  $Y$ is still positive.
\item At $s_2:= 4 m_\pi^2$  we have $\kappa=0$. For $s$ real and between $s_2$ and $s_3$ the function $\kappa$ is purely imaginary 
  and $Y$ is negative.
\item At $s_1:=(m_{\Sigma^*}-m_\Lambda)^2$ we have $\kappa=0$. For $s$ real and between $s_1$ and $s_2$ the function $\kappa$ is 
  real (and $Y$ is negative). 
\end{itemize}
For the case of a $\Sigma$ exchange we have $0< s_1 < s_2 < s_3 < s_4$. 
The function $K$ in (\ref{eq:Klog}) that enters finally (\ref{eq:tmandel}) is then defined on the relevant part of the 
real axis by
\begin{equation}
  \label{eq:properdefKl4}
  K(s) := g(s) - \frac{2 f(s)}{Y(s) \, \kappa^2(s)} + \frac{f(s)}{\kappa^3(s)} \, \log\frac{Y(s)+\kappa(s)}{Y(s)-\kappa(s)}
\end{equation}
for $s > s_4$, 
\begin{equation}
  \label{eq:properdefK34}
  K(s) := g(s)
  - \frac{2 f(s)}{Y(s) \, \kappa^2(s)} + \frac{2f(s)}{\kappa^2(s) \, \vert\kappa(s)\vert} \arctan\frac{\vert\kappa(s)\vert}{Y(s)} 
\end{equation}
for $s_3 < s < s_4$, and
\begin{eqnarray}
  K(s) &:=& g(s) - \frac{2 f(s)}{Y(s) \, \kappa^2(s)}  \nonumber \\
  && {}+ \frac{2f(s)}{\kappa^2(s) \, \vert\kappa(s)\vert} \left( \arctan\frac{\vert\kappa(s)\vert}{Y(s)} +\pi \right) 
  \label{eq:properdefK23}
\end{eqnarray}
for $s_2 < s < s_3$. 
Here the standard logarithm for positive real numbers is used and the arctan function with values between $-\pi/2$ and $\pi/2$.
Note that at the two-pion threshold $s=s_2$ the quantity $K(s)$ of (\ref{eq:properdefK23}) diverges
$\sim 2 \pi f(s)/(\kappa^2(s) \, \vert\kappa(s)\vert) \sim 1/\sigma^3(s)$, but the product
$K(s) \, \sin\delta(s)$ in (\ref{eq:tmandel}) remains finite due to $\sin\delta(s) \sim \sigma^3(s)$ for the p-wave
pion phase shift \cite{Colangelo:2001df,GarciaMartin:2011cn}. $K(s)$ also appears in the combination 
$K(s) \, p_{\rm c.m.}^3(s)$ in (\ref{eq:genGmdispelips}) which remains also finite. 

The second issue is the definition of $T^{\rm anom}$; see also the discussion in Appendix \ref{sec:anom-disp}. 
The branch points of the logarithm in (\ref{eq:Klog}) are defined by 
$Y^2(s)=\kappa^2(s)$. They are located at 
\begin{eqnarray}
  s_\pm &=& -\frac12 \, m_{\rm exch}^2 + \frac12 \left(m_{\Sigma^*}^2 + m_\Lambda^2 + 2 m_\pi^2 \right) \nonumber \\ && {}
  - \frac{m_{\Sigma^*}^2 \, m_\Lambda^2 - m_\pi^2 \, (m_{\Sigma^*}^2 + m_\Lambda^2) + m_\pi^4}{2m_{\rm exch}^2} 
  \nonumber \\ && {}
  \mp \frac{\lambda^{1/2}(m_{\Sigma^*}^2,m_{\rm exch}^2,m_\pi^2) \, \lambda^{1/2}(m_{\rm exch}^2,m_\Lambda^2,m_\pi^2)}{2m_{\rm exch}^2} \,.
  \phantom{mm}
  \label{eq:defspm}
\end{eqnarray}
We take $s_+$ as the solution that has a positive imaginary part for small values of $m_{\Sigma^*}^2$. 
If one replaces $m_{\Sigma^*}^2$ by $m_{\Sigma^*}^2 +i\epsilon$ and follows the motion 
of $s_+$ for increasing values of $m_{\Sigma^*}^2$, 
then $s_+$ moves towards the real axis and intersects with the unitarity cut where (\ref{eq:anomthr}) 
turns to an equality. For larger values of $m_{\Sigma^*}^2$ one finds $s_+$ in the lower half plane of the first Riemann sheet.
This is the situation for the physical value of $m_{\Sigma^*}^2$ for the case $m_{\rm exch}^2= m_\Sigma^2$. Thus we have
\begin{eqnarray}
  s_+ &=& -\frac12 \, m_{\Sigma}^2 + \frac12 \left(m_{\Sigma^*}^2 + m_\Lambda^2 + 2 m_\pi^2 \right) \nonumber \\ && {}
  - \frac{m_{\Sigma^*}^2 \, m_\Lambda^2 - m_\pi^2 \, (m_{\Sigma^*}^2 + m_\Lambda^2) + m_\pi^4}{2m_{\Sigma}^2} 
  \nonumber \\ && {}
  -i \, \frac{\lambda^{1/2}(m_{\Sigma^*}^2,m_{\Sigma}^2,m_\pi^2) \, \left(-\lambda(m_{\Sigma}^2,m_\Lambda^2,m_\pi^2)\right)^{1/2}}{2m_{\Sigma}^2} 
  \phantom{mm}
  \label{eq:defspconcr}
\end{eqnarray}
with positive square roots.

The anomalous contribution that enters \eqref{eq:tmandel} is then given by 
\begin{eqnarray}
  T^{\rm anom}(s) &=& \Omega(s) \, s \, \int\limits_0^1 \text{d}x \, \frac{\text{d}s'(x)}{\text{d}x}  \,
                      \frac{1}{s'(x)-s}  \nonumber \\
                  && \times 
                     \frac{2f(s'(x))}{(-\lambda(s'(x),m_{\Sigma^*}^2,m_\Lambda^2))^{1/2} \, \kappa^2(s'(x))}  \nonumber \\
                  && \times  \frac{t(s'(x))}{\Omega(s'(x)) \, s'(x)} \phantom{mm}
  \label{eq:tanom}
\end{eqnarray}
with the straight-line path
\begin{eqnarray}
  \label{eq:defsx}
  s'(x) := (1-x) s_+ + x\, 4m_\pi^2 
\end{eqnarray}
that connects the branch point (\ref{eq:defspconcr}) of the logarithm of (\ref{eq:Klog}) and the branch point $4m_\pi^2$ 
of the unitarity cut. 

One also needs the scattering amplitude $t(s)$ in the complex plane. 
Following \cite{Dax:2018rvs}, one could use an analytic continuation of the amplitude as constructed from $\chi$PT 
and unitarized by the inverse amplitude method.
This representation does not show a decent high-energy behavior. Therefore
we will use it only for the anomalous part. There the whole integration region is rather close to the two-pion threshold.
Therefore an expression from $\chi$PT or a unitarized version thereof should be sufficiently close to the
true scattering amplitude. We take from \cite{Dax:2018rvs} the following expressions (extended to the complex plane).
The approximation from $\chi$PT is given by
\begin{equation}
\label{tChPT1}
t_{\chi\rm PT}(s) \approx t_2(s)+t_4(s)
\end{equation}
and its unitarized version is
\begin{equation}
\label{tIAM1}
t_{\text{IAM}}(s)=\frac{t_2^2(s)}{t_2(s)-t_4(s)} 
\end{equation}
with
\begin{eqnarray}
  t_2(s) &=& \frac{s\sigma^2}{96\pi F_0^2}  \,, 
\end{eqnarray}
\begin{widetext}
\begin{eqnarray}
  t_4(s) &=& \frac{t_{2}(s)}{48\pi^2F_0^2}\bigg[s\left(\bar{l}+\frac{1}{3}\right)-\frac{15}{2} m_\pi^2
             -\frac{m_\pi^4}{2s}\Big(41-2L_\sigma\big(73-25\sigma^2\big)
             +3L_\sigma^2\big(5-32\sigma^2+3\sigma^4\big) \Big)\bigg] -\hat\sigma(s) \,t_2^2(s)   \,,
            \label{t2t4}
\end{eqnarray}
\end{widetext}
\begin{equation}
  \label{Abbrev_Lsig_sig}
  L_\sigma=\frac{1}{\sigma^2}\left(\frac{1}{2\sigma}\log\frac{1+\sigma}{1-\sigma}-1\right)   \,.
\end{equation}
The functions $\sigma(s)$ and $\hat\sigma(s)$ are defined in \eqref{eq:defsigma} and \eqref{eq:defsigmahat}, respectively.
Note that there is no square root ambiguity in the definition of $\sigma$ since all expressions are even
in $\sigma \to -\sigma$. The
square root appearing in the definition of the function $\hat\sigma$ has its cut on the negative real axis. Then the function
$\hat\sigma$ has the unitarity cut (and also a cut along the negative real axis). 

The value for the pion decay constant in the chiral limit $F_0$ is taken from the ratio $F_\pi/F_0=1.064(7)$, where
$F_\pi = 92.28(9)\,$MeV is the pion decay constant at the physical point. 
In the original paper \cite{Dax:2018rvs} the low-energy constant $\bar{l}=5.73(8)$ has been
adjusted such as to reproduce the position of the pole of the $\rho$-meson resonance on the second Riemann sheet. In this work instead we use $\bar{l}=6.47$ which is obtained by requiring agreement between the pion p-wave phase shifts from \eqref{tIAM1} and from \cite{GarciaMartin:2011cn} around the two-pion threshold. 

Finally, we provide the anomalous piece of the TFFs. As described in Appendix \ref{sec:anom-disp} one 
can relate the anomalous piece of the TFF to the anomalous piece of $T-K$. Therefore we obtain
\begin{eqnarray}
  G^{\rm anom}_m(q^2) &=& \frac{1}{12\pi} \, \int\limits_{0}^1 \text{d}x \, \frac{\text{d}s'(x)}{\text{d}x} \,
                          \frac{1}{s'(x)-q^2}
                          \nonumber \\ && \times
  \frac{f_m(s'(x)) \, s'(x) \, F^{V}_\pi(s'(x))}{-4\,  (-\lambda(s'(x),m_{\Sigma^*}^2,m_\Lambda^2))^{3/2}}  \,.
  \label{eq:Fanom-unsub}
\end{eqnarray} 
Note that the Omn\`es function \eqref{eq:omnesele} that enters the pion vector form factor \eqref{eq:FV-Omnes-alphaV} is defined everywhere on the first Riemann sheet via the
pion phase shift along the right-hand cut.
Therefore, \eqref{eq:Fanom-unsub} can be calculated without problems. 

Note that without any anomalous piece the TFF integral in \eqref{eq:genGmdispelips} would be real below the 
two-pion threshold. However, the TFF should be complex because the $\Sigma^*$ is unstable. The imaginary part emerges 
from the following process: Irrespective of the 
invariant mass of the photon, the $\Sigma^*$ can decay to a pion and a $\Sigma$. These states can rescatter into 
a $\Lambda$ and a real or virtual photon. The anomalous pieces take care of this physical process.

\subsection{Subtracted dispersion relations}
\label{subsec:sub}

Though the intermediate states not explicitly considered in \eqref{eq:genGmdispelips} might have a minor influence on the
{\em shape} of the TFFs at low energies, it is likely that they have an impact on the overall size; see e.g.\ the
discussion in \cite{Granados:2017cib,Hoferichter:2016duk,Leupold:2017ngs}. A way to enhance
the importance of the low-energy region in a dispersive integral is the use of a subtracted dispersion relation. The most
conservative approach that does not make use of any high-energy input is to start again from the unconstrained TFFs $F_i$. A subtracted dispersion relation reads
\begin{eqnarray}
  &&  F_i(q^2) =  F_i(0)  
     \nonumber \\
  && {}
     +  \frac{q^2}{12\pi} \, \int\limits_{4 m_\pi^2}^{\Lambda^2} \frac{\text{d}s}{\pi} \, 
     \frac{T_i(s) \, p_{\rm c.m.}^3(s) \, F^{V*}_\pi(s)}{s^{3/2} \, (s-q^2-i \epsilon)}
     + F^{\rm anom}_i(q^2)   \phantom{mm}
     \label{eq:dispbasic}  
\end{eqnarray}
for $i=1,2,3$. The last, ``anomalous'' piece will be specified below.

In principle, the scattering amplitudes $T_i$ are again given by \eqref{eq:tmandel} 
but now we need the amplitudes $K_i$, $i=1,2,3$ as input. They are obtained from $K_{+1,0,-1}$ in the same way as the
TFFs $F_i$ are obtained from $G_{+1,0,-1}$, i.e.
\begin{eqnarray}
  \label{eq:TiT0pm1}
  K_1(s) & = &\frac{K_{+1}(s)-K_{-1}(s)}{(m_{\Sigma^*}+m_\Lambda)^2-s}  \,, \\ 
  K_2(s) & = & \frac{2}{\lambda(m_{\Sigma^*}^2,m_\Lambda^2,s)}  
  \left[ -2s \, K_0(s) \right. \nonumber \\ 
  && \left. \phantom{nmmmmmmm} + (m_{\Sigma^*} \, m_\Lambda - m_\Lambda^2 + s) \, K_{+1}(s) \right.  \nonumber \\ 
  && \left. \phantom{nmmmmmmm}   + (m_{\Sigma^*}^2 - m_{\Sigma^*} \, m_\Lambda) \, K_{-1}(s) \right] \,, \nonumber \\
  K_3(s) & = & \frac{2}{\lambda(m_{\Sigma^*}^2,m_\Lambda^2,s)}  
  \left[ (m_{\Sigma^*}^2-m_\Lambda^2+s) \, K_0(s) \right.   \nonumber \\ 
  && \left. \phantom{mmmmmmmm} - m_{\Sigma^*}^2 \left(K_{+1}(s) + K_{-1}(s) \right)\right]  \,. \nonumber 
\end{eqnarray}

We have introduced a cutoff $\Lambda$ in \eqref{eq:dispbasic}. Since we have only the low-energy part under control where the
two-pion state dominates, it is not reasonable to extend the integral into the uncontrolled high-energy region. In practice,
the two-pion state dominates the isovector channel up to about 1 GeV. To explore the uncertainties of our low-energy
approximation we will vary the cutoff between 1 and 2 GeV. 

Finally we come back to the anomalous piece in \eqref{eq:dispbasic}:
\begin{eqnarray}
  F^{\rm anom}_i(q^2) &=& \frac{q^2}{12\pi} \, \int\limits_{0}^1 \text{d}x \, \frac{ds'(x)}{dx} \,
                          \frac{1}{s'(x)-q^2}   \nonumber \\ && \times
  \frac{f_i(s'(x)) \, F^{V}_\pi(s'(x))}{-4\, (-\lambda(s'(x),m_{\Sigma^*}^2,m_\Lambda^2))^{3/2}}  \,.
  \label{eq:Fanom}
\end{eqnarray}

The drawback of \eqref{eq:dispbasic} is that one needs experimental input for the three complex-valued(!) subtraction constants
$F_i(0)$. This is on top of the constants $P_m$ in \eqref{eq:tmandel}, which are ideally also fitted to experimental data.
At the moment such an amount of experimental information is not available.
Therefore we will explore an alternative in the next subsection. 

\subsection{Unsubtracted dispersion relations}
\label{subsec:unsub}

At large energies, the TFFs $F_i$ determined via \eqref{eq:dispbasic} approach a constant, 
in sharp contrast to the correct scaling behavior \eqref{eq:Qscaling}. The TFFs $G_m$ obtained from \eqref{eq:F-1def}, 
\eqref{eq:F0def}, \eqref{eq:F+1def} even diverge. All this is not a fundamental problem since by construction the 
representation \eqref{eq:dispbasic} 
is designed to be accurate at low energies only. Nonetheless, the representation \eqref{eq:dispbasic} requires the knowledge 
of several subtraction constants, all of them in principle complex, because the $\Sigma^*$ is unstable. Thus, it might be of 
advantage to explore the predictive power of an unsubtracted dispersion relation. 
As shown, e.g., in \cite{Hoferichter:2016duk,Granados:2017cib,Leupold:2017ngs}, one cannot expect to obtain completely correct 
values for the subtraction constants from the unsubtracted dispersion relations, if one uses only the two-pion 
intermediate states. However, it might be reasonable to use a simple effective pole to approximate the impact of all the other, 
higher-lying intermediate states on the low-energy quantities \cite{Hoferichter:2016duk,Hoferichter:2018dmo,Hoferichter:2018kwz}. 
The pole position might be varied in a reasonable range of masses of excited vector mesons \cite{pdg} while the residue can be 
chosen such that a more reasonable high-energy behavior is achieved. 

Enforcing a more realistic high-energy behavior provides an additional advantage. 
As already pointed out, one can then formulate simple dispersion relations also for the TFFs
$G_m$, $m=0,\pm 1$. In practice we write 
\begin{eqnarray}
  G_m(q^2) &=&  \frac{1}{12\pi} \, \int\limits_{4 m_\pi^2}^{\Lambda^2} \frac{\text{d}s}{\pi} \, 
     \frac{T_m(s) \, p_{\rm c.m.}^3(s) \, F^{V*}_\pi(s)}{s^{1/2} \, (s-q^2-i \epsilon)}  \nonumber \\
     && {} + G^{\rm anom}_m(q^2) + c_m \, \frac{M_V^2}{M_V^2-q^2}   \,,
     \label{eq:dispbasic-unsub}  
\end{eqnarray}
which is only valid for $q^2 \ll M_V^2$. The anomalous part is given in \eqref{eq:Fanom-unsub}. 
The dimensionless constant $c_m$ is adjusted such that 
\begin{eqnarray}
  \label{eq:asym-simple2}
  \lim\limits_{Q^2 \to \infty} Q^2 \, G_m(-Q^2) = 0  \,.
\end{eqnarray}
This leads to
\begin{eqnarray}
  \lefteqn{c_m M_V^2 = -\frac{1}{12\pi} \, \int\limits_{4 m_\pi^2}^{\Lambda^2} \frac{\text{d}s}{\pi} \, 
     \frac{T_m(s) \, p_{\rm c.m.}^3(s) \, F^{V*}_\pi(s)}{s^{1/2}} }  \nonumber \\
     && {} -\frac{1}{12\pi} \, \int\limits_{0}^1 \text{d}x \, \frac{\text{d}s'(x)}{\text{d}x} \, 
  \frac{f_m(s'(x)) \, s'(x) \, F^{V}_\pi(s'(x))}{-4 \, (-\lambda(s'(x),m_{\Sigma^*}^2,m_\Lambda^2))^{3/2}} \,. \phantom{m}
  \label{eq:detcm}
\end{eqnarray}
To explore the uncertainties of this approach one might vary the effective pole between the masses of the excited vector 
mesons \cite{pdg}, $1.4\,$GeV $< M_V < 1.7\,$GeV.
 
In practice, comparison to experimental results for $\Sigma^* \to \Lambda \, \gamma$ and
$\Sigma^* \to \Lambda \, e^+ e^-$ must show 
if \eqref{eq:dispbasic-unsub}, \eqref{eq:asym-simple2} is a reasonable approach
or if one has to resort to the subtracted dispersion relations \eqref{eq:dispbasic}. 
So far there are no Dalitz decay data available. In Section \ref{sec:results} we present numerical results
utilizing \eqref{eq:dispbasic-unsub}, \eqref{eq:detcm}.

\section{Input from chiral perturbation theory}
\label{sec:chiPT}

The leading-order (LO) chiral Lagrangian 
including the decuplet states 
is given by \cite{Jenkins:1991es,Pascalutsa:2006up,Ledwig:2014rfa,Holmberg:2018dtv} 
\begin{eqnarray}
  && {\cal L}_{\rm baryon}^{(1)} = {\rm tr}\left(\bar B \, (i \slashed{D} - m_{(8)}) \, B \right)  \nonumber \\ 
  && {}+ \bar T_{abc}^\mu \, ( i \gamma_{\mu\nu\alpha} D^\alpha - \gamma_{\mu\nu} \, m_{(10)}) \, (T^\nu)^{abc}
  \nonumber \\ 
  && {}+ \frac{D}{2} \, {\rm tr}(\bar B \, \gamma^\mu \, \gamma_5 \, \{u_\mu,B\}) 
  + \frac{F}{2} \, {\rm tr}(\bar B \, \gamma^\mu \, \gamma_5 \, [u_\mu,B])  \nonumber \\
  && {} + \frac{h_A}{2\sqrt{2}} \, 
  \left(\epsilon^{ade} \, \bar T^\mu_{abc} \, (u_\mu)^b_d \, B^c_e
  + \epsilon_{ade} \, \bar B^e_c \, (u^\mu)^d_b \, T_\mu^{abc} \right) \nonumber \\
  && {} - \frac{H_A}{4 m_R} \, \epsilon_{\mu\nu\alpha\beta} \, 
  \left(\bar T^\mu_{abc} \, (D^\nu T^\alpha)^{abd} \, (u^\beta)^c_d \right. \nonumber \\
  && \left. \phantom{mmmmmmmm} + (D^\nu \bar T^\alpha)_{abd} \, (T^\mu)^{abc} \, (u^\beta)^d_c \right)
  \label{eq:baryonlagr}
\end{eqnarray}
with tr denoting a flavor trace. 

We have introduced the totally antisymmetrized products of two and three 
gamma matrices\footnote{Throughout this work, when using the phrase ``gamma matrices'' we have the four gamma 
matrices $\gamma^\mu$, $\mu=0,1,2,3$, in mind, {\em not} $\gamma_5$.} \cite{pesschr},
\begin{eqnarray}
  \label{eq:defgammunu}
  \gamma^{\mu\nu} := \frac12 [\gamma^\mu,\gamma^\nu]
\end{eqnarray}
and 
\begin{eqnarray}
  \label{eq:defgammunual}
  \gamma^{\mu\nu\alpha}&:=& \frac16 
  \left(\gamma^\mu \gamma^\nu \gamma^\alpha + \gamma^\nu \gamma^\alpha \gamma^\mu + \gamma^\alpha \gamma^\mu \gamma^\nu
  \right.  \nonumber \\ && \left. \phantom{m} {}
    - \gamma^\mu \gamma^\alpha \gamma^\nu - \gamma^\alpha \gamma^\nu \gamma^\mu - \gamma^\nu \gamma^\mu \gamma^\alpha \right)
  \nonumber \\ 
  & = & \frac12 \{\gamma^{\mu\nu},\gamma^\alpha\} = +i\epsilon^{\mu\nu\alpha\beta} \gamma_\beta \gamma_5  \,,
\end{eqnarray}
respectively.
Our conventions are: $\gamma_5:=i \gamma^0 \gamma^1 \gamma^2 \gamma^3$ and 
\eqref{eq:epsilon}, the latter in agreement with \cite{pesschr} 
but opposite to \cite{Pascalutsa:2006up,Ledwig:2011cx}. If a formal manipulation program is used to calculate spinor traces and 
Lorentz contractions a good check for the convention for the Levi-Civita symbol is the last relation in (\ref{eq:defgammunual}).

The octet baryons are collected in ($B^a_b$ is the entry in the $a$th row, $b$th column)
\begin{eqnarray}
  \label{eq:baroct}
  B = \left(
    \begin{array}{ccc}
      \frac{1}{\sqrt{2}}\, \Sigma^0 +\frac{1}{\sqrt{6}}\, \Lambda 
      & \Sigma^+ & p \\
      \Sigma^- & -\frac{1}{\sqrt{2}}\,\Sigma^0+\frac{1}{\sqrt{6}}\, \Lambda
      & n \\
      \Xi^- & \Xi^0 
      & -\frac{2}{\sqrt{6}}\, \Lambda
    \end{array}   
  \right)  \,.  \phantom{mm}
\end{eqnarray}
The decuplet is expressed by a totally symmetric flavor tensor $T^{abc}$ 
with 
\begin{eqnarray}
  && T^{111} = \Delta^{++} \,, \quad T^{112} = \frac{1}{\sqrt{3}} \, \Delta^+  \,, \nonumber \\
  && T^{122} = \frac{1}{\sqrt{3}} \, \Delta^0  \,, \quad T^{222} = \Delta^- \,,    \nonumber \\
  && T^{113} = \frac{1}{\sqrt{3}} \, \Sigma^{*+}  \,, \quad T^{123} = \frac{1}{\sqrt{6}} \, \Sigma^{*0}  \,, \quad 
  T^{223} = \frac{1}{\sqrt{3}} \, \Sigma^{*-}  \,,  \nonumber \\
  && T^{133} = \frac{1}{\sqrt{3}} \, \Xi^{*0} \,, \quad T^{233} = \frac{1}{\sqrt{3}} \, \Xi^{*-} \,, \quad 
  T^{333} = \Omega \,. 
  \label{eq:tensorT}
\end{eqnarray}
The Goldstone bosons are encoded in
\begin{eqnarray}
  \Phi &=&  \left(
    \begin{array}{ccc}
      \pi^0 +\frac{1}{\sqrt{3}}\, \eta 
      & \sqrt{2}\, \pi^+ & \sqrt{2} \, K^+ \\
      \sqrt{2}\, \pi^- & -\pi^0+\frac{1}{\sqrt{3}}\, \eta
      & \sqrt{2} \, K^0 \\
      \sqrt{2}\, K^- & \sqrt{2} \, {\bar{K}}^0 
      & -\frac{2}{\sqrt{3}}\, \eta
    \end{array}   
  \right) 
  \,, \nonumber \\
  u^2 & := & U := \exp(i\Phi/F_\pi) \,, \quad u_\mu := i \, u^\dagger \, (\nabla_\mu U) \, u^\dagger = u_\mu^\dagger \,. \phantom{mm}
  \label{eq:gold}
\end{eqnarray}

The fields have the following transformation properties with respect to chiral 
transformations \cite{Jenkins:1991es,Scherer:2012xha}:
\begin{eqnarray}
  U \to L \, U \, R^\dagger & \,, &  u \to L \, u \, h^\dagger = h \, u \, R^\dagger  \,, \nonumber \\
  \label{eq:chiraltrafos}  
  u_\mu \to h \, u_\mu \, h^\dagger & \,, &  B \to h \, B \, h^\dagger \,,   \\
  T^{abc}_\mu \to h^a_{d} \, h^b_{e} \, h^c_{f} \, T^{def}_\mu & \,, & 
  \bar T_{abc}^\mu \to (h^\dagger)_a^{d} \, (h^\dagger)_b^{e} \, (h^\dagger)_c^{f} \, \bar T_{def}^\mu \,.  \nonumber 
\end{eqnarray}
In particular, the choice of upper and lower flavor indices is used to indicate that upper indices transform with $h$ 
under flavor transformations while the lower components transform with $h^\dagger$. 

For a (baryon) octet the chirally covariant derivatives are defined by
\begin{eqnarray}
  \label{eq:devder}
  D^\mu B := \partial^\mu B + [\Gamma^\mu,B]   \,,
\end{eqnarray}
for a decuplet $T$ by
\begin{eqnarray}
  (D^\mu T)^{abc} &:=& \partial^\mu T^{abc} + (\Gamma^\mu)^a_{a'} T^{a' bc} + (\Gamma^\mu)^b_{b'} T^{a b' c} \nonumber \\
  && {} + (\Gamma^\mu)^c_{c'} T^{a bc'}   \,,
  \label{eq:devderdec}
\end{eqnarray}
for an anti-decuplet by 
\begin{eqnarray}
  (D^\mu \bar T)_{abc} &:=& \partial^\mu \bar T_{abc} - (\Gamma^\mu)_a^{a'} \bar T_{a' bc} - (\Gamma^\mu)_b^{b'} \bar T_{a b' c}  
  \nonumber \\
  && {} - (\Gamma^\mu)_c^{c'} \bar T_{a bc'}   \,,
  \label{eq:devderantidec}
\end{eqnarray}
and for the Goldstone boson fields by
\begin{eqnarray}
  \label{eq:devderU}
  \nabla_\mu U := \partial_\mu U -i(v_\mu + a_\mu) \, U + i U \, (v_\mu - a_\mu)
\end{eqnarray}
with
\begin{eqnarray}
  \Gamma_\mu &:=&  \frac12 \, \left(
    u^\dagger \left( \partial_\mu - i (v_\mu + a_\mu) \right) u \right. \nonumber \\
    && \phantom{m} \left. {}+
    u \left( \partial_\mu - i (v_\mu - a_\mu) \right) u^\dagger
  \right) \,,
  \label{eq:defGammamu}
\end{eqnarray}
where $v$ and $a$ denote external sources. 

In (\ref{eq:baryonlagr}) $m_{(8)}$ ($m_{(10)}$) denotes the mass of the baryon octet (decuplet) in the chiral limit. 
For the next-to-leading-order (NLO) calculation that we perform in the present work we use the 
physical masses \cite{pdg} of all states. Indeed, for the octet and decuplet the 
flavor breaking terms that appear at NLO, cf.\ (\ref{eq:octetNLO}), (\ref{eq:decupletNLO}) below, 
are capable of splitting up the baryon masses such that they 
are sufficiently close to the physical masses; see, e.g.\ the corresponding discussion in \cite{Kubis:2000aa,Holmberg:2018dtv}.

For the coupling constants we use $D=0.80$, $F=0.46$ which implies for the pion-nucleon coupling constant $g_A=F+D =1.26$. The value for $h_A$ can be determined from
the partial decay width $\Sigma^* \to \pi \, \Lambda$ or $\Sigma^* \to \pi \, \Sigma$ 
yielding $h_A=2.3\pm 0.1$ \cite{Granados:2017cib}. For a large number of colors, $N_c$, one obtains the following relations 
for two or three flavors, respectively:
$h_A=3 g_A/\sqrt{2} \approx 2.67$ according to \cite{Pascalutsa:2005nd,Pascalutsa:2006up,Ledwig:2011cx} 
or $h_A = 2\sqrt{2} D \approx 2.26$ according to \cite{Dashen:1993as,Semke:2005sn}. 
Finally one has to specify $H_A$. In absence of a simple direct observable to pin it down we take estimates from large-$N_c$ 
considerations: $H_A = \frac95 \, g_A \approx 2.27$ \cite{Pascalutsa:2006up,Ledwig:2011cx} 
or $H_A = 9F -3D \approx 1.74$ \cite{Dashen:1993as,Semke:2005sn}.
Numerically we explore the range $H_A = 2.0 \pm 0.3$. 
We have checked explicitly that the sign of $H_A$ is in 
agreement with \cite{Pascalutsa:2006up,Ledwig:2011cx} and also with \cite{Semke:2005sn}. For quark-model estimates of these
coupling constants see \cite{Buchmann:1999ab,Buchmann:2013fxa}. 
For our purposes the interaction term proportional to
$H_A$ effectively reduces to
\begin{eqnarray}
  \label{eq:HAeff}
  + \frac{H_A}{2 m_R \, F_\pi} \, \epsilon_{\mu\nu\alpha\beta} \, 
  \bar T^\mu_{abc} \, \partial^\nu (T^\alpha)^{abd} \, \partial^\beta\Phi^c_d \,.
\end{eqnarray}

Working with relativistic spin-3/2 Rarita-Schwinger fields is plagued by ambiguities how to deal with the spurious spin-1/2 
components. In the present context the interaction term $\sim h_A$ causes not only the proper exchange of spin-3/2 resonances, 
but induces an additional contact interaction. This unphysical contribution can be avoided by constructing interaction terms 
according to the Pascalutsa description \cite{Pascalutsa:1999zz,Pascalutsa:2005nd,Pascalutsa:2006up,Ledwig:2011cx}. 
It boils down to the replacement
\begin{eqnarray}
  \label{eq:replace}
  T^\mu \to -\frac{1}{m_R} \, \epsilon^{\nu\mu\alpha\beta} \, \gamma_5 \, \gamma_\nu \, \partial_\alpha T_\beta
\end{eqnarray}
where $m_R$ denotes the resonance mass.
Strictly speaking this procedure induces an explicit flavor breaking, but such effects 
are anyway beyond leading order. In practice, we take the mass of the $\Sigma^*$ resonance. 
The $H_A$ term of (\ref{eq:HAeff}) is already constructed such that only the spin-3/2 components contribute.

We will explore both the standard interaction term $\sim h_A$ from (\ref{eq:baryonlagr}) and the corresponding one obtained 
by (\ref{eq:replace}). We will show explicitly that differences can be accounted for by contact interactions 
of the chiral Lagrangian at NLO and beyond. 
Quantitatively, it is interesting to see how much the contact terms $P_m$ in \eqref{eq:tmandel}
are changed when switching from the standard to the Pascalutsa interaction. This provides an uncertainty estimate if $P_m$
is not determined from a fit to form factor data. 
In principle we could do the same for the $H_A$ term and start instead with a simpler 
Lagrangian $\sim \bar T^\mu_{abc} [\slashed{u}]^c_d \gamma_5 T_\mu^{abd}$. But we refrain from this exercise.

Now we turn to the Lagrangian of second order in the chiral counting.
A complete and minimal NLO Lagrangian has been presented in \cite{Holmberg:2018dtv}. For our present purpose we need terms
that lift the mass degeneracies that hold at LO and we need terms that provide interactions for 
$\Sigma^* \pi \to \Lambda \pi$ (or formally $\Sigma^* \bar\Lambda \to 2\pi$) with the two pions in a p-wave. 

The relevant part of the NLO Lagrangian for the baryon octet sector reads \cite{Oller:2006yh,Frink:2006hx,Holmberg:2018dtv}
\begin{eqnarray}
  {\cal L}_{8}^{(2)} &=& b_{\chi,D} \, {\rm tr}(\bar B \, \{\chi_+,B\}) + b_{\chi,F} \, {\rm tr}(\bar B \, [\chi_+,B]) 
  \label{eq:octetNLO}
\end{eqnarray}
with $\chi_\pm = u^\dagger \chi u^\dagger \pm u \chi^\dagger u$ and $\chi = 2 B_0 \, (s+ip)$ 
obtained from the scalar source $s$ and the pseudoscalar source $p$. The low-energy constant $B_0$ is essentially the ratio of
the light-quark condensate and the square of the pion-decay constant; 
see, e.g.\ \cite{Gasser:1983yg,Gasser:1984gg,Scherer:2002tk,Scherer:2012xha}. 
While at LO all baryon octet states are degenerate in mass, the NLO terms of (\ref{eq:octetNLO}) lift this degeneracy and 
essentially move all masses to their respective physical values. Technically this is achieved if one replaces the 
scalar source $s$ by the quark mass matrix. Numerical results for the octet mass $m_{(8)}$ 
in (\ref{eq:baryonlagr}) and the splitting parameters $b_{\chi,D/\chi,F}$ in (\ref{eq:octetNLO}) 
are given, for instance, in \cite{Kubis:2000aa}. In practice we use the physical masses. Therefore we do not specify these 
parameters here.

The relevant part of the NLO Lagrangian for the baryon decuplet sector reads \cite{Holmberg:2018dtv}
\begin{eqnarray}
  {\cal L}_{10}^{(2)} &=& -d_{\chi,(8)} \bar T_{abc}^\mu \, (\chi_+)^c_d \, \gamma_{\mu\nu} \, (T^\nu)^{abd}  \,.
  \label{eq:decupletNLO}
\end{eqnarray}
It provides a mass splitting for the decuplet baryons such that 
$m_\Omega - m_{\Xi^*} = m_{\Xi^*} - m_{\Sigma^*} =  m_{\Sigma^*} - m_\Delta$, in good agreement with phenomenology \cite{pdg}. 
In the present work we only deal with the $\Sigma^*$. 
In practice we use the physical mass of the neutral $\Sigma^*$. In that way the physical thresholds are exactly reproduced. 

More concretely we use the following masses (in GeV): $m_\pi=0.13957$, $m_\Lambda=1.116$, $m_\Sigma=1.193$ and $m_{\Sigma^*}=1.384$.

For the formal reaction $\Sigma^* \bar\Lambda \to 2\pi$ the relevant part of the NLO Lagrangian \cite{Holmberg:2018dtv} 
is given by
\begin{eqnarray}
  \label{eq:NLOtrans}
  {\cal L}_{8-10}^{(2)} & \to & \frac{c_F}{2F_\pi^2} \, \bar \Lambda \gamma_\mu \gamma_5 \Sigma_\nu^{*0} 
  \left(\partial^\mu \pi^+\,\partial^\nu \pi^- - (\mu \leftrightarrow \nu) \right)  
 . \phantom{mm}
\end{eqnarray}
A vector-meson-dominance estimate for $c_F$ is provided in Appendix \ref{sec:vmd}.

\section{Results}
\label{sec:results}

\subsection{Matrix elements}
The first step is the calculation of the pion-hyperon tree-level amplitudes, i.e.\ $\chi$PT amplitudes up to (including) NLO.
In practice, the extraction of the reduced amplitudes is simplified and systemized by a projector formalism presented 
in Appendix \ref{sec:proj}. 

\begin{widetext}
The Feynman matrix element for the reaction $\Sigma^{*0} \bar \Lambda \to \pi^+ \pi^-$ up to (including) NLO is given by
\begin{eqnarray}
  && - \frac{D h_A}{6 \sqrt{2} F_\pi^2} \, \frac{1}{t-m_\Sigma^2+i\epsilon} \, 
  p^\mu_{\pi^+} g_{\mu\alpha} \, \bar v_\Lambda \, \slashed{p}_{\pi^-} \gamma_5 \, 
  (\slashed{p}_{\Sigma^*} - \slashed{p}_{\pi^+} + m_\Sigma) \, u^\alpha_{\Sigma^*} \nonumber \\
  && {} + \frac{D h_A}{6 \sqrt{2} F_\pi^2} \, \frac{1}{u-m_\Sigma^2+i\epsilon} \, 
  p^\mu_{\pi^-} g_{\mu\alpha} \, \bar v_\Lambda \, \slashed{p}_{\pi^+} \gamma_5 \, 
  (\slashed{p}_{\Sigma^*} - \slashed{p}_{\pi^-} + m_\Sigma) \, u^\alpha_{\Sigma^*}
  \nonumber \\
  && {}+ \frac{h_A H_A}{6 \sqrt{2} m_{\Sigma^*} F_\pi^2} \, i \epsilon^\lambda_{\phantom{\lambda}\nu\alpha\beta} \, 
  p^\nu_{\Sigma^*} \, p^\beta_{\pi^+} \, p^\mu_{\pi^-} \, \bar v_\Lambda S_{\mu\lambda}(p_{\Sigma^*}-p_{\pi^+}) \, u^\alpha_{\Sigma^*} 
  \nonumber \\
  && {} - \frac{h_A H_A}{6 \sqrt{2} m_{\Sigma^*} F_\pi^2} \, i \epsilon^\lambda_{\phantom{\lambda}\nu\alpha\beta} \, 
  p^\nu_{\Sigma^*} \, p^\beta_{\pi^-} \, p^\mu_{\pi^+} \, \bar v_\Lambda S_{\mu\lambda}(p_{\Sigma^*}-p_{\pi^-}) \, u^\alpha_{\Sigma^*} 
  \nonumber \\
  && {} + \frac{c_F}{2 F_\pi^2} \, (p^\mu_{\pi^+} p^\alpha_{\pi^-} - p^\alpha_{\pi^+} p^\mu_{\pi^-} ) \, g_{\alpha\beta} \, 
  \bar v_\Lambda \gamma_\mu \gamma_5 u^\beta_{\Sigma^*}   \,.
  \label{eq:feynlo+nlo}
\end{eqnarray}
Here $S_{\mu\nu}$ denotes the spin-3/2 propagator given in \eqref{eq:defspin32prop2}.

The $\Sigma$ and $\Sigma^*$ exchange diagrams yield the following amplitudes:
\begin{eqnarray}
  K_{+1} & = & \frac{D h_{A}}{6 \sqrt{2} F_{\pi }^2}( C_{+1} + D_{+1} \, R^{\rm oct.}_s)
  + \frac{h_{A} H_{A}}{6 \sqrt{2} F_{\pi }^2} \, ( E_{+1} + F_{+1} \, R^{\rm dec.}_s) \,, \nonumber \\
  K_{-1} & = & \frac{D h_{A}}{6 \sqrt{2} F_{\pi }^2}( C_{-1} + D_{-1} \, R^{\rm oct.}_s)
  + \frac{h_{A} H_{A}}{6 \sqrt{2} F_{\pi }^2} \, ( E_{-1} + F_{-1} \, R^{\rm dec.}_s)  \,, \nonumber \\
  K_{0} & = & \frac{D h_{A}}{6 \sqrt{2} F_{\pi }^2}( C_{0} + D_{0} \, R^{\rm oct.}_d)
  + \frac{h_{A} H_{A}}{6 \sqrt{2} F_{\pi }^2} \, ( E_{0} + F_{0} \, R^{\rm dec.}_d) 
  \label{eq:K-pw}  
\end{eqnarray}
with 
\begin{eqnarray}
  R^{\rm oct.}_s & = &
  \frac{-2Y_{\Sigma}}{\kappa^2} \left(1-\left(1-\frac{Y_{\Sigma}^2}{\kappa^2}\right) \frac{|\kappa|}{Y_{\Sigma}} 
    \left(\arctan\left(\frac{|\kappa|}{Y_{\Sigma}}\right)+\pi  \Theta (s_{3}-s)\right)\right) \,, \nonumber \\[0.8em]
  R^{\rm oct.}_d & = & \frac{4}{\kappa^2} \left(1-\frac{Y_{\Sigma}}{|\kappa|} 
    \left(\arctan\left(\frac{|\kappa|}{Y_{\Sigma}}\right)+\pi  \Theta (s_{3}-s)\right)\right)  \,, \nonumber \\[0.8em]
  R^{\rm dec.}_s & = & \frac{-2Y_{\Sigma^{*}}}{\kappa^2} \left(1-\left(1-\frac{Y_{\Sigma^{*}}^2}{\kappa^2}\right) 
    \frac{|\kappa|}{Y_{\Sigma^{*}}} \, \arctan\left(\frac{|\kappa|}{Y_{\Sigma^{*}}}\right) \right) \,, \nonumber  \\[0.8em]
  R^{\rm dec.}_d & = & \frac{4}{\kappa^2} \left(1-\frac{Y_{\Sigma^{*}}}{|\kappa|} 
    \, \arctan\left(\frac{|\kappa|}{Y_{\Sigma^{*}}}\right)\right)
  \label{eq:defRsd}
\end{eqnarray}
\end{widetext}
and
\begin{eqnarray}
  \label{eq:defABs}
  Y_{\Sigma} & = & 2 m_\Sigma^2-m_{\Sigma^*}^2-m_\Lambda^2-2 m_\pi^2+s \,, \\
  Y_{\Sigma^{*}} & = &  m_{\Sigma^*}^2-m_\Lambda^2-2 m_\pi^2+s \,,  \\
  \kappa^2 & = & \frac{1}{s} \, (s-4 m_\pi^2) \, \lambda(s,m_{\Sigma^*}^2,m_\Lambda^2)   \,, \\
  s_{3} & = & m_{\Sigma^*}^2+m_\Lambda^2+2 m_\pi^2-2 m_\Sigma^2  \,.
\end{eqnarray}
Note that $\kappa^2$ is negative in the range $4 m_\pi^2 < s < (m_{\Sigma^*} + m_\Lambda)^2$, i.e.\ $\vert \kappa \vert = \sqrt{-\kappa^2}$.
Only for negative $\kappa^2$ the expressions \eqref{eq:defRsd} are correct. For positive $\kappa^2$ one has log's instead of arctan's.

Finally the coefficient functions in \eqref{eq:K-pw} are given by
\begin{align}
  C_{+1} = {} & - \frac{2 \, (m_{\Sigma^*} - m_\Lambda) \, (m_\Lambda + m_\Sigma)}{s-\left(m_{\Sigma^*}-m_\Lambda\right){}^2}  \,,
  \\
  C_{-1} = {} & - \frac{6\, (m_{\Sigma^*} - m_\Lambda) \, (m_\Lambda + m_\Sigma)}{s-\left(m_{\Sigma^*}-m_\Lambda\right){}^2}
  \,, \\
  C_{0} = {} & \frac{(m_{\Sigma^*}+m_\Lambda) \, (m_{\Sigma^*}+m_\Sigma)}{s} \nonumber \\ 
  & {}-\frac{3 m_{\Sigma^*} (m_\Lambda + m_\Sigma)}{s-(m_{\Sigma^*}-m_\Lambda)^2}
\,,
\end{align}
\begin{widetext}
\begin{align}
  D_{+1} = {} & 3 m_\Sigma \, (m_\Lambda+m_\Sigma) + \frac{3 \, (m_{\Sigma^*}-m_\Lambda) \, (m_\Lambda+m_\Sigma) \, 
  (m_\pi^2 + m_{\Sigma^*} m_\Lambda - m_\Sigma^2)}{s-(m_{\Sigma^*}-m_\Lambda)^2}
  \,, \\
  D_{-1} = {} & \frac{3}{m_{\Sigma^*}} \, (m_\Lambda+m_\Sigma) \, (m_\pi^2 -m_{\Sigma^*}^2 + m_{\Sigma^*} m_\Sigma - m_\Sigma^2)
                + \frac{9 \, (m_{\Sigma^*}-m_\Lambda) \, (m_\Lambda+m_\Sigma) \, 
  (m_\pi^2 + m_{\Sigma^*} m_\Lambda - m_\Sigma^2)}{s-(m_{\Sigma^*}-m_\Lambda)^2}
  \,, \\
  D_{0} = {} & 3m_\Sigma(m_\Lambda+m_\Sigma)(m_{\Sigma^*}^2-m_{\Sigma^*}m_\Sigma-m_\pi^2+m_\Sigma^2)
               -\frac{9m_{\Sigma^*}(m_\Lambda+m_\Sigma)(m_{\Sigma^*}m_\Lambda+m_\pi^2-m_\Sigma^2)^2}{s-(m_{\Sigma^*}-m_\Lambda)^2} \nonumber \\
  & +\frac{3(m_{\Sigma^*}+m_\Lambda)(m_\Sigma+m_\Lambda)}{s}\Big(m_{\Sigma^*}^3m_\Lambda-m_\Sigma(m_{\Sigma^*}-m_\Lambda)(m_{\Sigma^*}^2+m_\pi^2)+2m_{\Sigma^*}^2m_\pi^2 \nonumber \\
              & \phantom{mmmmmmmmm}
 -m_\Sigma^2\big(m_{\Sigma^*}(m_{\Sigma^*}+m_\Lambda)+2m_\pi^2\big)+2m_{\Sigma^*}m_\Lambda m_\pi^2-m_\Sigma^3(m_\Lambda-m_{\Sigma^*})+m_\pi^4+m_\Sigma^4\Big)  \,,
\end{align}
\begin{align}
  E_{+1} = {} & \frac{(m_{\Sigma^*}-m_\Lambda) \left((m_{\Sigma^*}+m_\Lambda)^2- m_\pi^2\right)}%
  {3 m_{\Sigma^*} \left(s-\left(m_{\Sigma^*}-m_\Lambda\right){}^2\right)}
  \,, \\
  E_{-1} = {} & \frac{(m_{\Sigma^*}-m_\Lambda) \left((m_{\Sigma^*}+m_\Lambda)^2-m_\pi^2 \right)}{m_{\Sigma^*} (s-(m_{\Sigma^*}-m_\Lambda)^2)} \,, \\
  E_{0} = {} & - \frac{( m_{\Sigma^*}+m_\Lambda) (2m_{\Sigma^*}^2+2m_{\Sigma^*}m_\Lambda -m_\pi^2)}{6 m_{\Sigma^*}\, s}
 + \frac{(m_{\Sigma^*}+m_\Lambda)^2-m_\pi^2}
 {2 (s-(m_{\Sigma^*}-m_\Lambda )^2 )}  \,,
\end{align}
\begin{align}
  F_{+1} = {} & -\frac{3s}{2}-\frac{m_\pi^2(2m_{\Sigma^*}+3m_\Lambda)}{2m_{\Sigma^*}}+\frac{5(m_{\Sigma^*}+m_\Lambda)^2}{2}
  +\frac{(m_{\Sigma^*}-m_\Lambda)((m_{\Sigma^*}+m_\Lambda)^2-m_\pi^2)(m_{\Sigma^*}^2-m_{\Sigma^*}m_\Lambda-m_\pi^2)}{2m_{\Sigma^*}(s-(m_{\Sigma^*}-m_\Lambda)^2)}  \,, \\
  F_{-1} = {} & \frac{3s}{2} +\frac{m_\pi^2( m_{\Sigma^*}^2+ m_{\Sigma^*} m_\Lambda-m_\Lambda^2)+m_\pi^4}{2 m_{\Sigma^*}^2} -\frac{5( m_{\Sigma^*}+m_\Lambda)^2}{2} \nonumber \\
  & +\frac{3(m_{\Sigma^*}-m_\Lambda)((m_{\Sigma^*}+m_\Lambda)^2-m_\pi^2)(m_{\Sigma^*}^2-m_{\Sigma^*}m_\Lambda-m_\pi^2)}{2m_{\Sigma^*}(s-(m_{\Sigma^*}-m_\Lambda)^2)}   \,, \\
  F_{0} = {} & \frac{3m_{\Sigma^*}^2\, s}{2} -\frac{m_\pi^2(7m_{\Sigma^*}^2-2m_{\Sigma^*}m_\Lambda+2m_\Lambda^2)+m_{\Sigma^*}^2(m_{\Sigma^*}+m_\Lambda)^2}{2}+m_\pi^4  \nonumber \\
  & +\frac{4m_{\Sigma^*}^2m_\pi^2(m_{\Sigma^*}-2m_\Lambda)(m_{\Sigma^*}+m_\Lambda)^2-m_\pi^4(2m_{\Sigma^*}^3+m_{\Sigma^*}^2m_\Lambda+m_\Lambda^3)+m_\pi^6(m_{\Sigma^*}+m_\Lambda)}{2m_{\Sigma^*} \, s} \nonumber \\
  & +\frac{3((m_{\Sigma^*}+m_\Lambda)^2-m_\pi^2)(m_{\Sigma^*}(m_\Lambda-m_{\Sigma^*})+m_\pi^2)^2}{2(s-(m_{\Sigma^*}-m_\Lambda)^2)}   \,.
\end{align}
\end{widetext}

The explicit expressions for the polynomial terms are
\begin{eqnarray}
  P_{+1} & = & \frac{D h_{A}}{6 \sqrt{2} F_{\pi }^2} 2
  + \frac{h_{A} H_{A}}{6 \sqrt{2} F_{\pi }^2} \, \frac{5 \, (m_{\Sigma^*}+m_\Lambda)}{6m_{\Sigma^*}}   \,, \nonumber \\
  P_{-1} & = & \frac{D h_{A}}{6 \sqrt{2} F_{\pi }^2} \frac{2 \, (m_{\Sigma^*}-m_\Lambda-m_\Sigma)}{m_{\Sigma^*}} \nonumber \\
               && {} + \frac{h_{A} H_{A}}{6 \sqrt{2} F_{\pi }^2} 
               \, \frac{s-2 m_\pi^2 - (m_{\Sigma^*}+m_\Lambda)(6m_{\Sigma^*}-m_\Lambda)}{6 m_{\Sigma^*}^2}
               \nonumber \\
         &\approx& \frac{D h_{A}}{6 \sqrt{2} F_{\pi }^2} \frac{2 \, (m_{\Sigma^*}-m_\Lambda-m_\Sigma)}{m_{\Sigma^*}} \nonumber \\
               && {} + \frac{h_{A} H_{A}}{6 \sqrt{2} F_{\pi }^2} \,
               \frac{-(m_{\Sigma^*}+m_\Lambda)(6m_{\Sigma^*}-m_\Lambda)}{6 m_{\Sigma^*}^2}
               \,, \nonumber \\
  P_{0} & = & \frac{D h_{A}}{6 \sqrt{2} F_{\pi }^2} 
  + \frac{h_{A} H_{A}}{6 \sqrt{2} F_{\pi }^2} \, \frac{3 m_{\Sigma^*}-m_\Lambda}{6 m_{\Sigma^*} }  \,.
  \label{eq:P-pw}  
\end{eqnarray}
For $P_{-1}$ we dropped terms which are suppressed by two orders in the chiral counting. 

The $\Sigma^*\Lambda\pi^+\pi^-$ contact diagram produces the following polynomials:
\begin{equation}\begin{split}
P^{\rm NLO\,\chi PT}_{+1}&=c_F \frac{m_{\Sigma^*}+m_\Lambda}{2 F_\pi^2}   \,, \\
P^{\rm NLO\,\chi PT}_{0}&=c_F \frac{m_{\Sigma^*}}{2 F_\pi^2}  \,,  \\
P^{\rm NLO\,\chi PT}_{-1}&=c_F \frac{s-m_\Lambda(m_{\Sigma^*}+m_\Lambda)}{2 F_\pi^2 m_{\Sigma^*}} \nonumber \\ &\approx
- c_F \frac{m_\Lambda(m_{\Sigma^*}+m_\Lambda)}{2 F_\pi^2 m_{\Sigma^*}} \,.
\end{split}\end{equation} 

The amplitudes \eqref{eq:feynlo+nlo} become slightly different when the Pascalutsa prescription is used:
new contact terms pop up but the pole terms and therefore \eqref{eq:K-pw} are not affected. In particular we have:
\begin{equation}\begin{split}
    P^{P}_{+1}&= P_{+1}  \\
    & {}+\frac{h_A H_A}{18\sqrt{2}F_\pi^2m_{\Sigma^*}^2}((m_\Lambda+m_{\Sigma^*})(2m_{\Sigma^*}+3m_\Lambda)-3s)   \,, \\
    P^{P}_{0} &= P_{0}
    -\frac{h_A H_A}{18\sqrt{2}F_\pi^2}   \,, \\
    P^{P}_{-1}&=P_{-1}  \\
    & {}-\frac{h_A H_A}{18\sqrt{2}F_\pi^2m_{\Sigma^*}^2}((m_\Lambda+m_{\Sigma^*})(3m_{\Sigma^*}+2m_\Lambda)-2s).
\end{split}\end{equation} 
As expected the $\Sigma$ exchange diagrams do not get any contribution since the external $\Sigma^*$ hyperon is on-shell. 

It is illuminating to translate these contact interactions to the $i=1,2,3$ amplitudes defined in \eqref{eq:TiT0pm1}.
One obtains
\begin{eqnarray}
  \label{eq:NLO123}
  P^{\rm NLO \, \chi PT}_1 = \frac{c_F}{2 m_{\Sigma^*} \, F_\pi^2} \,, \quad P^{\rm NLO \, \chi PT}_{2,3} = 0 
\end{eqnarray}
and
\begin{eqnarray}
  P^{P}_1 &=& P_1 + 5 \, \frac{h_A H_A}{18\sqrt{2}F_\pi^2m_{\Sigma^*}^2} \,, \nonumber \\
  P^{P}_2 &=& P_2 - 6 \, \frac{h_A H_A}{18\sqrt{2}F_\pi^2m_{\Sigma^*}^2} \,, \nonumber \\
  P^{P}_3 &=& P_3  \,. 
  \label{eq:pascal123}
\end{eqnarray}
Thus the NLO contact term can be used to compensate for the difference between naive and Pascalutsa interaction concerning 
the $i=1$ amplitude structure, but not for $i=2$. In fact, there is a one-to-one correspondence between the contact terms of the
pion-hyperon scattering amplitudes and the constraint-free TFFs introduced at the very beginning in \eqref{eq:defTFF2}.
Chiral power counting shows that in $\chi$PT,
the TFF $F_i$ receives tree-level contributions starting at
chiral order $i+1$. At an NLO accuracy, one has only full access to $F_1$. Correspondingly, the NLO contact interaction for
the pion-hyperon amplitudes contributes only to $P_1$ as shown explicitly in \eqref{eq:NLO123}. To compensate the difference
between naive and Pascalutsa interaction concerning the $i=2$ amplitude structure, one needs a contact term from the
next-to-next-to-leading-order Lagrangian. This is beyond the scope of the present work.

\subsection{Numerical results}
\label{sec:Plots}

The results below have been obtained using Pascalutsa amplitudes. They consist in unsubtracted dispersion relations for the TFFs $G_m$ \eqref{eq:dispbasic-unsub}, evaluated at the photon point ($q^2=0$), 
followed by the corresponding radii:
\begin{equation}
  \langle r^2_m \rangle := \frac{6}{G_m(0)} \left. \frac{\text{d} G_m(q^2)}{\text{d}q^2} \right\vert_{q^2=0}  \,.
  \label{eq:radii}
\end{equation}
Other quantities of interest are the integrated decay rate for $\Sigma^*\rightarrow \Lambda e^+e^-$ and
the decay width for $\Sigma^*\rightarrow \Lambda\gamma$. 

We start by fixing the input parameters $h_A$, $H_A$ and $M_V$ to the respective central value. We will explore later the impact of their uncertainties on the final results, while we will not vary $D$ nor the pion phase shift since they are better constrained. We also want to investigate the dependence on the cutoff $\Lambda$, which we will take equal to 1 and 2 GeV, respectively. 
Furthermore recall that in order to account for the anomalous contribution \eqref{eq:tanom} to the scattering amplitudes, one needs to know the pion scattering amplitude $t(s)$ in the complex plane. We will explore two options: an approximation from $\chi$PT  \eqref{tChPT1}, denoted by $t_{\rm\chi PT}$, and its unitarized version  \eqref{tIAM1}, $t_{\mathrm{IAM}}$.

Our strategy is to adjust the dimensionless constants $c_m$'s according to \eqref{eq:detcm} and fix the NLO low-energy constant $c_F$ to the experimental value of the decay width $\Sigma^*\rightarrow\Lambda\gamma$ which is 0.452 MeV \cite{pdg}. In doing so one has two possible values ($c_F=-6.33$ GeV$^{-1}$ and $c_F=2.39$ GeV$^{-1}$) to choose from. We pick the first,\footnote{The results corresponding to the other choice for $c_F$ reflect the fact that the two possible values have opposite signs. Otherwise the results are qualitatively similar.} being closer to the VMD estimate \eqref{eq:estimate-cF}. The chosen value of $c_F$ is kept unchanged throughout the whole analysis, while the constants $c_m$'s are adjusted by \eqref{eq:detcm} each time any other parameter is varied. For completeness we report the $c_m$ values obtained when choosing $\Lambda=2$ GeV, $t_{\mathrm{IAM}}$, central values for $h_A,\, H_A$ and $M_V$: $c_{-1}=-0.59-0.04\,i,\, c_{0}=1.05 -0.10\, i,\, c_{+1}=0.96 -0.05\, i$.
This scenario gives rise to the results of Table \ref{tab:cutoff}, right column. 

From Table \ref{tab:cutoff} we get the encouraging message that varying the cutoff has a rather small impact.  
In Table \ref{tab:t} we compare the choice of using $t_{\rm\chi PT}$ versus $t_{\mathrm{IAM}}$.  As expected both approaches lead essentially to the same results. Finally we study the changes of the $G_m(0)$'s, the radii, the partial widths $\Gamma_{\Sigma^*\rightarrow\Lambda e^+e^-}$ and $\Gamma_{\Sigma^*\rightarrow\Lambda \gamma}$ when varying $h_A,\, H_A$ and $M_V$, one at a time, as shown in Table \ref{tab:ha}. The uncertainties related to $h_A,\, H_A$ and $M_V$ turn out to be moderate and comparable. It is satisfying  to observe that the $G_m(0)$ values are not subject to large changes and the radii are even less sensitive to these variations. In fact the dispersive machinery is supposed to work better for the radii since by definition they receive a suppressed contribution from the high-energy region.

As previously stated we stick to Pascalutsa amplitudes here, but the very same analysis can be carried out using the naive couplings instead. Note that it would then be necessary to refit $c_F$ since the meaning of the contact interaction changes based on which three-point coupling is used.\footnote{Again, the final results show similar qualitative behavior as in the Pascalutsa case.}
\begin{widetext}

\begin{table}[H]
  \centering
  \begin{tabular}{|c|c|c|}
    \hline
    quantity & $\Lambda = 1$ GeV & $\Lambda = 2$ GeV\\  \hline \hline
    $G_0(0)$ & $-3.5-0.0 \,i$ & $-3.7-0.0\, i$ \\  \hline 
    $\langle r_0^2 \rangle$ [GeV$^{-2}$] & $21.5+ 7.1\,i$ & $21.0+6.8 \,i$ \\  \hline 
    $G_{+1}(0)$ & $-4.5-0.0\,i $ & $-4.8-0.0 \, i$ \\  \hline 
    $\langle r_{+1}^2 \rangle$ [GeV$^{-2}$] & $16.9+ 1.3\,i$ & $16.7+1.3 \,i$  \\  \hline 
    $G_{-1}(0)$ & $3.2- 0.0\,i$ & $3.5-0.0\, i$ \\  \hline 
    $\langle r_{-1}^2 \rangle$ [GeV$^{-2}$] & $16.8- 1.2\,i$ & $16.5-1.2\,i$\\  \hline 
    $\Gamma_{\Sigma^*\rightarrow\Lambda e^+ e^-}$ [keV] & $3.0$ & $3.4$ \\  \hline 
    $\Gamma_{\Sigma^*\rightarrow\Lambda \gamma}$ [MeV] & $0.39$ & $0.45$ \\  \hline 
  \end{tabular}
  \caption{Comparison of the results for various observables using $t_{\mathrm{IAM}}$, $c_F=-6.33$ GeV$^{-1}$, central values for $h_A,\, H_A$ and $M_V$ and varying the cutoff $\Lambda$.}
  \label{tab:cutoff}
\end{table}
\begin{table}[H]
  \centering
  \begin{tabular}{|c|c|c|}
    \hline
    quantity & $t_{\rm\chi PT}$  & $t_{\mathrm{IAM}}$ \\  \hline \hline
    $G_0(0)$ & $-3.7-0.0 \,i$ & $-3.7-0.0\, i$ \\  \hline 
    $\langle r_0^2 \rangle$ [GeV$^{-2}$] & $20.8+ 6.7\,i$ & $21.0+6.8 \,i$ \\  \hline 
    $G_{+1}(0)$ & $-4.8-0.0\,i $ & $-4.8-0.0 \, i$ \\  \hline 
    $\langle r_{+1}^2 \rangle$ [GeV$^{-2}$] & $16.7+ 1.3\,i$ & $16.7+1.3 \,i$  \\  \hline 
    $G_{-1}(0)$ & $3.5- 0.0\,i$ & $3.5-0.0 i$ \\  \hline 
    $\langle r_{-1}^2 \rangle$ [GeV$^{-2}$] & $16.5- 1.2\,i$ & $16.5-1.2\,i$\\  \hline 
    $\Gamma_{\Sigma^*\rightarrow\Lambda e^+ e^-}$ [keV] & $3.4$ & $3.4$ \\  \hline 
    $\Gamma_{\Sigma^*\rightarrow\Lambda \gamma}$ [MeV] & $0.45$ & $0.45$ \\  \hline 
  \end{tabular}
  \caption{Same as Table \ref{tab:cutoff} for the comparison between  $t_{\rm\chi PT}$ and $t_{\mathrm{IAM}}$ using $\Lambda =2\,$GeV, $c_F=-6.33\,$GeV$^{-1}$, central values for $h_A$, $H_A$ and $M_V$.}
  \label{tab:t}
\end{table}

\begin{table}[H]
  \centering
  \begin{tabular}{|c|c|c|c|c|c|c|}
    \hline
    quantity & $h_A = 2.2$ & $h_A = 2.4$ & $H_A=1.7$ & $H_A=2.3$ & $M_V=1.4\,$GeV & $M_V=1.7\,$GeV \\  \hline \hline
    $G_0(0)$ & $-3.7-0.0 \,i$ & $-3.6-0.0\,i$ & $-3.7-0.0\,i$ & $-3.6-0.0\,i$ & $-3.4-0.0\,i$ & $-3.8+0.0\,i$ \\  \hline 
    $\langle r_0^2 \rangle$ [GeV$^{-2}$] & $20.6+6.5\,i$ & $21.3+7.1\, i$ & $21.1+6.8\,i$& $20.9+6.8\,i$ & $22.1+7.2\,i$ & $20.2+6.6\,i$ \\  \hline 
    $G_{+1}(0)$ & $-4.9-0.0\,i$ & $-4.7-0.0\,i$ & $-5.1-0.0\,i$ & $-4.6-0.0\,i$ & $-4.6-0.0\,i$ & $-5.0+0.0\,i$ \\  \hline 
    $\langle r_{+1}^2 \rangle$ [GeV$^{-2}$] & $16.5+1.2\,i$ & $16.9+1.3\,i$ & $16.3+1.2\,i$& $17.1+1.3\,i$ & $17.2+1.3\,i$ & $16.3+1.2\,i$  \\  \hline 
    $G_{-1}(0)$ & $3.6-0.0\,i$ & $3.4-0.0\,i$ & $3.8-0.0\,i$ & $3.2-0.0\,i$ & $3.4-0.0\,i$ & $3.6+0.0\,i$ \\  \hline 
    $\langle r_{-1}^2 \rangle$ [GeV$^{-2}$] & $16.3-1.1\,i$ & $16.8-1.3\,i$ & $16.0-1.1\,i$& $17.2-1.3\,i$ & $17.0-1.2\,i$ & $16.2-1.2\,i$\\  \hline 
    $\Gamma_{\Sigma^*\rightarrow\Lambda e^+ e^-}$ [keV] & $3.5$ & $3.3$ & $3.8$& $3.0$ & $3.1$ & $3.6$\\  \hline  
    $\Gamma_{\Sigma^*\rightarrow\Lambda \gamma}$ [MeV] & $0.47$ & $0.43$ & $0.51$ & $0.40$ & $0.41$ & $0.48$ \\  \hline 
  \end{tabular}
  \caption{Same as Table \ref{tab:cutoff} using $t_{\mathrm{IAM}}$, $\Lambda = 2 \,$GeV, 
    $c_F=-6.33\,$GeV$^{-1}$ and varying $h_A$, $H_A$ and $M_V$ one at a time.}
  \label{tab:ha}
\end{table}
\end{widetext}

\begin{figure}
  \centering
      \includegraphics[keepaspectratio,width=0.5\textwidth]{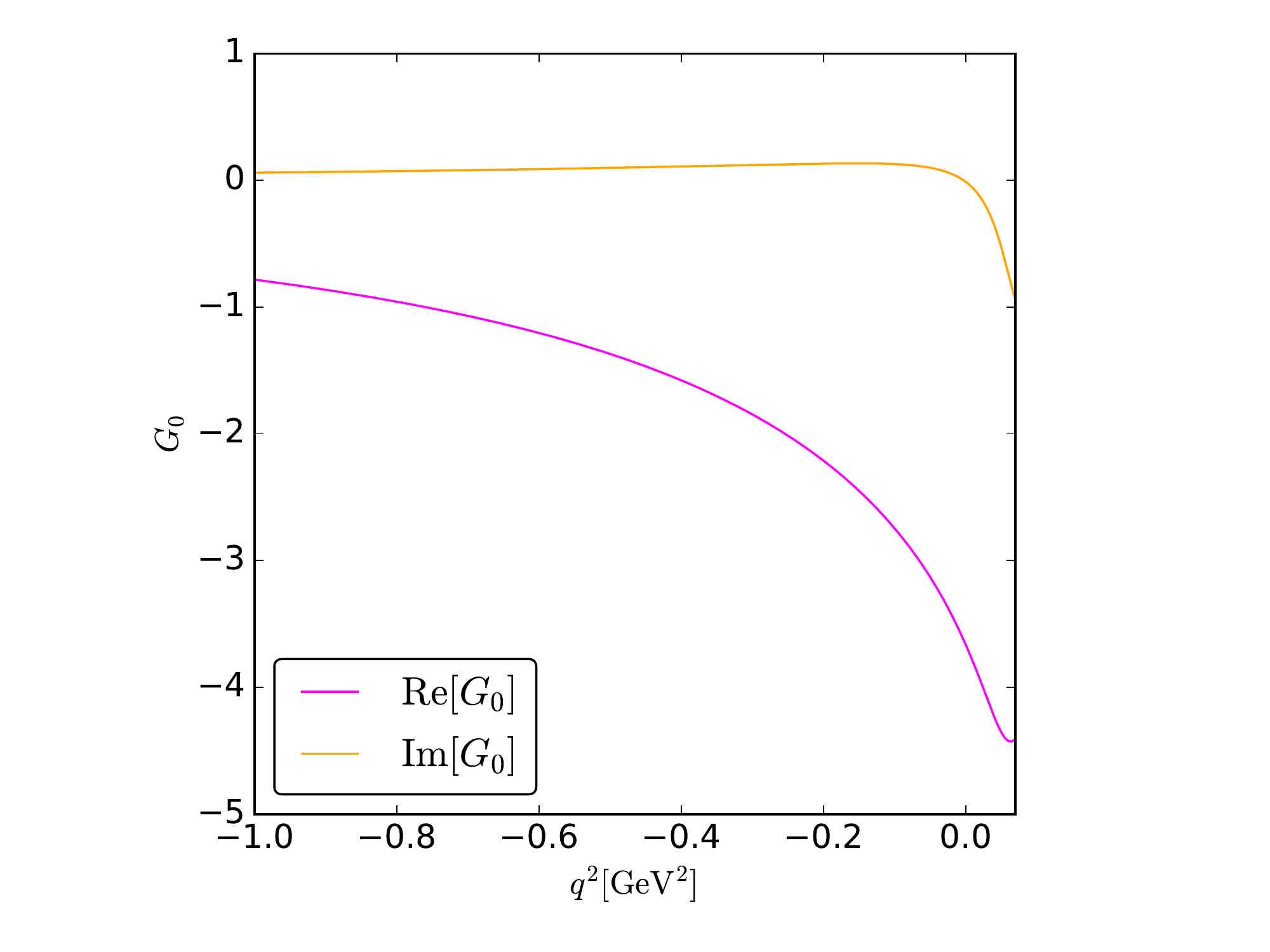}
  \caption{Real and imaginary part of $G_0(q^2)$.}
  \label{G0}
\end{figure}

\begin{figure}
  \centering
      \includegraphics[keepaspectratio,width=0.5\textwidth]{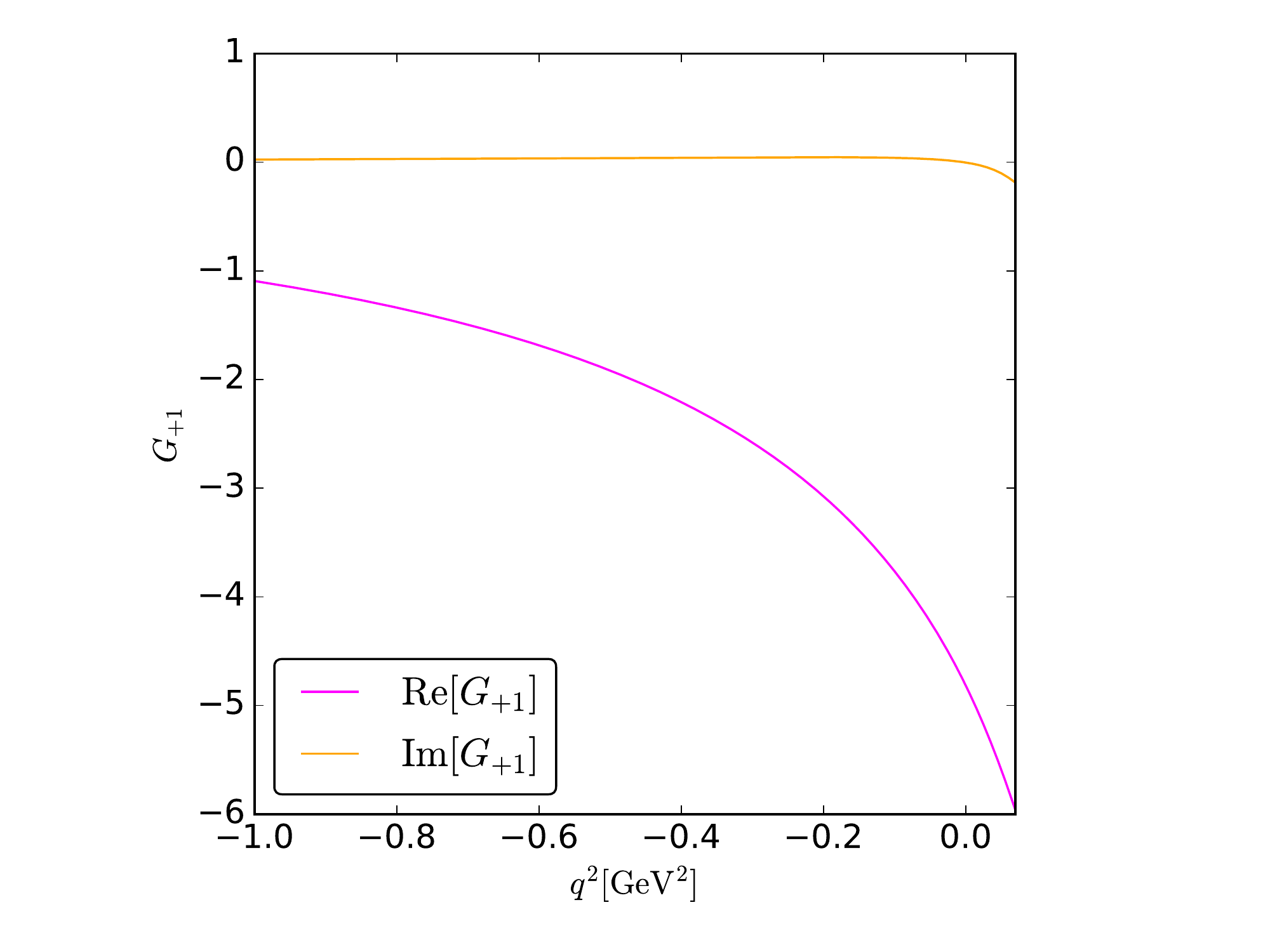}
  \caption{Real and imaginary part of $G_{+1}(q^2)$.}
  \label{Gp1}
\end{figure}

\begin{figure}
  \centering
      \includegraphics[keepaspectratio,width=0.5\textwidth]{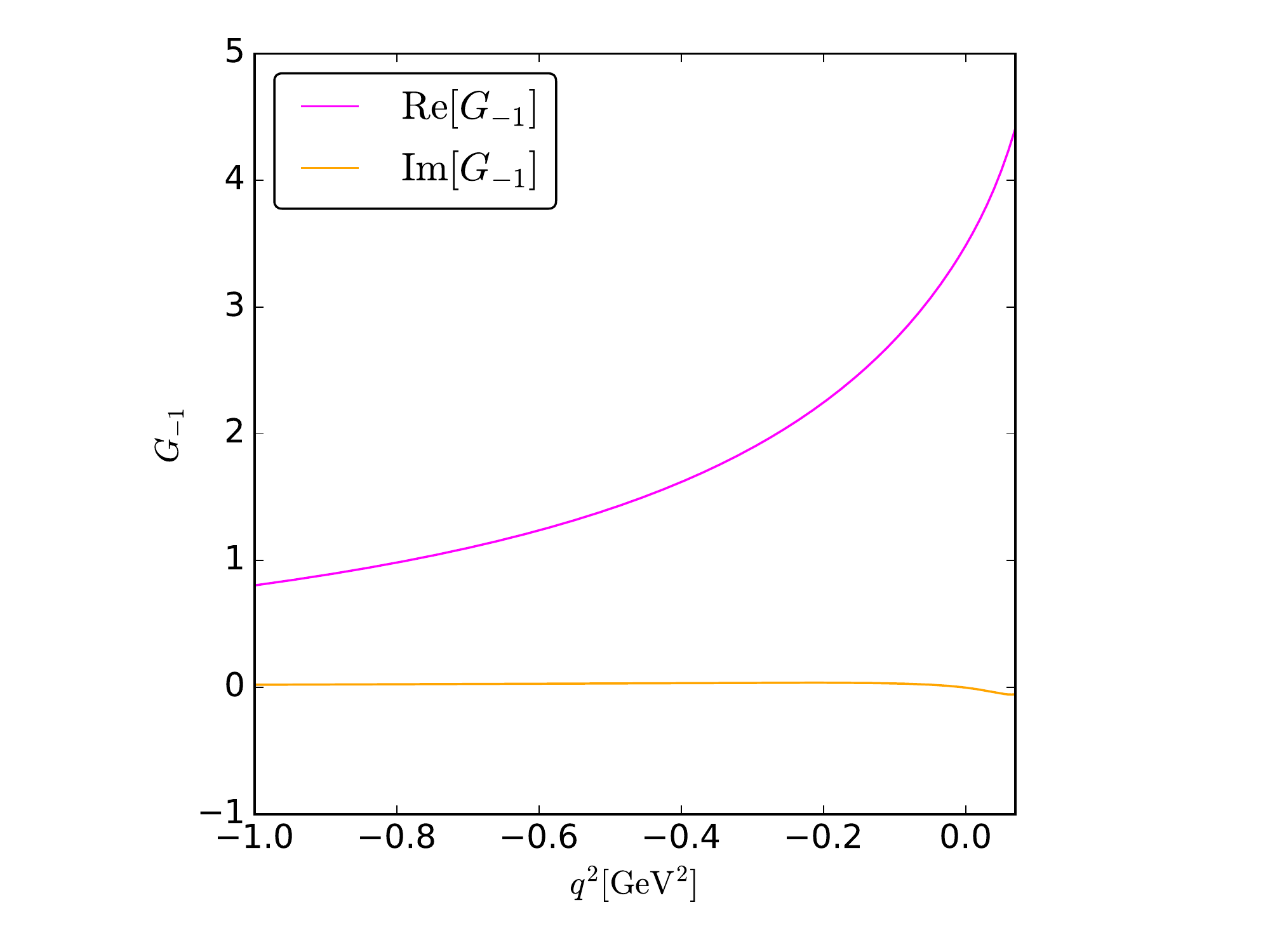}
  \caption{Real and imaginary part of $G_{-1}(q^2)$.}
  \label{Gm1}
\end{figure}

Still using $t_{\mathrm{IAM}}$, central values for $h_A,\, H_A$ and $M_V$ and cutoff $\Lambda=2$ GeV,  we plot the real and imaginary part of the TFFs  $G_m(q^2)$ \eqref{eq:dispbasic-unsub} in the space- and timelike regions, up to $q^2=(m_{\Sigma^*}-m_{\Lambda})^2$. As shown in Figs.\ \ref{G0}, \ref{Gp1}, \ref{Gm1}, all three functions are complex, already below the two-pion threshold. Technically this is a consequence of the additional anomalous cut located on the first Riemann sheet.      

\begin{figure}[H]
  \centering
      \includegraphics[keepaspectratio,width=0.5\textwidth]{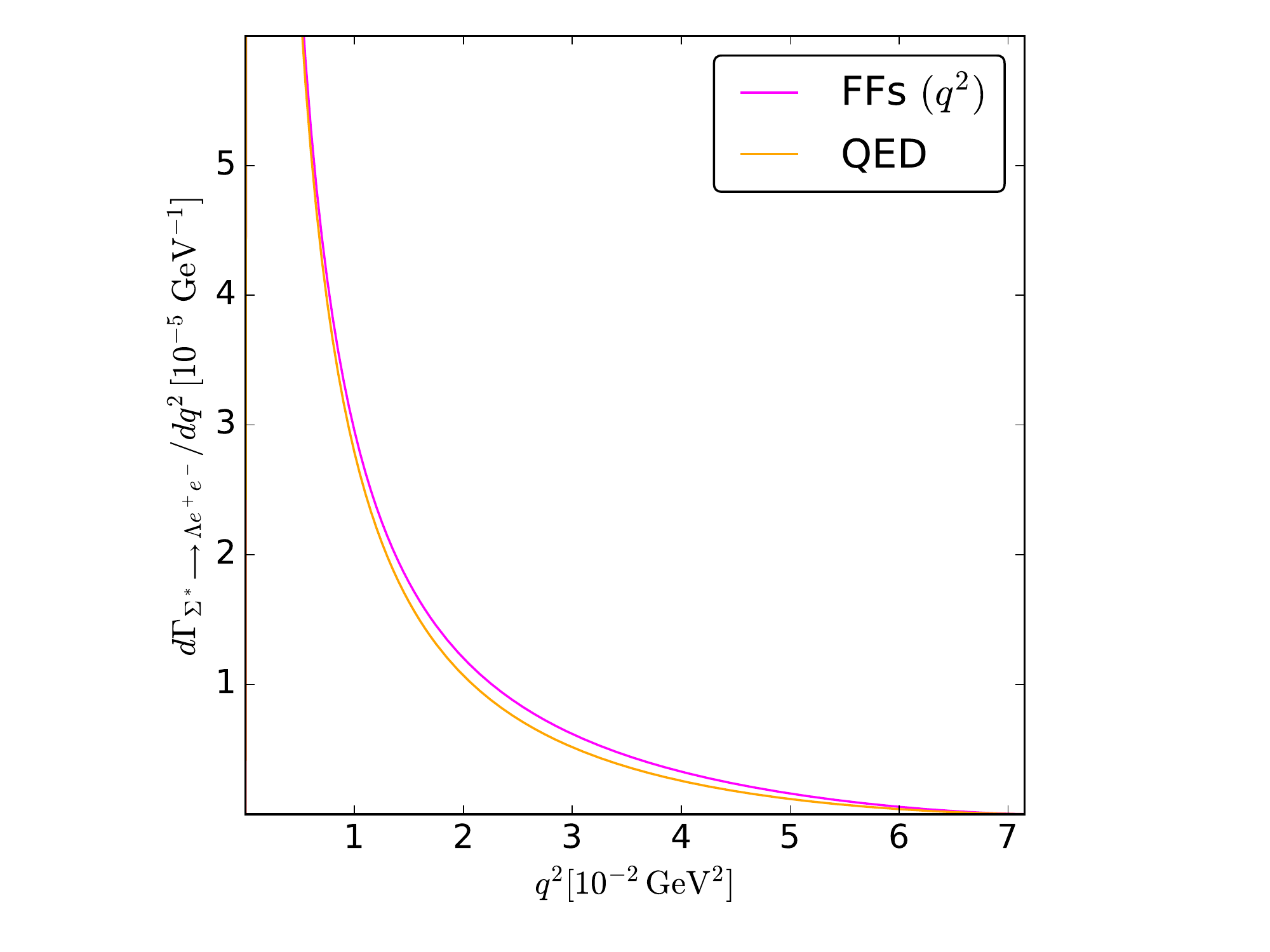}
  \caption{Single-differential decay width for the $\Sigma^*\rightarrow \Lambda e^+ e^-$ Dalitz decay. The curve labeled ``FFs$\,(q^2)$'' is the angular integral of \eqref{eq:Dalitzdistr}. The other curve is the QED analogue, given by \eqref{eq:QEDcase}.}
  \label{fig:Gamma}
\end{figure}

We plot the single differential decay width $\mathrm{d}\Gamma/\mathrm{d}q^2$ for the Dalitz decay $\Sigma^*\rightarrow \Lambda e^+ e^-$, i.e. the angular integral of \eqref{eq:Dalitzdistr}, in the Dalitz region $4m_e^2 < q^2 < (m_{\Sigma^*}-m_\Lambda)^2$. In particular in Fig.\ \ref{fig:Gamma} we show a comparison with the corresponding QED case \eqref{eq:QEDcase}, for which the $q^2$-dependence of the TFFs is not resolved. 
The two curves show a slight off-set in the central region, but essentially coincide over the whole range. This implies that high resolution is needed from the experimental side in order to appreciate this discrepancy and gain new insight on the internal structure of hyperons. 

\begin{figure}[H]
  \centering
      \includegraphics[keepaspectratio,width=0.5\textwidth]{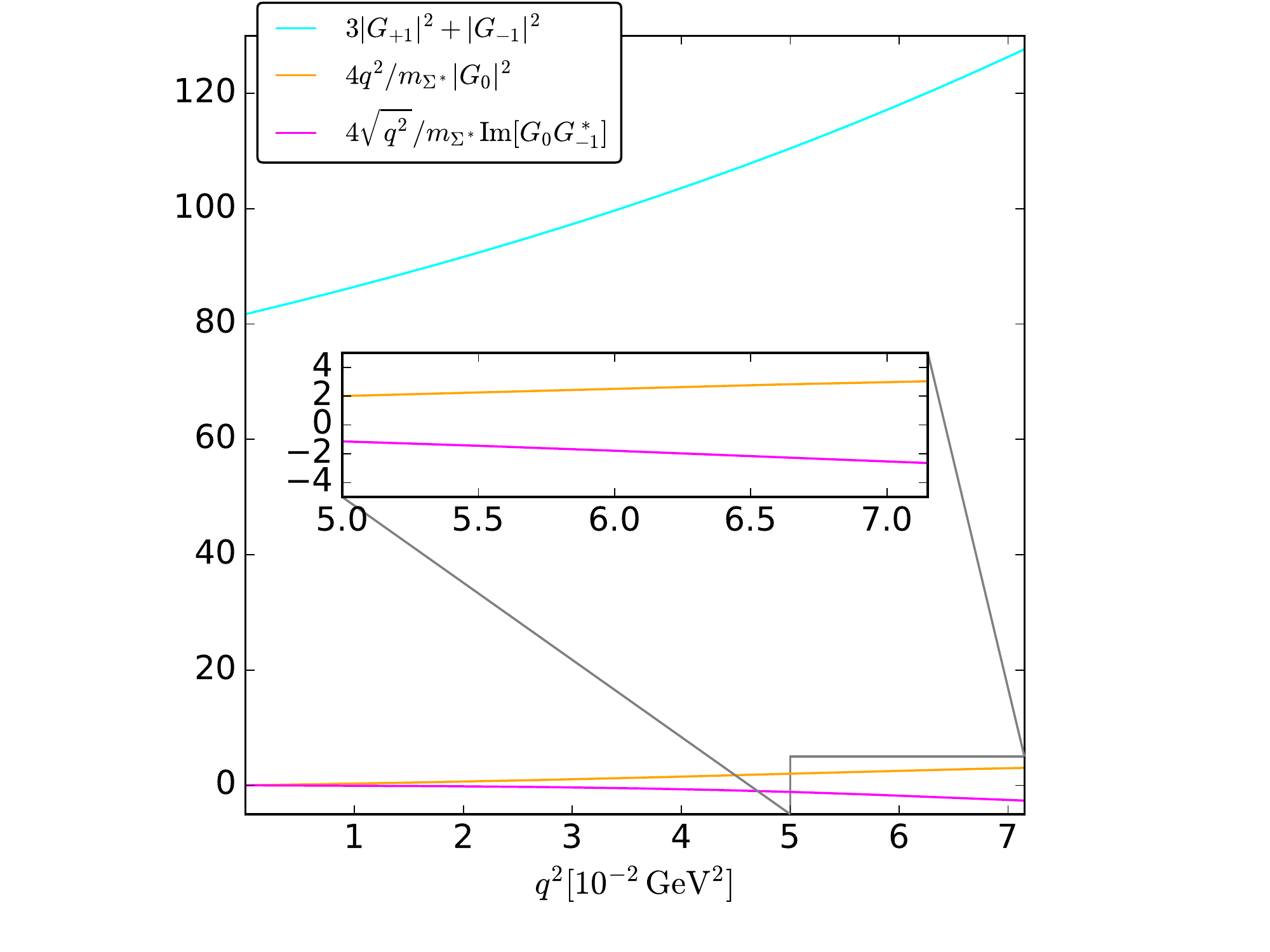}
  \caption{A comparison of the three combinations of TFFs in front of the trigonometric functions in \eqref{eq:M2av-withweak} for the $\Sigma^*\rightarrow p\pi^- e^+ e^-$ decay.}
  \label{fig:PreFact}
\end{figure}

In the Dalitz region it is also meaningful to plot the three combinations of TFFs that appear in front of the trigonometric functions in the four-body decay expression \eqref{eq:M2av-withweak}, in order to compare their magnitude. Fig.\ \ref{fig:PreFact} shows that one of them, the linear combination of $\lvert G_{+1}\rvert^2$ and $\lvert G_{-1}\rvert^2$ is dominant, making it very challenging to extract information on the individual TFFs. Yet with sufficient statistics and angular resolution for the four-body
decay $\Sigma^* \to \Lambda \, e^+ e^- \to p \pi^- \, e^+ e^-$ one might get access to the smaller form factor combinations.
The situation might be compared to the history of the experimental determination of the pion-to-photon TFF and the corresponding radius from Dalitz decays $\pi^0\rightarrow\gamma e^+e^-$ as documented in the citations of \cite{pdg}. Also there one had to establish first the mere existence of the decay, then the approximate agreement with the QED case and finally with much higher experimental efforts the existence of a non-trivial form factor. We are looking forward to this future endeavor for the hyperon sector.

\begin{acknowledgements}
  SL thanks Martin Hoferichter and Bastian Kubis for very valuable discussions on anomalous thresholds.
\end{acknowledgements}

\appendix

\section{Meson vs baryon dynamics}
\label{sec:disc}

The purpose of this appendix is to discuss the different physical aspects that are contained in a dispersive determination of the
low-energy TFFs. As an integral part of the main text it might distract the reader too much from the presentation of
conceptual developments and results. Therefore we dedicate this appendix to this discussion.

\begin{figure}[H]
  \centering
  \includegraphics[keepaspectratio,width=0.25\textwidth]{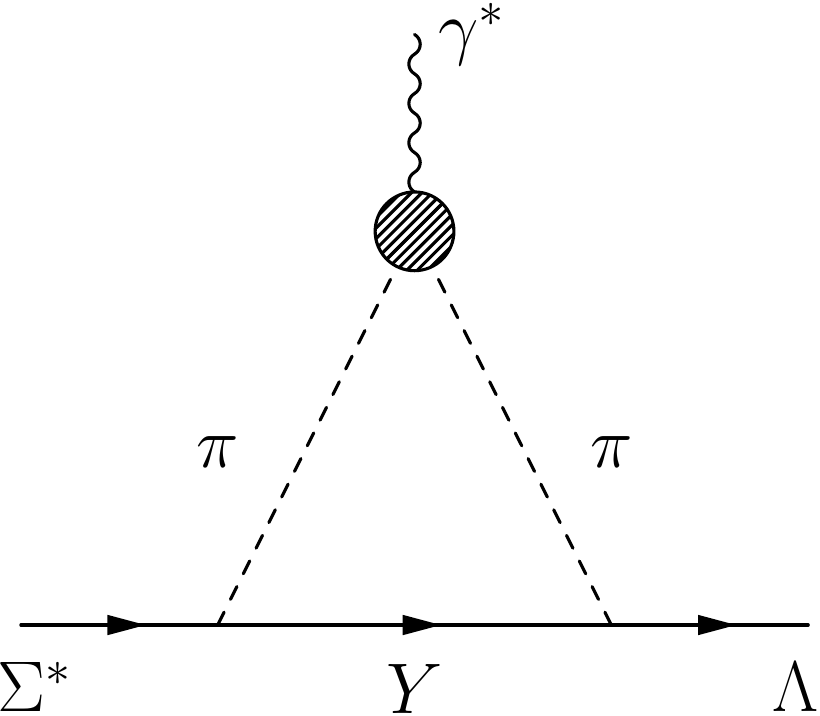}
  \hfill
  \includegraphics[keepaspectratio,width=0.25\textwidth]{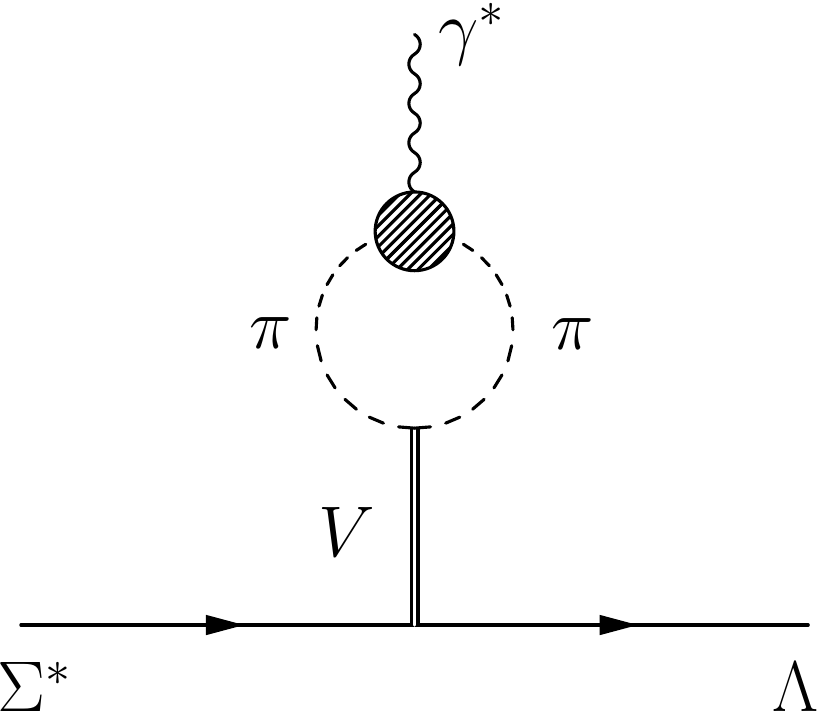}
  \hfill
  \includegraphics[keepaspectratio,width=0.25\textwidth]{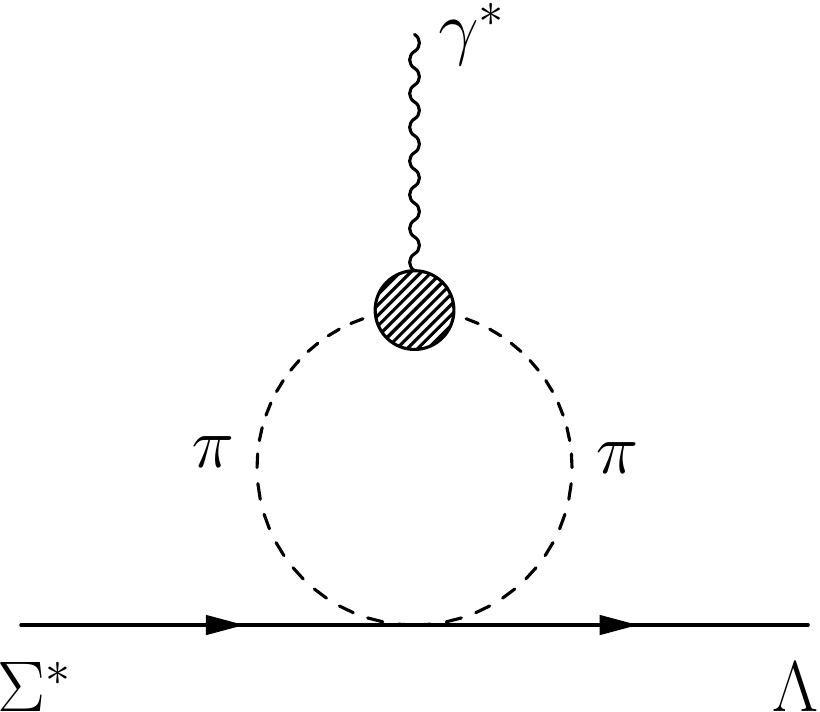}
  \caption{One-loop diagrams contained in our approach. The shaded blob denotes the pion vector form factor. The first diagram
    leads to the amplitude $K$ in \eqref{eq:tmandel}.}
  \label{fig:one-loop-diag}
\end{figure}
To understand the physical content of our approach, it might be illuminating to study a form factor on the one-loop level. 
This is displayed in Fig.\ \ref{fig:one-loop-diag}. 
Before discussing these diagrams, we stress that the dispersive approach contains more than these one-loop diagrams by including
in \eqref{eq:tmandel} the complete rescattering of pions via their measured pion phase shift. 
The first diagram in Fig.\ \ref{fig:one-loop-diag} displays the exchange of a hyperon $Y$ in the crossed channels. 
The second diagram shows the formation of a vector meson $V$. The third diagram contains a contact interaction between the 
hyperons and the pions. A contact interaction is without structure. It contributes with a polynomial to the hyperon-pion 
scattering amplitude. Thus the contact interaction provides a contribution to the polynomial $P$ in \eqref{eq:tmandel}. For 
the following discussion we call this contribution $P_c$. 

Next we want to specify the relevant exchange hadrons $Y$ and $V$. If such a hadron is very heavy, its pole and cut structures 
caused by its propagator are not resolved. 
It contributes effectively like a contact interaction. Thus what is not covered (at the one-loop level) 
by the third diagram of Fig.\ \ref{fig:one-loop-diag} are exchanges of light hadrons. Concerning the baryon exchange diagrams,
we have included explicitly the relevant lightest baryon states from the octet and decuplet. We call the impact of these 
processes on the form factors the ``aspect of baryon dynamics''. Below we will show a calculation that focuses only on this 
aspect. The second and third diagram of Fig.\ \ref{fig:one-loop-diag} couple the external baryons directly to mesons. Therefore
we call the impact of these processes on the form factors the ``aspect of meson dynamics''. 
This part might be linked to the notion of vector meson dominance \cite{sakuraiVMD}. Below we will also show a calculation 
that focuses only on this aspect. 

\begin{figure}[H]
  \centering
  \includegraphics[keepaspectratio,width=0.3\textwidth]{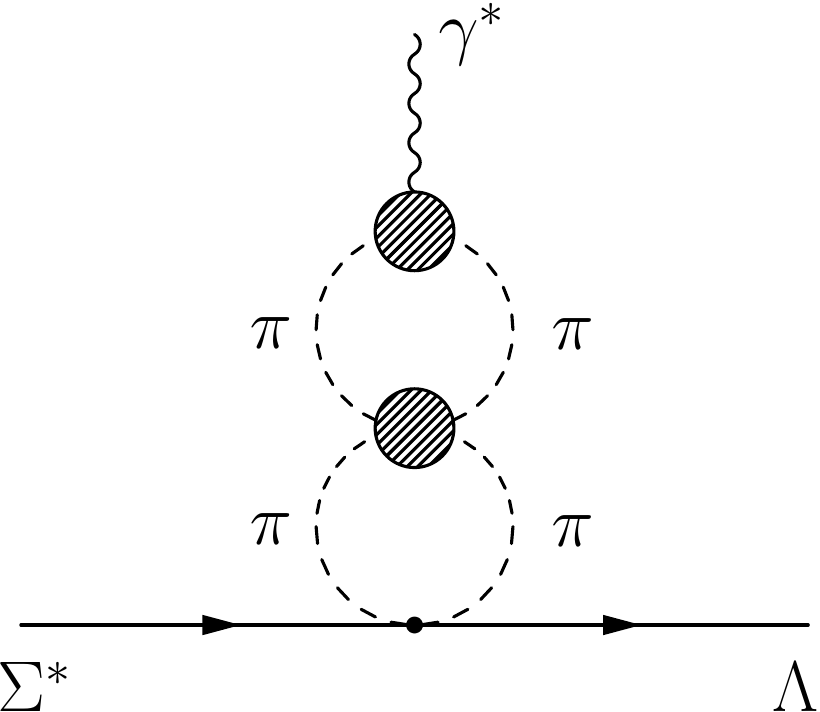}
  \caption{Diagrammatic representation of all processes that do not contain the cross-channel exchange of light baryons. The 
    shaded blob with four pion legs represents the S-matrix of pion scattering. The black dot contains the contact interaction
    of the third diagram of Fig.\ \ref{fig:one-loop-diag} and the strength mediated by the vector meson of the second diagram
    of Fig.\ \ref{fig:one-loop-diag}. This black dot leads to the polynomial $P$ in \eqref{eq:tmandel}.}
  \label{fig:overarch-meson}
\end{figure}

Finally, let us look at the second diagram of Fig.\ \ref{fig:one-loop-diag} in more detail. The dynamics of the 
lightest vector meson, 
the $\rho$-meson, is automatically contained in the measured pion phase shift because the $\rho$-meson couples 
essentially with 100\% \ 
to a two-pion state. Diagrammatically the second and the third diagram of 
Fig.\ \ref{fig:one-loop-diag} are covered by the diagram of Fig.\ \ref{fig:overarch-meson}. 

What is not automatically covered is the initial coupling strength with which the vector meson $V$ couples the pions to the 
hyperons. Schematically
\begin{eqnarray}
  && i g_{BV} \, \bar B \gamma_\mu \gamma_5 T_\nu V^{\mu\nu} + i G_V \, [u_\mu,u_\nu] \, V^{\mu\nu} \nonumber \\ && \to 
  \frac{g_{BV} \, G_V}{M_V^2} \, \bar B \gamma_\mu \gamma_5 T_\nu \, [u^\mu,u^\nu]  \,,
  \label{eq:VMD-lagr-schem}
\end{eqnarray}
which leads to 
\begin{eqnarray}
  \label{eq:PcV}
  P = P_c + P_V  \quad \mbox{with} \quad P_V \sim \frac{g_{BV} \, G_V}{M_V^2}  \,.
\end{eqnarray}
In Appendix \ref{sec:vmd} and in \cite{Ecker:1988te}, respectively, the flavor structure of \eqref{eq:VMD-lagr-schem} 
is specified, which is, however, of no concern for our qualitative discussion. 
We will show below in more detail how the dynamics contained in the second diagram of Fig.\ \ref{fig:one-loop-diag} emerges 
from the dispersive framework by translating and simplifying this framework to the vector meson dominance language. 

The result of the present discussion is that our dispersive framework contains all processes of Fig.\ \ref{fig:one-loop-diag} 
if the contact interaction strength $\sim P$ of Fig.\ \ref{fig:overarch-meson} is determined by a fit to experiment. Without 
further theory input, this needs to be done separately for each form factor. 
If we need to estimate the size of $P$ on the theory side, we must include the influence of vector mesons as shown 
in \eqref{eq:PcV} and carried out in Appendix \ref{sec:vmd}. In this context, we note that a pion-hyperon contact term of a 
given order in $\chi$PT leads to a contribution of the same order for the form factor. To be concrete, the 
TFFs $F_i$ of \eqref{eq:defTFF2} start at second, third and fourth chiral order for $i=1,2,3$, respectively.  
Correspondingly, to fully account for the contribution of the $\rho$-meson to the TFF $F_i$ requires a 
pion-hyperon contact interaction from the chiral Lagrangian of $(i+1)$th order. With our present NLO input, we have a full 
coverage of $F_1$ only. In turn, $F_1$ constitutes the leading contribution to the TFFs $G_{\pm 1}$ 
in \eqref{eq:F-1def}, \eqref{eq:F+1def}. In addition, our formalism contains the pertinent contributions to 
all TFFs from the baryon dynamics induced by the first diagram of Fig.\ \ref{fig:one-loop-diag}. 

Baryon form factors are influenced by meson dynamics and by baryon dynamics. Therefore it might
be illuminating to disentangle the meson and the baryon dynamics by switching off one of the two aspects. This will be discussed
in the following two subsections. Yet we would like to stress that both cases miss part of the physics. 

\subsection{Pure meson dynamics}
\label{subsec:mesonaspect}

To focus on the impact of pion rescattering, we switch off the aspect of baryon dynamics, i.e.\ we put $K \to 0$ for the
calculation of the hyperon-pion scattering amplitude $T$ in \eqref{eq:tmandel}. Consequently, we put $f \to 0$
in \eqref{eq:tanom} and \eqref{eq:Fanom-unsub}. 
For simplicity we use the unsubtracted dispersion relation \eqref{eq:dispbasic-unsub}.
Since we want to focus on the low-energy aspects, we also leave out the effective-pole term, i.e.\ $c_m \to 0$. Thus we obtain
finally 
\begin{equation}
  G_{\rm pure \,meson}(q^2) =  \frac{P}{12\pi} \, \int\limits_{4 m_\pi^2}^\infty \frac{\text{d}s}{\pi} \, 
     \frac{\Omega(s) \, p_{\rm c.m.}^3(s) \, F^{V*}_\pi(s)}{s^{1/2} \, (s-q^2-i \epsilon)}   \,.
      \label{eq:dispbasic-unsub-vmdlike}  
\end{equation}
Since the subtraction constant = contact interaction strength $P$ is just a number that does not influence the $s$ dependence
of the integrand, we have not specified the TFF by any label. 
We show its generic form in Fig.\ \ref{fig:pureMes}, in the unphysical region between the two-pion threshold and 1 GeV$^2$.
Obviously, the TFF displays the influence of the $\rho$-meson, i.e.\ the mesonic aspects are very well covered. An unphysical aspect emerges from the fact that the imaginary part of the TFF vanishes below the two-pion
threshold. In reality, the TFF is complex everywhere, since the $\Sigma^*$ is unstable. This is, however, hardly visible in the full results for $G_{-1}$ in Fig.\ \ref{fig:totGm1unphy}, since the imaginary part at $q^2\approx 4m_\pi^2$ becomes extremely tiny. 
From the comparison of Figs.\ \ref{fig:pureMes}, \ref{fig:totGm1unphy}, one sees that if one adjusted the $\rho$-peak of the imaginary parts to the same size, then the peak of the real part of the full calculation will be somewhat smaller than the one of the pure-meson calculation. 
Moreover at low energies the curvature in the real part of the full $G_{-1}$ is milder with respect to the pure-meson calculation.

A relation to strict vector meson dominance can be deduced from \eqref{eq:dispbasic-unsub-vmdlike}. Suppose that the width
of the vector meson is small. Essentially this means that the pion phase shift is zero below the vector meson mass and
$\pi$ above. This leads to $\Omega(s)=m_\rho^2/(m_\rho^2-s)$. With a slight refinement,
$m_\rho^2-s \to m_\rho^2-s - i m_\rho \Gamma_\rho$, one obtains $ \Omega(s) \, F^{V*}_\pi(s) \sim \delta(s-m_\rho^2)/\Gamma_\rho$
and therefore $G_{\rm pure \,meson}(q^2) \sim P/\Gamma_\rho \cdot 1/(m_\rho^2-q^2)$. The appearance of the ratio $P/\Gamma_\rho$ has
a natural interpretation: In a vector meson dominance picture the
contact term $\sim P$ for the hyperon-pion scattering amplitude emerges from integrating out the vector meson,
see \eqref{eq:VMD-lagr-schem}, \eqref{eq:PcV}. This leads to $P \sim g_{B\rho} \, G_V$.
On the other hand, in strict vector meson dominance, the coupling of the vector meson to the pions
must be adjusted such that the correct electric charge of the pion emerges that is independent of strong-interaction
coupling constants, i.e.\ $G_V \sim 1/F_V$ where $F_V$ denotes the coupling strength with which the photon couples to the vector
meson. Thus one finds $P/\Gamma_\rho \sim P/G_V^2 \sim g_{B\rho}/G_V \sim g_{B\rho} F_V$. This is exactly
what one expects as the coefficient of a form factor obtained in the vecor meson dominance framework, i.e.\ in full analogy to \eqref{eq:VMD-lagr-schem} one finds
\begin{eqnarray}
  && i g_{BV} \, \bar B \gamma_\mu \gamma_5 T_\nu V^{\mu\nu} + F_V \, f_+^{\mu\nu} \, V_{\mu\nu} \nonumber \\ && \to 
  i \, \frac{g_{BV} \, F_V}{M_V^2} \, \bar B \gamma_\mu \gamma_5 T_\nu \, f_+^{\mu\nu} \,.
  \label{eq:VMD-lagr-schem2}
\end{eqnarray}

\begin{widetext}
\hphantom{m}
  \begin{figure}[H]
  {\centering
     \begin{subfigure}[b]{0.49\textwidth}
        \includegraphics[width=\textwidth]{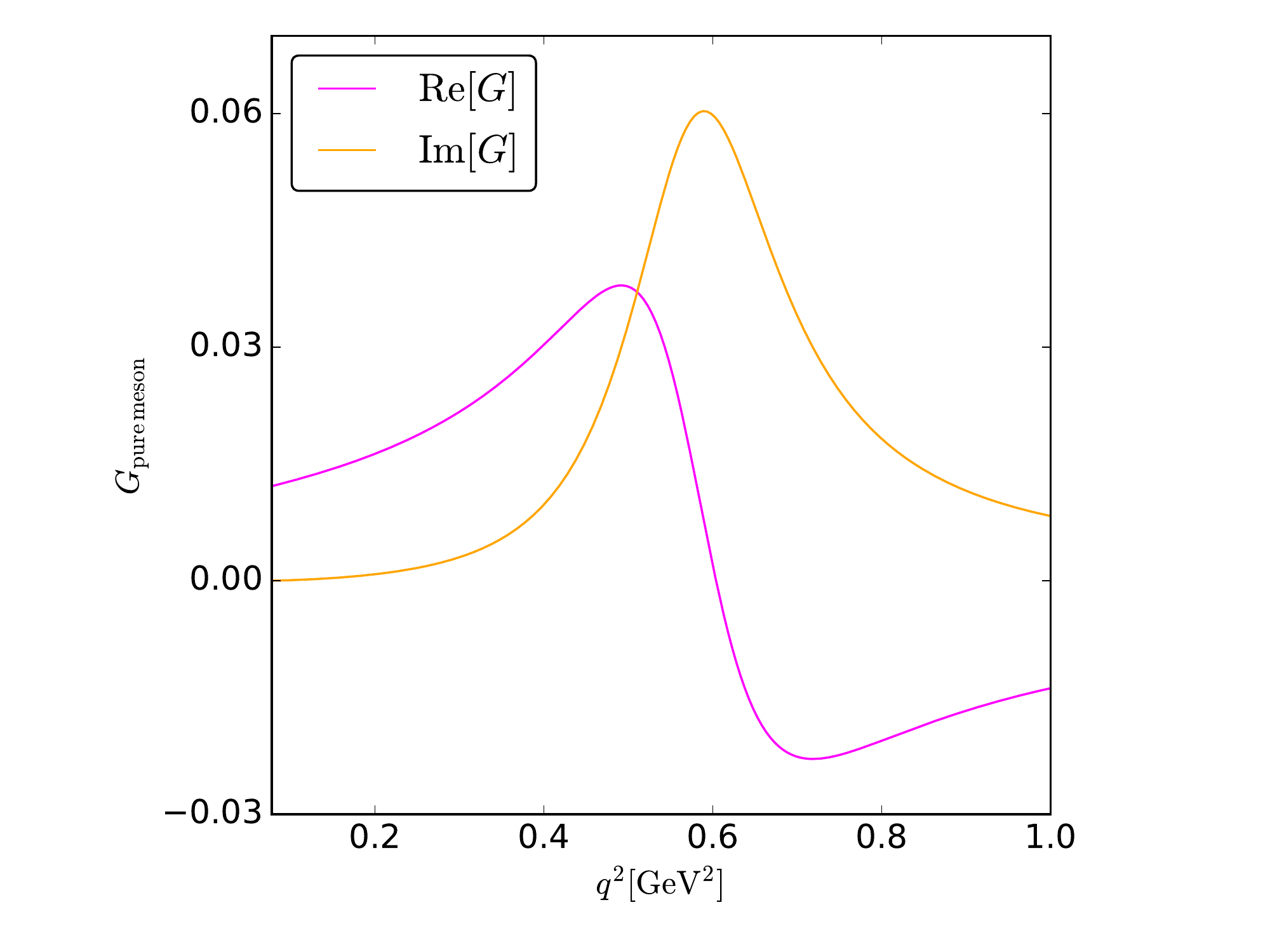}
        \caption{}
        \label{fig:pureMes}
    \end{subfigure}
 	\begin{subfigure}[b]{0.49\textwidth}
        \includegraphics[width=\textwidth]{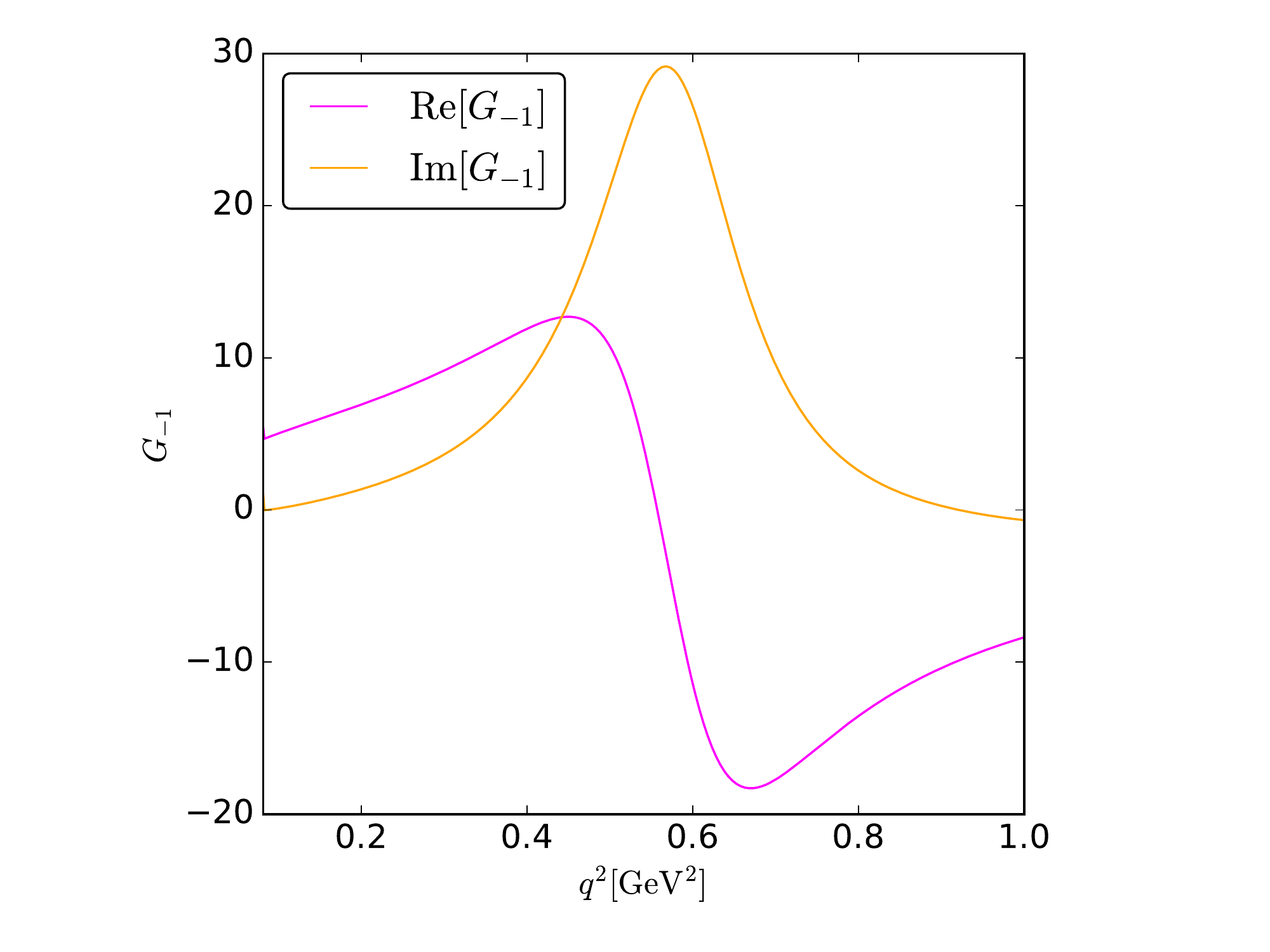}
        \caption{}
        \label{fig:totGm1unphy}
    \end{subfigure}}
  \caption{Pure mesonic contribution to the TFFs (a) in arbitrary units, compared with the full $G_{-1}(q^2)$ described by meson and baryon dynamics together (b), in the unphysical timelike region $4m_\pi^2<q^2<1$ GeV$^2$.}
  \label{fig:MesComp}
\end{figure}
\end{widetext}
 
Thus the dispersive
framework reproduces and refines the vector meson dominance aspects. The only
input one needs is the initial strength (contact interaction) with which the pions couple to the baryons. In the vector meson
dominance setup this is obtained by integrating out the vector meson. One can also rephrase it in the following way:
If one integrates out the vector meson for the interaction terms between the vector meson and the baryons and between the
vector meson and the photon, one obtains a photon-baryon coupling right away. The dispersive framework produces the same
with the pions as intermediate agents. Of course, the dispersive framework based on data for the pion phase shift and the pion
vector form factor is more acurate than the schematic and model-dependent vector meson dominance scenario but it covers
qualitatively the same physics. In addition, the dispersive framework presented in this paper contains also the
aspects of baryon dynamics that is completely missing in the vector meson dominance approach.

\subsection{Pure baryon dynamics}
\label{subsec:baryonaspect}

To focus on the impact of the processes where baryons are exchanged in the cross channel, we switch off the contact interaction 
and the pion rescattering. For simplicity we use an unsubtracted dispersion relation.
Thus we put $F_\pi^V \to 1$ in \eqref{eq:dispbasic-unsub} and \eqref{eq:Fanom-unsub}. Since we want to focus on the low-energy
aspects, we also leave out the effective-pole term, i.e.\ $c_m \to 0$. 
For the calculation of the scattering amplitude $T$ we put $P \to 0$ and $\delta \to 0$ in \eqref{eq:tmandel}. This also
implies that the anomalous piece vanishes. In simple terms: $T_m \to K_m$ in \eqref{eq:dispbasic-unsub}. This leads to
\begin{eqnarray}
  G^{\rm pure \, baryon}_m(q^2) &=& \frac{1}{12\pi} \, \int\limits_{4 m_\pi^2}^{\Lambda^2} \frac{\text{d}s}{\pi} \, 
   \frac{K_m(s) \, p_{\rm c.m.}^3(s)}{s^{1/2} \, (s-q^2-i \epsilon)}
    \nonumber \\  && {}
                     + \frac{1}{12\pi} \, \int\limits_{0}^1 \text{d}x \, \frac{ds'(x)}{dx} \, \frac{1}{s'(x)-q^2}
                     \nonumber \\  && \times
  \frac{f_m(s'(x)) \, s'(x)}{-4 \, (-\lambda(s'(x),m_{\Sigma^*}^2,m_\Lambda^2))^{3/2}}   \,. \phantom{mm}
  \label{eq:gpurebaryon}
\end{eqnarray}
%

In Fig.\ \ref{fig:pureBar} this contribution is plotted for the TFF $G_{-1}(q^2)$, in the range $-1<q^2 \,[\mathrm{GeV}^2]<1$.
As expected the form factor has an imaginary part for all values of $q^2$, even if very tiny for $q^2<0$. The baryon exchange diagrams contain the physical
aspect that the $\Sigma^*$ is unstable. What is missing, of course, is the influence of the $\rho$-meson, i.e.\ the mesonic
aspects. 
For comparison we show in Fig.\ \ref{fig:totGm1phys} also the complete result for $G_{-1}(q^2)$, taking into account the contributions of both meson and baryon dynamics, again across the space- and timelike regions. 
Note that the same quantity has been previously plotted (Fig.\ \ref{fig:totGm1unphy}) but in a different range. This time we include the negative $q^2$ physical region and focus on the region around $q^2=0$. There we notice in Fig.\ \ref{fig:BarComp}a
a steep rise in the real part, which is mitigated in Fig.\ \ref{fig:BarComp}b by the $\rho$-meson tail. 
In summary, even if the $\rho$-meson dictates in general the shape and size of the form factor, the low-energy behavior is
significantly influenced by the baryon dynamics. 
\begin{widetext}
\hphantom{m}
  \begin{figure}[H]
  \centering
     \begin{subfigure}[b]{0.49\textwidth}
        \includegraphics[width=\textwidth]{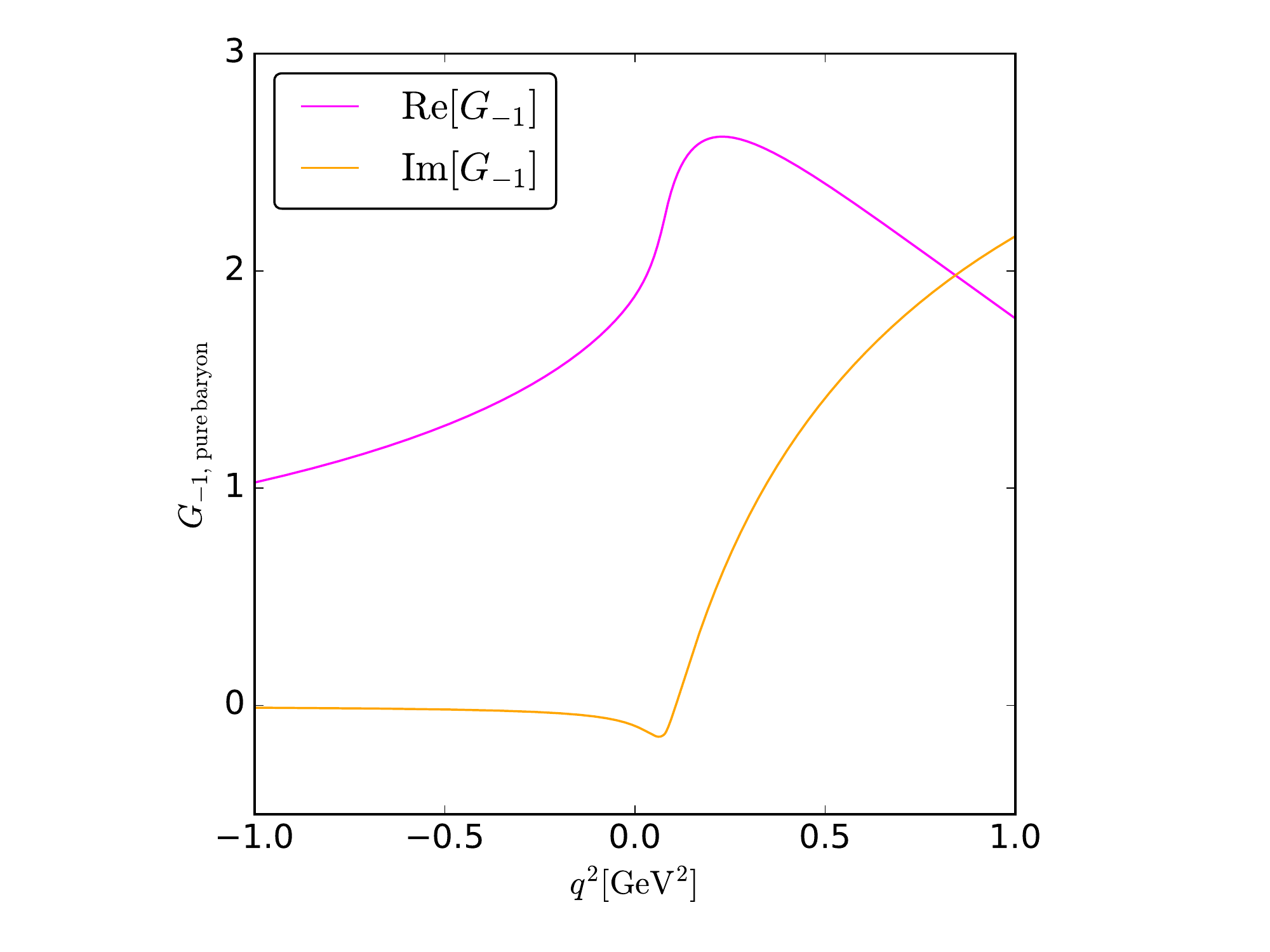}
        \caption{}
        \label{fig:pureBar}
    \end{subfigure}
 	\begin{subfigure}[b]{0.49\textwidth}
        \includegraphics[width=\textwidth]{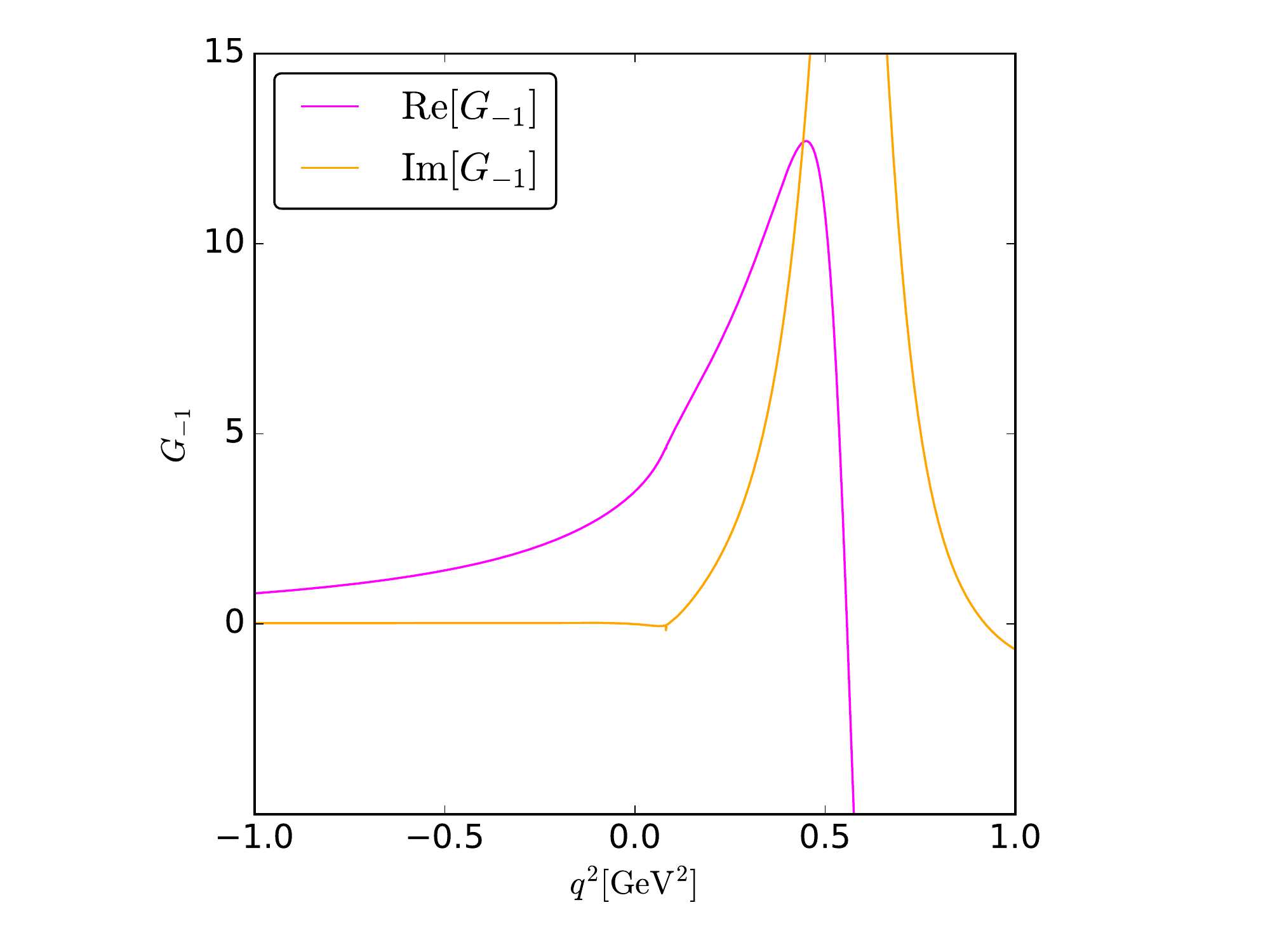}
        \caption{}
        \label{fig:totGm1phys}
    \end{subfigure}
  \caption{Pure baryonic contribution to $G_{-1}(q^2)$ (a) as compared to the full result for $G_{-1}(q^2)$ described by meson and baryon dynamics together (b), in the range $-1<q^2\,[\mathrm{GeV}^2]<1$.}
  \label{fig:BarComp}
\end{figure}
\end{widetext}

\section{Projector formalism for helicity amplitudes}
\label{sec:proj}

Spin-3/2 objects can be obtained from the coupling of spin-1/2 and spin-1 states. 
Thus we construct a spin-3/2 vector-spinor \cite{Rarita:1941mf} by 
\begin{eqnarray}
  \label{eq:spin32112}
  u^\mu(p,\sigma) = \sum\limits_{\rho,\lambda} \left(\frac32,\sigma \right\vert 1,\rho \,; \left. \frac12,\lambda \right)
  \, u(p,\lambda) \, \varepsilon^\mu(p,\rho)  \phantom{mm}
\end{eqnarray}
with a spin-1/2 spinor $u$, a spin-1 polarization vector $\varepsilon^\mu$ and a Clebsch-Gordan coefficient 
$(J,M \vert j_1,m_1;j_2,m_2)$. Here in slight contrast to the rest of this work the spin projections on a given quantization axis
(and not the helicities) are denoted by $\sigma$, $\lambda$ and $\rho$, respectively. Yet if one chooses the quantization axis 
along the flight direction (as we will do in a moment) then helicity and spin projection coincide.

It is useful to provide (\ref{eq:spin32112}) in an explicit form:
\begin{eqnarray}
  \label{eq:expl32spinor}
  u^\mu(p,\pm 3/2) &=& u(p,\pm 1/2) \, \varepsilon^\mu(p,\pm 1)  \,, \nonumber \\
  u^\mu(p,\pm 1/2) &=& \frac{1}{\sqrt{3}} \, u(p,\mp 1/2) \, \varepsilon^\mu(p,\pm 1) \nonumber \\
  && {} +\frac{\sqrt{2}}{\sqrt{3}} \, u(p,\pm 1/2) \, \varepsilon^\mu(p,0)   \,.
\end{eqnarray}

For the spin-1/2 spinors we use the conventions of \cite{pesschr}. For the spin-1 polarization vectors for massive 
states we provide here only their explicit form for the case 
where the $z$-direction constitutes both the spin quantization axis and the direction of motion of the 
particle \cite{Penner:1999jia,Stoica:2011cy}:
\begin{eqnarray}
  \label{eq:expl1pol}
  \varepsilon^\mu(p_z,\pm 1) = \frac{\mp 1}{\sqrt{2}} \, (0,1,\pm i,0) \,, \nonumber \\
  \varepsilon^\mu(p_z,0) = (p_z/m,0,0,E/m)
\end{eqnarray}
where $m$ denotes the mass of the particle and $E$ its energy. Note that the coefficient $\mp 1$ does not show up in the 
definitions of \cite{pesschr}. 

Irrespective of flight direction and spin quantization axis the spin-3/2 vector-spinors satisfy 
\begin{eqnarray}
  \label{eq:complspin32}
  \sum\limits_\sigma u_\mu(p,\sigma) \, \bar u_\nu(p,\sigma) = - (\slashed{p}+m) \, P^{3/2}_{\mu\nu}(p)
\end{eqnarray}
where $p^0=\sqrt{m^2+\vec p^2}$ denotes the energy of the particle described by the vector-spinor and $m$ its mass. 
The projector on spin 3/2 is defined by
\begin{eqnarray}
  \label{eq:defproj32aa}
  P^{3/2}_{\mu\nu}(p) := g_{\mu\nu} - \frac13 \, \gamma_\mu \gamma_\nu 
    - \frac{1}{3 p^2} \, (\slashed{p} \, \gamma_\mu \, p_\nu + p_\mu \, \gamma_\nu \, \slashed{p})  \,. \phantom{mm}
\end{eqnarray}
Note that for (\ref{eq:complspin32}) the scalar product $p^2$ appearing in (\ref{eq:defproj32aa}) can be replaced by $m^2$.

For our Lagrangian (\ref{eq:baryonlagr}) a spin-3/2 (vector-spinor) field $\psi_\mu(x)$ has the following 
propagator \cite{deJong:1992wm,Hacker:2005fh}
\begin{eqnarray}
  \label{eq:defspin32prop1}
  \langle 0 \vert {\rm T} \psi_\mu(x) \, \bar\psi_\nu(y) \vert 0 \rangle 
  = \int \frac{\text{d}^4 p}{(2\pi)^4} \, i \, S_{\mu\nu}(p) \, e^{-ip(x-y)} \phantom{mm}
\end{eqnarray}
with
\begin{eqnarray}
  \label{eq:defspin32prop2}
  S_{\mu\nu}(p) & := & - \frac{\slashed{p}+m}{p^2-m^2+ i\epsilon} \, P^{3/2}_{\mu\nu}(p) 
  + \frac{2}{3m^2} \, (\slashed{p}+m) \, \frac{p_\mu p_\nu}{p^2}   \nonumber \\ 
  && {} - \frac{1}{3m} \, \frac{p_\mu \, p^\alpha \, \gamma_{\alpha\nu} + \gamma_{\mu\alpha} \, p^\alpha \, p_\nu}{p^2}  \,. 
\end{eqnarray}
Note that for the propagator of (\ref{eq:defspin32prop1}), (\ref{eq:defspin32prop2}) the scalar product $p^2$ appearing 
in (\ref{eq:defproj32aa}) {\em cannot} be replaced by $m^2$. The propagator (\ref{eq:defspin32prop2}) describes propaging modes
of spin 3/2 and frozen modes of spin 1/2 \cite{deJong:1992wm}. 

The scattering amplitudes for the reaction $\Sigma^* \, \bar \Lambda \to \pi^+ \pi^-$ have the following structure:
\begin{eqnarray}
  \label{eq:genstrucscatt}
  \bar v_\Lambda(p_\Lambda,\lambda) \, M_\mu(p_{\Sigma^*},p_\Lambda,k) \, u^\mu_{\Sigma^*}(p_{\Sigma^*},\sigma)
\end{eqnarray}
with $k:=p_{\pi^+}-p_{\pi^-}$. Feynman rules can provide a quite lengthy expression for the spinor $4\times 4$ matrix $M^\mu$. 
Therefore we aim at a projector formalism where (\ref{eq:genstrucscatt}) is related to scalar quantities $a_i$ and a pre-defined 
set of spinor objects such that only the scalar quantities depend on the explicit form of $M^\mu$, i.e. 
\begin{eqnarray}
  \label{eq:genM-Mi}
  \bar v_\Lambda \, M_\mu \, u^\mu_{\Sigma^*} = \sum\limits_i a_i \; \bar v_\Lambda \, M_i^\mu \, g_{\mu\nu} \, u^\nu_{\Sigma^*} \,.
\end{eqnarray}
The tasks are to construct a complete set of linearly independent structures $M_i^\mu$ and to find a convenient way to determine
$a_i$ from an arbitrary $M^\mu$. Such an endeavor is similar in spirit to \cite{Stoica:2011cy}.

Due to parity symmetry we can focus on the case where the $\bar \Lambda$ has positive helicity, $\lambda=+1/2$. 
Then we need four pre-defined spinor objects corresponding to the possible values for the helicity of the $\Sigma^+$ baryon,
$\sigma = +3/2, +1/2, -1/2, -3/2$. It is convenient to introduce the following four-vectors:
\begin{eqnarray}
  \label{eq:def-4-vec}
  && q := p_{\Sigma^*}+p_\Lambda \,, \qquad \bar k := p_{\Sigma^*}-p_\Lambda \,, \nonumber \\
  && r := \bar k - \frac{\bar k \cdot q}{q^2} \, q  \,, \qquad k_\perp := k - \frac{k \cdot r}{r^2} \, r \,.
\end{eqnarray}
In the center-of-mass frame with the three-momentum of the $\Sigma^*$ 
pointing in the $z$-direction and the reaction taking place in 
the $x$-$z$ plane one finds that $q$ has only a zeroth component, $r$ has only a third ($z$) component and $k_\perp$ 
has only a first ($x$) component. 

We are looking now for four independent structures of type $M^\mu$ in (\ref{eq:genstrucscatt}). 
In general, $M^\mu$ contains products of
$\gamma$ matrices and exactly one $\gamma_5$.\footnote{Note that 
alternatively to a $\gamma_5$ one might involve a Levi-Civita symbol. However, this can be related to one $\gamma_5$ and
a product of $\gamma$ matrices.} All $\gamma$ matrices that are contracted with $p_\Lambda$, $p_{\Sigma^*}$ or the spin-3/2 
spinor $u_{\Sigma^*}$ can be moved towards $\bar v_\Lambda$ or $u_{\Sigma^*}$ and eliminated by equations of motion. This results 
in structures with less many $\gamma$ matrices. If two $\gamma$ matrices are contracted with each other or with the very same 
four-momentum, then one can also simplify the expression. 

This whole procedure leaves us with four independent structures of $M^\mu$ type:
\begin{eqnarray}
  \gamma_5 \, k_{\perp}^\mu \,, \quad \gamma_5 \, p^\mu_\Lambda \,, \quad \slashed{k}_\perp \gamma_5 \, p^\mu_\Lambda \,,
  \quad \slashed{k}_\perp \gamma_5 \, k^\mu_\perp \,.
  \label{eq:indep1}
\end{eqnarray}
It is simpler, however, to use the following linear combinations:
\begin{eqnarray}
  M^\mu_1 & := & \left(q^2 - (m_{\Sigma^*}+m_\Lambda)^2 \right) \gamma_5 \, k_{\perp}^\mu 
  - m_{\Sigma^*} \, \slashed{k}_\perp \gamma_5 \, p^\mu_\Lambda \,,   \nonumber  \\
  M^\mu_2 & := & \gamma_5 \, p^\mu_\Lambda \,, \nonumber \\
  M^\mu_3 & := & \slashed{k}_\perp \gamma_5 \, p^\mu_\Lambda \,, \nonumber \\
  M^\mu_4 & := & \left(q^2 - (m_{\Sigma^*}-m_\Lambda)^2 \right) \slashed{k}_\perp \gamma_5 \, k_{\perp}^\mu 
  - m_{\Sigma^*} \, k_\perp^2 \gamma_5 \, p^\mu_\Lambda \,.  \nonumber \\ &&
  \label{eq:indep2}
\end{eqnarray}
They are constructed such that in the center-of-mass frame they satisfy 
\begin{eqnarray}
  \label{eq:ortho-i-sigma}
  \bar v_\Lambda(p_\Lambda,+1/2) \, M^\mu_i \, g_{\mu\nu} \, u^\nu_{\Sigma^*}(p_{\Sigma^*},\sigma) \sim \delta_{i \, i_{\Sigma^*}}
\end{eqnarray}
with $i_{\Sigma^*} := 5/2-\sigma$. In other words, each $M_i^\mu$ contributes only to one helicity amplitude. Thus the sum in 
(\ref{eq:genM-Mi}) reduces to only one term.  

The remaining task is to find the scalar quantity $a_i$ for a given $M^\mu$. What makes the task non-trivial is the fact that 
different $M^\mu$ lead to the same $a_i$ because of the equations of motion for $\bar v_\Lambda$ and $u^\nu_{\Sigma^*}$. Therefore 
we construct on- and off-shell projectors to decompose a completely general $M^\mu$. Since $M^\mu$ is a $4 \times 4$ spinor 
matrix with $\mu$ ranging from 0 to 3 we need in general a basis of 64 Lorentz-spinor structures. Due to parity symmetry 
we can restrict ourselves to 32 structures. For the first 4 terms we use
\begin{eqnarray}
  \label{eq:def-first4}
  T_\mu^i := P_{\rm on}^\Lambda \, M_i^\nu \, P_{\rm on}^{\Sigma^*} \, P^{3/2}_{\nu\mu}
\end{eqnarray}
introducing the projectors \cite{deJong:1992wm}
\begin{eqnarray}
  \label{eq:projonoff}
  P_{\rm on}^\Lambda & := & \frac{1}{2m_\Lambda} \, (m_\Lambda - \slashed{p}_\Lambda) \,, \quad
  P_{\rm off}^\Lambda := \frac{1}{2m_\Lambda} \, (m_\Lambda + \slashed{p}_\Lambda) \,, \phantom{m}  \nonumber \\
  P_{\rm on}^{\Sigma^*} & := & \frac{1}{2m_{\Sigma^*}} \, (m_{\Sigma^*} + \slashed{p}_{\Sigma^*}) \,,   \nonumber \\
  P_{\rm off}^{\Sigma^*} & := & \frac{1}{2m_{\Sigma^*}} \, (m_{\Sigma^*} - \slashed{p}_{\Sigma^*}) \,, \phantom{m}  \nonumber \\
  P^{1/2}_{\mu\nu} & := & \frac13 \, \gamma_\mu \gamma_\nu 
  + \frac{1}{3 p_{\Sigma^*}^2} \, (\slashed{p}_{\Sigma^*} \gamma_\mu \, g_{\nu\alpha} + g_{\mu\alpha} \, \gamma_\nu \, \slashed{p}_{\Sigma^*}) 
  \, p_{\Sigma^*}^\alpha \,, \nonumber \\
  P^{3/2}_{\mu\nu} & := & g_{\mu\nu}- P^{1/2}_{\mu\nu}  \,. 
\end{eqnarray}
The other 28 structures are obtained from (\ref{eq:def-first4}) by exchanging $P_{\rm on}^\Lambda$ by $P_{\rm off}^\Lambda$ 
and/or $P_{\rm on}^{\Sigma^*}$ by $P_{\rm off}^{\Sigma^*}$ and/or $P^{3/2}_{\nu\mu}$ by $P^{1/2}_{\nu\mu}$. We do not specify how we enumerate
these structures from $i=5$ to $i=32$ because we will not need them in the end. We also introduce the Dirac adjoint 
structures 
\begin{eqnarray}
  \label{eq:defbarT}
  \bar T_\mu^i := \gamma_0 \, (T_\mu^i)^\dagger \, \gamma_0  \,, \quad \mbox{for} \quad i=1,\ldots,32  \,.
\end{eqnarray}
Provided that the $T_\mu^i$ form 32 linearly independent structures we can decompose any $M_\mu$ as 
\begin{eqnarray}
  \label{eq:decompMgen32}
  M_\mu = \sum\limits_{i=1}^{32} a_i \, T_\mu^i
\end{eqnarray}
with
\begin{eqnarray}
  \label{eq:get-ai}
  a_i = \sum\limits_{j=1}^{32} (C^{-1})_{ij} \, {\rm Tr}(\bar T_\mu^j M^\mu)
\end{eqnarray}
and the $32 \times 32$ matrix $C$ with elements
\begin{eqnarray}
  \label{eq:defCmatr}
  C_{ij} := {\rm Tr}(\bar T_\mu^i \, T_\nu^j) \, g^{\mu\nu}  \,.
\end{eqnarray}
Here Tr denotes the spinor trace. 
We have checked explicitly that det$C \neq 0$ which shows that the 32 structures $T^i_\mu$ are linearly independent, 
i.e.\ form a basis to construct the most general $M_\mu$.

Inserting (\ref{eq:decompMgen32}) in (\ref{eq:genstrucscatt}) and using the equations of motion for the spinors shows
\begin{eqnarray}
  \label{eq:genM-Mifin}
  \bar v_\Lambda \, M_\mu \, u^\mu_{\Sigma^*} = \sum\limits_{i=1}^4 a_i \; \bar v_\Lambda \, M_i^\mu \, g_{\mu\nu} \, u^\nu_{\Sigma^*} \,.
\end{eqnarray}
Thus we only need to determine the four scalar quantities $a_i$ with $i=1,2,3,4$ from (\ref{eq:get-ai}). Since the projectors in
(\ref{eq:projonoff}) are pairwise orthogonal one finds 
\begin{eqnarray}
  \label{eq:ijnoneq}
  {\rm Tr}(\bar T_\mu^i T_\nu^j) \, g^{\mu\nu} = 0 \quad \mbox{for} \quad i>4, \; j=1,2,3,4 \,.
\end{eqnarray}
In addition, we have checked by an explicit calculation 
\begin{eqnarray}
  \label{eq:orthogT}
  C_{ij} \sim \delta_{i j} \quad \mbox{for} \quad i,j=1,2,3,4 \,. 
\end{eqnarray}
A result that one could have anticipated already from (\ref{eq:ortho-i-sigma}). 
Finally (\ref{eq:get-ai}) simplifies to
\begin{eqnarray}
  \label{eq:get-aifin}
  a_i = \frac{{\rm Tr}(\bar T_\mu^i M^\mu)}{C_i} 
\end{eqnarray}
with
\begin{eqnarray}
  \label{eq:defCi}
  C_i := {\rm Tr}(\bar T_\mu^i \, T_\nu^i) \, g^{\mu\nu}   \,.
\end{eqnarray}
Explicit expressions are given by
\begin{eqnarray}
  C_1 & := & \frac{k_\perp^2}{4 \, m_{\Sigma^*} \, m_\Lambda} \, \left((m_{\Sigma^*} + m_\Lambda)^2-q^2 \right) \, 
  \lambda(q^2,m^2_{\Sigma^*},m^2_\Lambda)   \,,  \nonumber \\
  C_2 & := & \frac{-1}{12 \, m^3_{\Sigma^*} \, m_\Lambda} \, \left((m_{\Sigma^*} - m_\Lambda)^2-q^2 \right) \, 
  \lambda(q^2,m^2_{\Sigma^*},m^2_\Lambda)    \,, \nonumber \\
  C_3 & := & \frac{k_\perp^2}{12 \, m^3_{\Sigma^*} \, m_\Lambda} \, \left((m_{\Sigma^*} + m_\Lambda)^2-q^2 \right) \, 
  \lambda(q^2,m^2_{\Sigma^*},m^2_\Lambda)    \,, \nonumber \\
  C_4 & := & \frac{-(k_\perp^2)^2}{4 \, m_{\Sigma^*} \, m_\Lambda} \, \left((m_{\Sigma^*} - m_\Lambda)^2-q^2 \right) \, 
  \lambda(q^2,m^2_{\Sigma^*},m^2_\Lambda)      \nonumber \\
  && 
  \label{eq:Cexpl}
\end{eqnarray}
with the K\"all\'en function defined in (\ref{eq:kallenfunc}). 
In the center-of-mass frame 
one finds 
\begin{eqnarray}
  \label{eq:kperpsq}
  k_\perp^2 = - 4 \, p_{\rm c.m.}^2 \, \sin^2 \theta
\end{eqnarray}
with the center-of-mass momentum of the pions $p_{\rm c.m.} := \sqrt{q^2-4m_\pi^2}/2$ and $\theta$ denoting the angle between
the three-momenta of $\Sigma^*$ and $\pi^+$. 

To summarize, for a given amplitude structure $M_\mu$ and a given helicity $\sigma$ we find in the center-of-mass frame
\begin{eqnarray}
  && \bar v_\Lambda(p_\Lambda,+1/2) \, M_\mu \, u^\mu_{\Sigma^*}(p_{\Sigma^*},\sigma)  \nonumber \\
  && = \frac{{\rm Tr}(\bar T_\alpha^i M^\alpha)}{C_i} \; 
  \bar v_\Lambda(p_\Lambda,+1/2) \, M_i^\mu \, g_{\mu\nu} \, u^\nu_{\Sigma^*}(p_{\Sigma^*},\sigma)  \phantom{mm}
  \label{eq:finvmmuumu}
\end{eqnarray}
with $i=5/2-\sigma$. Note in particular that in (\ref{eq:finvmmuumu}) there is {\em no} implicit summation over $i$, 
it is fixed by the choice of $\sigma$, the helicity of the $\Sigma^*$.

In the main text we have introduced reduced amplitudes (\ref{eq:def-redampl}) for the dispersive representation of the 
TFFs.
To make contact with these reduced amplitudes we present here the ratios 
\begin{eqnarray}
  && \frac{\bar v_\Lambda(-p_z,+1/2) \, M_1^\mu \, g_{\mu\nu} \, u^\nu_{\Sigma^*}(p_z,+3/2)}%
  {\bar v_\Lambda(-p_z,+1/2) \, \gamma_5 \, u^1_{\Sigma^*}(p_z,+3/2) \, p_{\rm c.m.} }  \nonumber \\
  && = -2 \, \sin \theta \left(q^2-(m_{\Sigma^*}+m_\Lambda)^2\right)  \,,  \nonumber \\[1em]
  && \frac{\bar v_\Lambda(-p_z,+1/2) \, M_2^\mu \, g_{\mu\nu} \, u^\nu_{\Sigma^*}(p_z,+1/2)}%
  {\bar v_\Lambda(-p_z,+1/2) \, \gamma_5 \, u^3_{\Sigma^*}(p_z,+1/2)\, p_{\rm c.m.} }    \nonumber \\
  && =  \frac{2 \, q^2}{m_{\Sigma^*}^2-m_\Lambda^2+q^2} \, \frac{p_z}{p_{\rm c.m.}} \,,  \nonumber \\[1em]
  && \frac{\bar v_\Lambda(-p_z,+1/2) \, M_3^\mu \, g_{\mu\nu} \, u^\nu_{\Sigma^*}(p_z,-1/2)}%
  {\bar v_\Lambda(-p_z,+1/2) \, \gamma_5 \, u^1_{\Sigma^*}(p_z,-1/2)\, p_{\rm c.m.} }   \nonumber \\
  && = -2 \, \sin \theta \; \frac{q^2-(m_{\Sigma^*}+m_\Lambda)^2}{m_{\Sigma^*}}
  \label{eq:threerat}
\end{eqnarray}
with $p_z :=\lambda^{1/2}(q^2,m^2_{\Sigma^*},m^2_\Lambda)/(2\sqrt{q^2})$ denoting the center-of-mass momentum 
of $\Sigma^*$ and $\bar \Lambda$. Note that for the $M_2$ case (non-flip amplitude) there will always be a 
factor $p_z \, p_{\rm c.m.}$ from the partial-wave projection of ${\rm Tr}(\bar T_\alpha^2 M^\alpha)$. 
Together with the last ratio on the right-hand side of
the corresponding equation in (\ref{eq:threerat}) this leads to an expression for the reduced amplitude without any square roots.

In practice the whole task of dealing with a Feynman scattering amplitude for given helicities is reduced to the 
calculation of one spinor trace ${\rm Tr}(\bar T_\alpha^i M^\alpha)$.


\section{Dispersive representations, cuts and anomalous thresholds}
\label{sec:anom-disp}

This appendix has two purposes. First, we provide a detailed discussion of the analytic structure of a scalar triangle diagram.
This resembles the first diagram shown in Fig.\ \ref{fig:one-loop-diag}, except that one deals with p-waves there and with
s-waves in the scalar case. But the appearance of anomalous thresholds has the very same pattern. Therefore we use the scalar
triangle as a test case to check that we include all bits and pieces in the correct way for our TFF calculations.
The second purpose is the derivation of \eqref{eq:tanom}, \eqref{eq:Fanom-unsub}, and \eqref{eq:Fanom}. 

\begin{widetext}
Consider the result of a triangle loop diagram \cite{Karplus:1958zz,tHooft:1978jhc,Lucha:2006vc,Hoferichter:2013ama}, 
\begin{eqnarray}
  C(s) &=& \frac{1}{i\pi^2} \int \! \text{d}^4l \, \frac{1}{(l^2-m_{\rm exch}^2)\left((l+p_1)^2-m_\pi^2\right)\left((l-p_2)^2-m_\pi^2\right)}  \,,
  \label{eq:triangle}
\end{eqnarray}
which can be calculated directly when rewritten as
\begin{eqnarray}
     C(s) & = & \int\limits_0^1 \! \text{d}x_1 \, dx_2 \, dx_3 \, \delta(1-x_1-x_2-x_3)
       \left[ x_1 x_3 m_1^2 + x_2 x_3 m_2^2 + x_1 x_2 s - x_1 m_\pi^2 - x_2 m_\pi^2 - x_3 m_{\rm exch}^2 \right]^{-1}
       \label{eq:CDirectCalc}
\end{eqnarray}
\end{widetext}
with $s:=(p_1+p_2)^2$, $m_1^2:=p_1^2$ and $m_2^2:=p_2^2$. We consider first the case that $m_{\rm exch}^2$ is large enough and 
$m_1^2$ and $m_2^2$ are small enough. A quantitative specification will follow later. 
In this case, the imaginary part of $C$ (for real values of $s$) is just given by 
cutting \cite{Cutkosky:1960sp} the two pion lines of the Feynman diagram. The result is
\begin{equation}
  \label{eq:imC}
  {\rm Im}C(s) = -\pi \, \frac{\sigma(s)}{\kappa(s)} \, \log\frac{Y(s)+\kappa(s)}{Y(s)-\kappa(s)} \, \Theta(s-4m_\pi^2)
\end{equation}
where 
\begin{eqnarray}
  \label{eq:defY}
  Y(s) := s + 2 m_{\rm exch}^2 - m_1^2 - m_2^2 - 2 m_\pi^2  \,,
\end{eqnarray}
\begin{eqnarray}
  \label{eq:defKacser}
  \kappa(s) := \lambda^{1/2}(s,m_1^2,m_2^2) \, \sigma(s) \,,
\end{eqnarray}
and 
\begin{eqnarray}
  \label{eq:defsigma}
  \sigma(s) := \sqrt{1-\frac{4m_\pi^2 }{s}}  \,.
\end{eqnarray}
We use the log and the square root function both with a cut on the real negative axis.

The triangle function $C$ can be represented by a dispersive integral 
in the variable $s$ ranging from the two-pion threshold to infinity (unitarity cut):
\begin{eqnarray}
  \label{eq:Cdispinitial}
  C(z) = \int\limits_{4m_\pi^2}^\infty \frac{ds'}{\pi} \frac{{\rm Im}C(s')}{s'-z} 
  = \int\limits_{4m_\pi^2}^\infty \frac{ds'}{\pi} \frac{\sigma(s') \, l(s')}{s'-z}
\end{eqnarray}
with 
\begin{eqnarray}
  \label{eq:defhlog}
  l(s) := -\frac{\pi}{\kappa(s)} \, \log\frac{Y(s)+\kappa(s)}{Y(s)-\kappa(s)}   \,.
\end{eqnarray}
Here $z$ is an arbitrary complex number that does not lie on the unitarity cut, i.e.\ $z \not\in [4m_\pi^2,+\infty[$. 
 
It should be possible to find a dispersive 
representation of the function $C$ for any values of the masses. But it is necessary to study the cut structure of the 
logarithm in \eqref{eq:defhlog}. 
If this cut intersects with the unitarity cut, one needs a proper analytic continuation of the logarithm along the unitarity cut
and one picks up an anomalous contribution. 

To understand these statements, we consider first the case where \eqref{eq:Cdispinitial} works. In this case, $l(s)$ from 
\eqref{eq:defhlog} is a smooth function along and in the vicinity of the unitarity cut. Concerning the function $\sigma(s)$, it 
has a cut for $s \in [0,4m_\pi^2]$. It is convenient to define a function that has a cut 
along the unitarity cut \cite{Moussallam:2011zg}:
\begin{eqnarray}
  \label{eq:defsigmahat}
  \hat\sigma(z) := \sqrt{\frac{4 m_\pi^2}{z}-1}  \,. 
\end{eqnarray}
For $s \in [4m_\pi^2,+\infty[$ it satisfies 
\begin{eqnarray}
  \label{eq:hatsigma-sigma}
  \hat\sigma(s \pm i \epsilon) = \mp i \sigma(s)  \,.
\end{eqnarray}

By construction, the function 
$C(z)$ is defined via \eqref{eq:Cdispinitial} in the whole complex plane except for the unitarity cut. 
This cut defines a second Riemann sheet. We construct 
a function $C^{\rm II}(z)$ that constitutes an analytic continuation of $C$ through the cut. 
For $s \in [4m_\pi^2,+\infty[$ we demand 
\begin{eqnarray}
  C^{\rm II}(s+i\epsilon) &\stackrel{!}{=}& C(s-i\epsilon) = C(s+i\epsilon) - 2i \sigma(s) \, l(s)  \nonumber \\
  &=& C(s+i\epsilon) + 2 \hat\sigma(s+i\epsilon) \, l(s+i\epsilon)  \,. 
  \label{eq:RiemannIIcond}
\end{eqnarray}
In the last step we have used \eqref{eq:hatsigma-sigma} and the assumption that $l$ is a smooth function around the 
unitarity cut. This assumption will be critically reviewed below. 

For the case at hand, we can use \eqref{eq:RiemannIIcond} 
to define an analytic continuation of $C$ on the second Riemann sheet:
\begin{eqnarray}
  \label{eq:defCIIRiemann}
  C^{\rm II}(z) := C(z) + 2 \hat\sigma(z) \, l(z)   \,.
\end{eqnarray}
The cut structure of $C^{\rm II}$ originates from the unitarity cut, from the additional cut of $\hat\sigma$ along the negative 
real axis and from the cut of the logarithm in the expression \eqref{eq:defhlog} for the function $l$. 
We note that the square root functions that define $\kappa$ in \eqref{eq:defKacser} and therefore enter \eqref{eq:defhlog} 
do {\em not} cause an additional cut because $l$ is an even function in $\kappa$. 

Let us first focus on the unitarity cut. For $s \in [4m_\pi^2,+\infty[$, we find
\begin{eqnarray}
  C^{\rm II}(s-i\epsilon) &=& C(s-i\epsilon) + 2 \hat\sigma(s-i\epsilon) \, l(s) \nonumber \\
  &=& C(s+i\epsilon)  \,.
  \label{eq:linkbackI}
\end{eqnarray}
Thus the unitarity cut connects just the two Riemann sheets. 

\begin{figure}[H]
  \centering
      \includegraphics[keepaspectratio,width=0.45\textwidth]{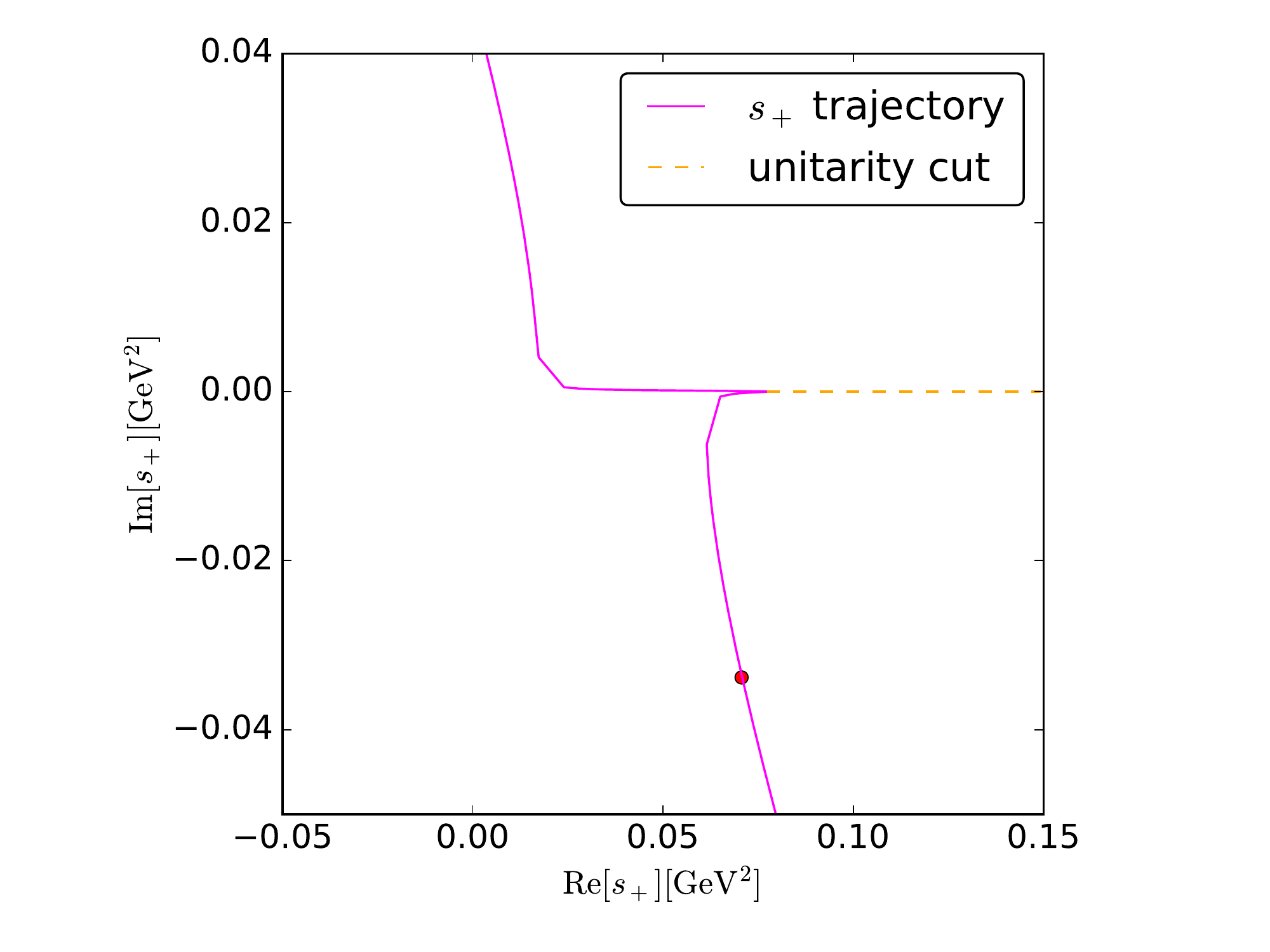}
  \caption{Real and imaginary part of $s_+$ obtained by varying $m_1^2$. The red dot corresponds to $m_1^2=m_{\Sigma^*}^2$, which is our case of interest.}
  \label{fig:sTrajec}
\end{figure}

Next we focus on the log function. 
The branch points of the logarithm in \eqref{eq:defhlog} are given by 
$Y^2(s)=\kappa^2(s)$. They are located at 
\begin{eqnarray}
  s_\pm &=& -\frac12 \, m_{\rm exch}^2 + \frac12 \left(m_1^2 + m_2^2 + 2 m_\pi^2 \right) \nonumber \\
        && {}
           - \frac{m_1^2 \, m_2^2 - m_\pi^2 \, (m_1^2 + m_2^2) + m_\pi^4}{2m_{\rm exch}^2} 
           \nonumber \\
        && {}
           \mp \frac{\lambda^{1/2}(m_1^2,m_{\rm exch}^2,m_\pi^2) \, \lambda^{1/2}(m_2^2,m_{\rm exch}^2,m_\pi^2)}{2m_{\rm exch}^2}
           \nonumber \\ 
        &=& \frac{1}{4 m_{\rm exch}^2} \Big[ (m_1^2-m_2^2)^2  \nonumber \\
        &&  \phantom{mmmm}  {}
           - \Big(\lambda^{1/2}(m_1^2,m_{\rm exch}^2,m_\pi^2) \nonumber \\
        &&  \phantom{mmmmmm}  {}
           \pm \lambda^{1/2}(m_2^2,m_{\rm exch}^2,m_\pi^2) \Big)^2 \Big] 
           \,.
  \label{eq:defspmapp}
\end{eqnarray}
The problem is that as a function of the masses, the values of $s_\pm$ move through the complex plane that constitutes the second
Riemann sheet. If any of the two branch points hits the unitarity cut then this branch point moves on the physical (=first) Riemann sheet.
To be specific, we take $s_+$ as the solution that has a positive imaginary part for small values of $m_1^2$. 
If one replaces $m_1^2$ by $m_1^2 +i\epsilon$ and follows the motion 
of $s_+$ for increasing values of $m_1^2$, 
then $s_+$ moves towards the real axis and could intersect with the unitarity cut.\footnote{For completeness we note that 
$s_-$ does not intersect with the unitarity cut and therefore does not enter the first Riemann sheet.}
Fig.\ \ref{fig:sTrajec} shows the trajectory of $s_+$ in the complex plane obtained by varying $m_1^2$, having fixed $m_{\mathrm{exch}}^2=m_\Sigma^2$ and $m_2^2=m_\Lambda^2$. Note the intersection with the unitarity cut, which implies that an additional cut must be located on the first Riemann sheet. The red dot indicates the actual position of $s_+$ for the physical choice $m_1^2=m_{\Sigma^*}^2$.
Indeed, for 
\begin{equation}
  \label{eq:intersect}
  m_1^2 + m_2^2 - 2 m_\pi^2 - 2 m_{\rm exch}^2 = 0  \quad \mbox{(cross point)}
\end{equation}
the two K\"all\'en functions in \eqref{eq:defspmapp} become identical and one finds at this point
\begin{eqnarray}
  \left. s_+ \right\vert_{\rm cross \, point} = 4m_\pi^2 \,, \quad 
  \left. \frac{\partial s_+}{\partial m_1^2} \right\vert_{\rm cross \, point} = 0 \,, \nonumber \\
  \left. \frac{\partial^2 s_+}{\partial (m_1^2)^2} \right\vert_{\rm cross \, point} 
  = \frac{2 m_\pi^2}{\lambda(m_2^2,m_{\rm exch}^2,m_\pi^2)}  \,.  \phantom{mm}
  \label{eq:valspders}
\end{eqnarray}
Therefore we obtain
\begin{eqnarray}
  && \left. s_+(m_1^2+i\epsilon,\ldots) \right\vert_{\rm cross \, point} \approx  \nonumber \\ 
  && \left[ s_+(m_1^2,\ldots) 
     + i\epsilon \, \frac{\partial s_+(m_1^2,\ldots) }{\partial m_1^2} \right.
     \nonumber \\ && \left. {}
    - \frac12 \epsilon^2 \, \frac{\partial^2 s_+(m_1^2,\ldots) }{\partial (m_1^2)^2} \right]_{\rm cross \, point}  \nonumber \\
   && = 4 m_\pi^2 - \epsilon^2 \, \frac{m_\pi^2}{\lambda(m_2^2,m_{\rm exch}^2,m_\pi^2)}  \,.
  \label{eq:spval}
\end{eqnarray}
In other words, the motion of $s_+$ just turns around (vanishing derivative) at the two-pion threshold. $s_+$ intersects with the 
unitarity cut if 
\begin{eqnarray}
  \label{eq:secondcond}
  \lambda(m_2^2,m_{\rm exch}^2,m_\pi^2) < 0  \,.
\end{eqnarray}

One can already see in the original expression \eqref{eq:defhlog} that something goes wrong if $m_1^2$ becomes so large that 
\eqref{eq:intersect} is satisfied. On the real axis, the log in \eqref{eq:defhlog} is ill-defined for $Y(s)=0$. From 
\eqref{eq:defY} we see that this zero of $Y$ is small as long as $m_1^2$ and $m_2^2$ are small and $m_{\rm exch}^2$ is large. 
But the zero of $Y$ reaches the unitarity cut, i.e.\ the branch point at the two-pion threshold for \eqref{eq:intersect}. 
For even larger values of $m_1^2$, i.e.\ for
\begin{eqnarray}
  \label{eq:afterintersect}
  m_1^2 + m_2^2 - 2 m_\pi^2 - 2 m_{\rm exch}^2 > 0  \,,
\end{eqnarray}
one needs a smooth analytic continuation of the logarithm along the unitarity cut. Otherwise, 
the dispersive representation \eqref{eq:Cdispinitial} does not make sense. In addition, \eqref{eq:Cdispinitial} is incomplete,
because one has to circumvent also the branch point $s_+$, which is now on the physical Riemann sheet. It is convenient
to choose the branch that starts at $s_+$ such that it intersects with the unitarity cut just at its own branch point at the 
two-pion threshold \cite{Hoferichter:2013ama}. 

The two conditions for $s_+$ being located on the first Riemann sheet are \eqref{eq:secondcond} and \eqref{eq:afterintersect}.
The dispersive representation of the triangle function \eqref{eq:triangle} is then
given by 
\begin{eqnarray}
  C(s) &=& \frac{1}{2\pi i} \int \! \text{d}s' \, \frac{{\rm disc}C(s')}{s'-s} \nonumber \\
  &=& \frac{1}{2\pi i} \int\limits_{4m_\pi^2}^\infty \! \text{d}s' \, \frac{{\rm disc}_{\rm unit}C(s')}{s'-s} \nonumber \\ && {}
  + \frac{1}{2\pi i} \int\limits_0^1 \! \text{d}x \, \frac{\text{d}z(x)}{\text{d}x} \, \frac{{\rm disc}_{\rm anom}C(z(x))}{z(x)-s}
  \phantom{mm}
  \label{eq:Cdispfinal}
\end{eqnarray}
with
the straight-line path connecting $s_+$ and the two-pion threshold,
\begin{eqnarray}
  \label{eq:defsx2}
  z(x) := (1-x) s_+ + x\, 4m_\pi^2 \,,
\end{eqnarray}
the function 
\begin{eqnarray}
  \label{eq:discanomC}
  {\rm disc}_{\rm anom}C(z) = -\frac{4 \pi^2 i}{(-\lambda(z,m_1^2,m_2^2))^{1/2}}  \,,
\end{eqnarray}
and a piecewise defined function given by (cf.\ \eqref{eq:imC})
\begin{eqnarray}
  && \frac{{\rm disc}_{\rm unit}C(s)}{2i} = \\
  && -\pi \, \frac{\sigma(s)}{\kappa(s)} \, \left[ \log\frac{Y(s)+\kappa(s)}{Y(s)-\kappa(s)} 
     + 2\pi i \Theta\left( (m_1-m_2)^2-s \right)  \right]  \nonumber 
 \end{eqnarray}
 for $\lambda(s,m_1^2,m_2^2) > 0$ while it is given by 
\begin{eqnarray}
  && \frac{{\rm disc}_{\rm unit}C(s)}{2i} = \nonumber \\
  && -2\pi \frac{\sigma(s)}{\tilde \kappa(s)} \, \left[ \arctan\frac{\tilde\kappa(s)}{Y(s)} + \pi \Theta(-Y(s)) \right]
  \label{eq:imCtrue}
\end{eqnarray}
for $\lambda(s,m_1^2,m_2^2) < 0$.
This function is continuous along the 
unitarity cut except if $s=(m_1-m_2)^2$ lies on the cut; there one has an integrable divergence.
We have introduced 
\begin{eqnarray}
  \label{eq:deftildekacser}
    \tilde\kappa(s) := (-\lambda(s,m_1^2,m_2^2))^{1/2} \, \sigma(s) \,.
\end{eqnarray}

In Fig.\ \ref{fig:direct} the real and imaginary part of the triangle function \eqref{eq:triangle} are plotted using $m_1=m_{\Sigma^*}$, 
  $m_{\rm exch}=m_\Sigma$ and $m_2=m_\Lambda$.
We have checked that the dispersive representation \eqref{eq:Cdispfinal} for $s+i\epsilon$ with arbitrary real $s$ fully agrees with the direct calculation \eqref{eq:CDirectCalc}. We want to stress that ignoring the integration along the anomalous cut produces a very incomplete result, shown in Fig.\ \ref{fig:dispWrong}. 

Having established the correct analytic structure, we leave the case of the scalar triangle behind and turn to our TFFs, which
have a different partial-wave structure and include the full pion rescattering.

\begin{widetext}
  \hphantom{m}
\begin{figure}[H]
  \centering
  \begin{subfigure}[b]{0.45\textwidth}
        \includegraphics[width=\textwidth]{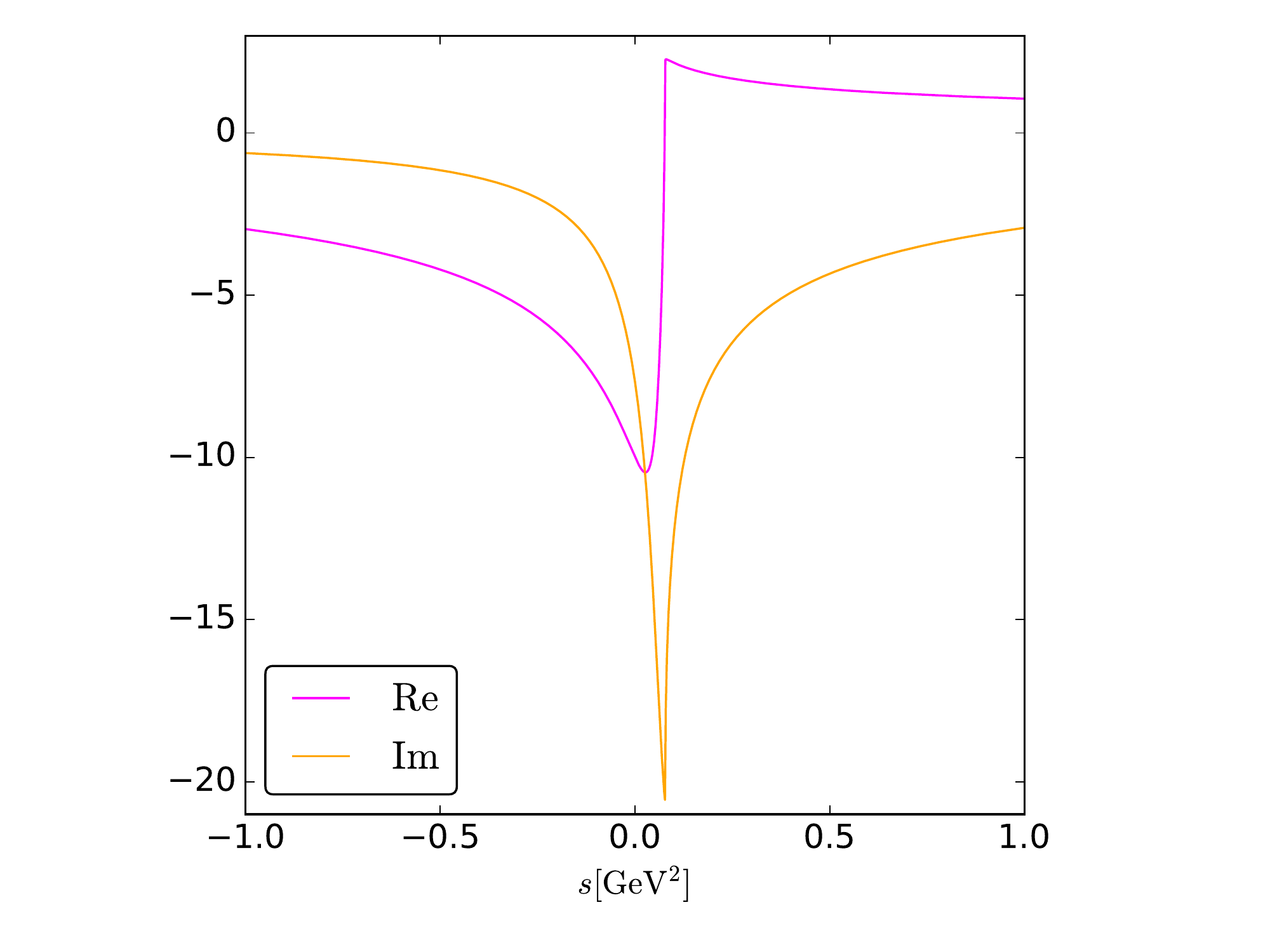}
        \caption{}
        \label{fig:direct}
    \end{subfigure}
    ~ 
    \begin{subfigure}[b]{0.45\textwidth}
        \includegraphics[width=\textwidth]{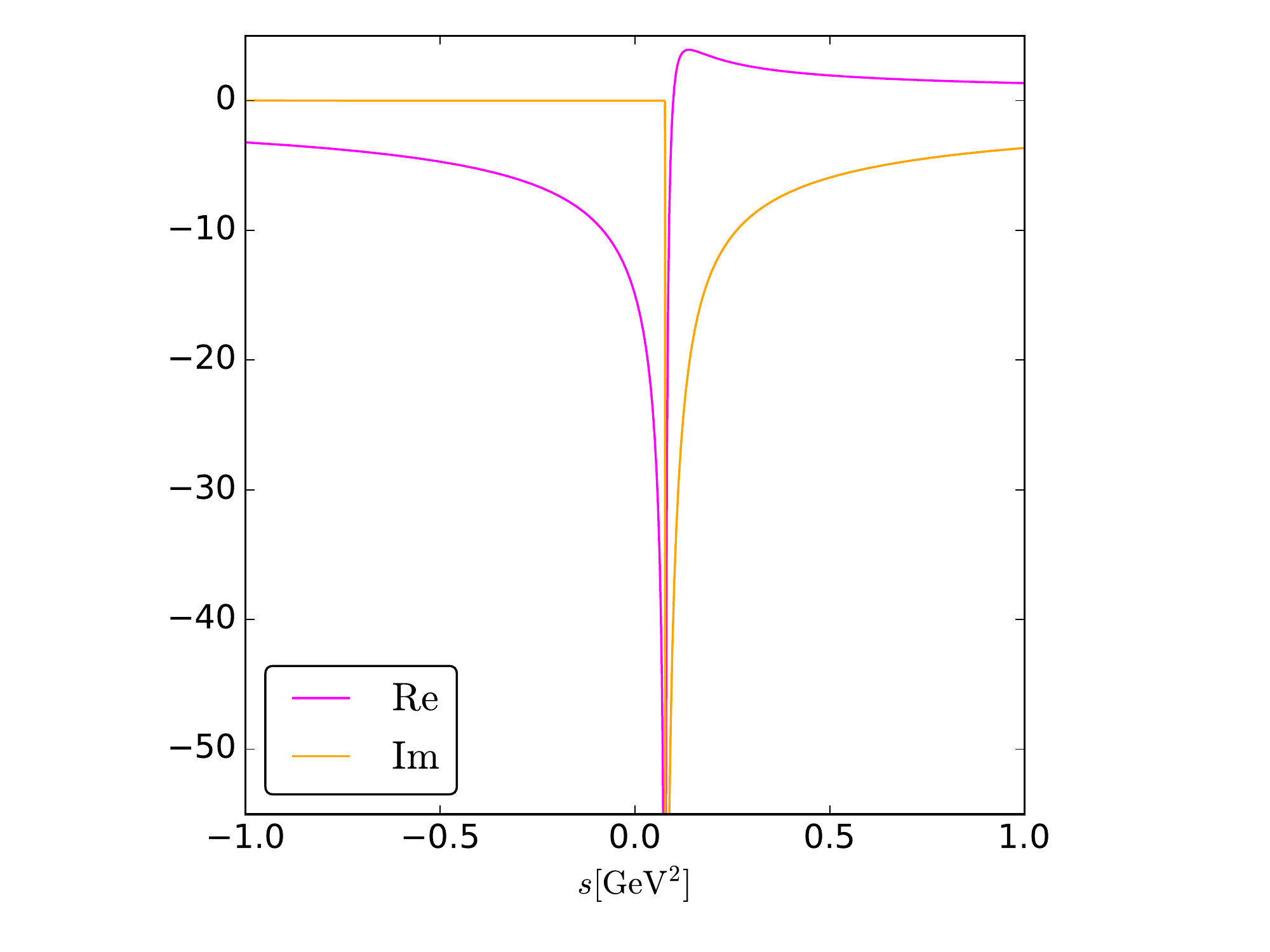}
        \caption{}
        \label{fig:dispWrong}
    \end{subfigure}
  \caption{Comparison between (a) the triangle function \eqref{eq:triangle}, obtained by either \eqref{eq:CDirectCalc} or \eqref{eq:Cdispfinal}, and (b) its incomplete dispersive representation \eqref{eq:Cdispinitial}, where only the unitarity cut has been taken into account, neglecting the presence of the anomalous cut. The masses involved here are $m_1=m_{\Sigma^*}$, 
  $m_{\rm exch}=m_\Sigma$ and $m_2=m_\Lambda$.}
  \label{fig:compLoop}
\end{figure}
\end{widetext}

For triangle diagrams with full two-pion rescattering we extend the usual formulae to allow for the presence of the anomalous
cuts. We introduce the values of a function $A$ to the left ($A_+$) and to the right ($A_-$) of a (directed) cut line. The 
discontinuity of $A$ is then defined by
\begin{eqnarray}
  \label{eq:defdiscgen}
  {\rm disc}A := A_+ - A_-  \,.
\end{eqnarray}
For a cut along the real axis this yields the well-known relations
\begin{eqnarray}
  {\rm disc}A(s) &=& A(s+i\epsilon)-A(s-i\epsilon) \nonumber \\
                 &=& A(s+i\epsilon) - A^*(s+i\epsilon) \nonumber \\
                 &=& 2i {\rm Im} A(s+i\epsilon) \,.
  \label{eq:discim}
\end{eqnarray}

The optical theorem that leads to \eqref{eq:dispbasic} and \eqref{eq:tmandel} generalizes to 
\begin{eqnarray}
  \label{eq:discFF}
  {\rm disc}F = 2i \frac{1}{24\pi} T_+ \sigma p^2_{\rm cm} F_{\pi -}^V
\end{eqnarray}
and
\begin{eqnarray}
  \label{eq:discT-K}
  {\rm disc}(T-K) = 2i T_+ \sigma t_-
\end{eqnarray}
with the p-wave pion scattering amplitude $t$. Along the unitarity cut, the amplitude $t$ is given by 
$t = \sin\delta \, e^{i\delta}/\sigma$. 

We recall how \eqref{eq:discT-K} is solved \cite{Omnes:1958hv}. The Omn\`es function is introduced as a solution of 
\begin{eqnarray}
  \label{eq:discOmnes}
  {\rm disc}\Omega = 2i \Omega_+ \sigma t_-  \,.
\end{eqnarray}
This allows to calculate
\begin{eqnarray}
  {\rm disc}\frac{T-K}{\Omega}
  & = & \frac{(T-K)_+ \Omega_- - (T-K)_- \Omega_+}{\Omega_+ \Omega_-} \nonumber \\
  &=& \frac{(T-K)_+ \Omega_- - (T-K)_+ \Omega_+ }{\Omega_+ \Omega_-}  \nonumber \\
  && {} + \frac{(T-K)_+ \Omega_+ - (T-K)_- \Omega_+}{\Omega_+ \Omega_-}  \nonumber \\
  & = & \frac{2 i K_+ \sigma t_-}{\Omega_-} \,.
  \label{eq:longcalc}
\end{eqnarray}
The product $K \sigma$ is essentially proportional to the discontinuity of the triangle function $C$. 
The proportionality factor $h$ is a rational 
function of $s$, i.e.\ has no cuts. With the previous construction of disc$C$ we have achieved that the two cuts (unitarity cut 
and anomalous cut) do not intersect. Therefore we can write for the discontinuity \eqref{eq:longcalc} along the unitarity cut
\begin{eqnarray}
  \label{eq:longcalcunit}
  {\rm disc}\frac{T-K}{\Omega} = \frac{2 i K \sigma t_-}{\Omega_-} = \frac{2i K \sin\delta}{\vert\Omega\vert}
\end{eqnarray}
because here $K$ is by construction a continuous function and $\Omega$ has the same phase as $t$. This leads to the standard 
dispersive part for $T-K$ explicitly given in \eqref{eq:tmandel}. Along the anomalous cut we have 
\begin{eqnarray}
  \label{eq:longcalcanom}
  {\rm disc}\frac{T-K}{\Omega} = 2 i \, h \, {\rm disc}_{\rm anom}C \, \frac{t_-}{\Omega_-} \,,
\end{eqnarray}
which leads to \eqref{eq:tanom}. 

Finally we have to solve \eqref{eq:discFF}. For the unitarity cut we can just integrate the right-hand side of \eqref{eq:discFF}.
For the anomalous cut we use \eqref{eq:discT-K} and find 
\begin{eqnarray}
  \label{eq:anomdiscF}
  {\rm disc}_{\rm anom}F = \frac{1}{24\pi} \, {\rm disc}_{\rm anom}(T-K) \, p^2_{\rm cm} \, \frac{F_{\pi -}^V}{t_-}  \,. 
\end{eqnarray}
Since we have a dispersive representation for $T-K$ in \eqref{eq:tmandel}, \eqref{eq:tanom} we just need to read off the 
discontinuity along the anomalous cut. This leads to \eqref{eq:Fanom}. 

If one compares the expressions \eqref{eq:tanom} and \eqref{eq:Fanom}, one notices that \eqref{eq:tanom} looks more complicated 
with $\Omega$ appearing outside and inside of the integral. Isn't it possible to write \eqref{eq:tanom} in a simpler way? 
After all, $\Omega$ is continuous along the anomalous cut. From \eqref{eq:longcalcanom} one sees that the discontinuity of 
$T-K$ along the anomalous cut is indeed just $2 i \, h \, {\rm disc}_{\rm anom}C \, t$. The same can be obtained from 
\eqref{eq:tanom}. But the expression \eqref{eq:tanom} inherits from $\Omega$ also a discontinuity along the unitarity cut. 
Therefore a direct dispersive representation of $T-K$ instead of the ratio $(T-K)/\Omega$ leads to an integral where in the 
integrand the integral of \eqref{eq:tanom} appears. For the form factor we have this situation of a double integral anyway in 
\eqref{eq:dispbasic} where the integral expressions for $T$ enter in the integrand. But for $T$ itself one can avoid the double
integral representation if one lives with $\Omega$ appearing outside and inside of the integrals.

\section{Estimate for the NLO four-point pion-baryon coupling constant}
\label{sec:vmd}

Ideally the low-energy constant $c_F$ from (\ref{eq:NLOtrans}) should be determined from experiment. To have a rough estimate
for its size we apply a vector-meson-dominance (resonance-saturation) assumption \cite{sakuraiVMD,Ecker:1988te,Ecker:1989yg,Meissner:1997hn}. 
To get a feeling for its accuracy we will make the same estimate for the octet sector. To this end, we consider the following 
part of the NLO Lagrangian of \cite{Holmberg:2018dtv}:
\begin{eqnarray}
  \label{eq:partNLO-4VMD}
  {\cal L}^{(2)}_{V} &:=& 
  i \, c_M \, ({\cal O}_{\mu\nu})^b_d \, (f_+^{\mu\nu})^d_b  + {\rm h.c.} \nonumber \\ 
  && {} + \frac14 \, c_F \, ({\cal O}_{\mu\nu})^b_d \, ([u^\mu,u^\nu])^d_b + {\rm h.c.} \nonumber \\
  && {} +
  b_{M,D} \, {\rm tr}(\bar{B}\{ f_+^{\mu\nu},\sigma_{\mu\nu} B\})
  \nonumber \\
  && {}+\frac{i}{2} \, b_{3,2} \, {\rm tr}(\bar{B}\{[u^\mu,u^\nu],\sigma_{\mu\nu} B\}) 
\end{eqnarray}
with 
\begin{eqnarray}
  \label{eq:def-calO}
  ({\cal O}_{\mu\nu})^b_d := \epsilon_{ade} \, \bar B^e_c \, \gamma_\mu \gamma_5 \, T_\nu^{abc}  \,.
\end{eqnarray}
Estimates for $c_M$ and $b_{M,D}$ have been provided in \cite{Holmberg:2018dtv}, based on fits to experimental data on 
radiative decays and magnetic moments, respectively: $\vert c_M \vert \approx 1.9\,$GeV$^{-1}$ and 
$b_{M,D} \approx 0.32\,$GeV$^{-1}$. 

Vector-meson dominance \cite{sakuraiVMD} implies that the coupling strengths of hadrons to two pions (in a p-wave) and to photons 
are correlated. In the $\chi$PT framework this might be rephrased as the statement that the two building 
blocks $[u^\mu,u^\nu]$ and $f_+^{\mu\nu}$ appear in a fixed combination, i.e.\ as the chiral field strength \cite{Ecker:1988te,Ecker:1989yg} 
\begin{eqnarray}
  \label{eq:supprDD2}
  \Gamma^{\mu\nu} := \frac14 \, [u^\mu,u^\nu] - \frac{i}{2} \, f_+^{\mu\nu}    \,.
\end{eqnarray}
Under this assumption, we obtain the following estimates: $c_F \approx -2 \, c_M$ and $b_{3,2} \approx b_{M,D}$. 
In \cite{Meissner:1997hn}, based on a resonance-saturation approach, the vector-meson contribution to the parameter $b_{3,2}$ 
(denoted by $b_{10}$ therein) has been estimated to $\approx$$\,0.5\,$GeV$^{-1}$, 
i.e.\ about 50\% larger than our value for $b_{M,D}$. Therefore, we use as an estimate:
\begin{eqnarray}
  \label{eq:estimate-cF}
  \vert c_F \vert = (4.8 \pm 1.2 ) \, \mbox{GeV}^{-1} \quad \mbox{and} \quad \frac{c_F}{c_M} < 0 \,. 
\end{eqnarray}

\bibliography{lit}{}
\end{document}